# Combining physical models with dynamically acquired experimental information for the optimization of multicomponent NASICON fast ionic conductors in a self-driving laboratory

Bernardus Rendy[1,2,3], Yuxing Fei[1,2], Tanjin He[4], Xiaochen Yang[1,2], Andrea Giunto[1,2], Lauren N. Walters[5,1], David Milsted[2], Hao Qiu[1], Matthew J. McDermott[2], Bin Ouyang[2,6], Yan Zeng[2,7,*], Gerbrand Ceder[1,2,3,*]

## Abstract

Elemental substitution within existing structural frameworks is a widely applied strategy for developing advanced materials. Yet, optimizing target properties while maintaining phase purity usually demands extensive trial-and-error, which becomes substantially inefficient when navigating a complex design space. Here, we introduce a strategy that simultaneously and dynamically assesses composition-dependent synthetic accessibility and target properties via aggregated cost functions that guide autonomous experimentation in a truly self-driving and self-learning mode. Specifically, we developed a cost-guided autonomous solid-state synthesis (CASS) framework and demonstrate its application in the discovery of Na superionic conductor (NASICON) solid electrolytes. CASS successfully optimizes ionic conductivity and phase purity, leading to the identification of 18 promising compositions in 78 trials conducted in an autonomous laboratory, the A-Lab. Among these, we identified two fast-conducting NASICONs yielding total (bulk) ionic conductivity of 0.7 (3.96) and 0.3 (3.17) mS/cm. The successful deployment of CASS reinforces the potential of coupling self-driving autonomous laboratories with physics-informed generative models to accelerate materials discovery. Moreover, the interpretability of the design factors, informed by outcomes from both successful and failed syntheses, enables model refinement and generation of new chemical insights.

[1] Department of Materials Science & Engineering, University of California Berkeley, Berkeley, CA 94720, USA
[2] Materials Sciences Division, Lawrence Berkeley National Laboratory, Berkeley, CA 94720, USA
[3] Energy Storage Research Alliance, Argonne National Laboratory, 9700 South Cass Avenue, Lemont, IL 60439, USA
[4] Argonne National Laboratory, 9700 South Cass Avenue, Lemont, IL 60439, USA
[5] Bakar Institute of Digital Materials for the Planet, UC Berkeley, Berkeley, CA 94720, USA
[6] Department of Chemical and Biomolecular Engineering, Vanderbilt University, Nashville, TN 37235, USA
[7] Department of Mechanical Engineering, Vanderbilt University, Nashville, TN 37235, USA
[*] Corresponding authors: gceder@berkeley.edu (G.C.), yan.zeng@vanderbilt.edu (Y.Z.)

## 1. Introduction

Elemental substitution constitutes a fundamental strategy in materials science, enabling the continuous tuning of functional properties to meet complex performance criteria[1]. Almost all technological materials in engineering have very specific compositions adjusted for target properties, even defining very different materials classes within a given solid-solution space. Examples can be found in the $LiNi_xMn_yCo_zO_2$ (NMC) family of Li-ion cathode materials[2], Ti[3] and Al[4] alloys, stainless steels[5], and others[6–11]. While novel compound discovery is often the focus of academic research, this compositional tuning to satisfy multiple (often opposing) property constraints forms the bread and butter of most technological materials innovation. This vast compositional flexibility is also essential for discovering the diverse portfolio of candidate materials required to address the multi-objective challenges in next-generation energy storage[12]. Specifically, for all-solid-state sodium-ion batteries, a sustainable alternative to lithium-ion battery technologies, the development of solid-state electrolytes (SSEs) must simultaneously optimize ionic conductivity, stability, and processability[13–19]. For this purpose, oxide-based phosphosilicate Na superionic conductors (NASICONs) with the general formula $Na_wM_2(PO_4)_{3-z}(SiO_4)_z$ offer a versatile SSE platform with good stability and wide electrochemical windows[20,21]. This structure allows for systematic property tuning via extensive cation (M) and polyanion ($PO_4$, $SiO_4$) substitution, followed by sodium (Na) content adjustment to maintain charge balance[21–26]. Given their structural flexibility to accommodate vast chemical variations, accelerating the discovery of optimal NASICONs requires exploiting established compositional design rules. Strategies such as aliovalent cation doping to optimize charge-carrier concentration, and anion substitution to tune the lattice bottleneck size, are well-established approaches for designing fast-conducting NASICON-type materials[27–29]. However, synthesizing these tailored NASICONs remains experimentally challenging due to the formation of impurity or competing phases that can impede ionic conductivity[29,30].

To tackle this synthesis challenge, materials discovery campaigns often rely on either static computational funnels[31] or Bayesian optimization (BO) for systematic screening and decision-making. Incorporating these methods has become increasingly important, particularly given the proven acceleration offered by self-driving and automated laboratories[25,32–40]. However, static funnels cascade candidates through rigid filters[25,36] without adapting to mid-campaign experimental feedback, thereby limiting the design space. Conversely, while BO has successfully tuned NASICON compositions[41,42], these campaigns are largely restricted to the close compositional vicinity of known materials. Within such constrained single-phase solid-solution regimes, elemental substitutions often yield smooth and stationary property landscapes well-suited to standard Gaussian process kernels[43,44]. However, navigating beyond these local regimes to map new phase boundaries exposes a highly rugged and non-stationary global property landscape[45], characterized by vast regions of competing phases punctuated by the abrupt emergence of the target phase. We demonstrate in this paper that by utilizing experimental feedback to update prior models in an autonomous way, it is possible to rapidly identify target phase domains[45–47].

To practically implement this dynamic prior approach and navigate the non-stationary synthesis landscapes, we established the Cost-guided Autonomous Solid-state Synthesis (CASS) active learning framework. CASS initializes the design space using model estimates of synthetic accessibility and ionic conductivity, which serve as the foundational prior model. To efficiently identify the optimal composition within a complex chemical space, the framework couples an evolutionary algorithm with a gradient-based optimizer. Rather than relying on static computational filters, CASS submits candidate compositions to an autonomous synthesis laboratory, the A-Lab[32], to acquire mid-campaign experimental feedback. Crucially, the framework utilizes experimental characterization results, specifically the synthetically accessible compositions inferred from observed lattice parameters and the presence and characterization of impurity phases, to actively recalibrate the design space. This closed-loop process iteratively steers the system toward synthesizable candidates that adhere to the targeted ionic conductivity design rule. We demonstrate that CASS successfully navigated high-dimensional chemical spaces and strategically designed and synthesized promising fast conducting NASICONs. This success lays the foundation for integrating additional design rules into the framework, enabling the simultaneous execution of multiple predictive models alongside probabilistic synthetic accessibility assessments.

## 2. Results

### 2.1. Overview of the CASS framework

Discovering functional materials in multicomponent chemical spaces requires decision-making strategies that can jointly account for target properties and synthetic accessibility. While it is possible to rely on models to predict properties, any optimization campaign needs to account for the physical reality of laboratory observations which may be inconsistent with prior models. To meet this need, we established the CASS framework, which combines prior physical models with iterative experimental feedback in a closed-loop design process. We use this framework here to identify NASICON solid electrolytes with high ionic conductivity. **Figure 1a** shows how CASS explores a predefined chemical space of NASICON by minimizing an aggregated cost function over candidate compositions. Each candidate is represented by its cation composition (excluding Na) and polyanion stoichiometry. To navigate the resulting search space efficiently, CASS combines global exploration by a genetic algorithm followed by gradient-based refinement for precise local optimization (**Methods**). The selected compositions are then synthesized through automated experimentation in the A-Lab[32], characterized by X-ray diffraction (XRD), and the resulting phase information is returned to the framework to inform the next design iteration.

**Figure 1b** shows how the aggregated cost function evolves over the course of the closed-loop search. The first design iteration is guided by static cost functions ($\Pi_{Static}$) derived from prior physical knowledge, specifically synthetic accessibility and ionic conductivity. In subsequent iterations, dynamic cost functions ($\Pi_{Dynamic}$) inferred from experimental characterization outcomes (phase identification and refined lattice parameters of the NASICON phase) are added to update the total cost landscape. This allows CASS to move from prior-guided exploration to feedback-informed redesign as the campaign progresses.

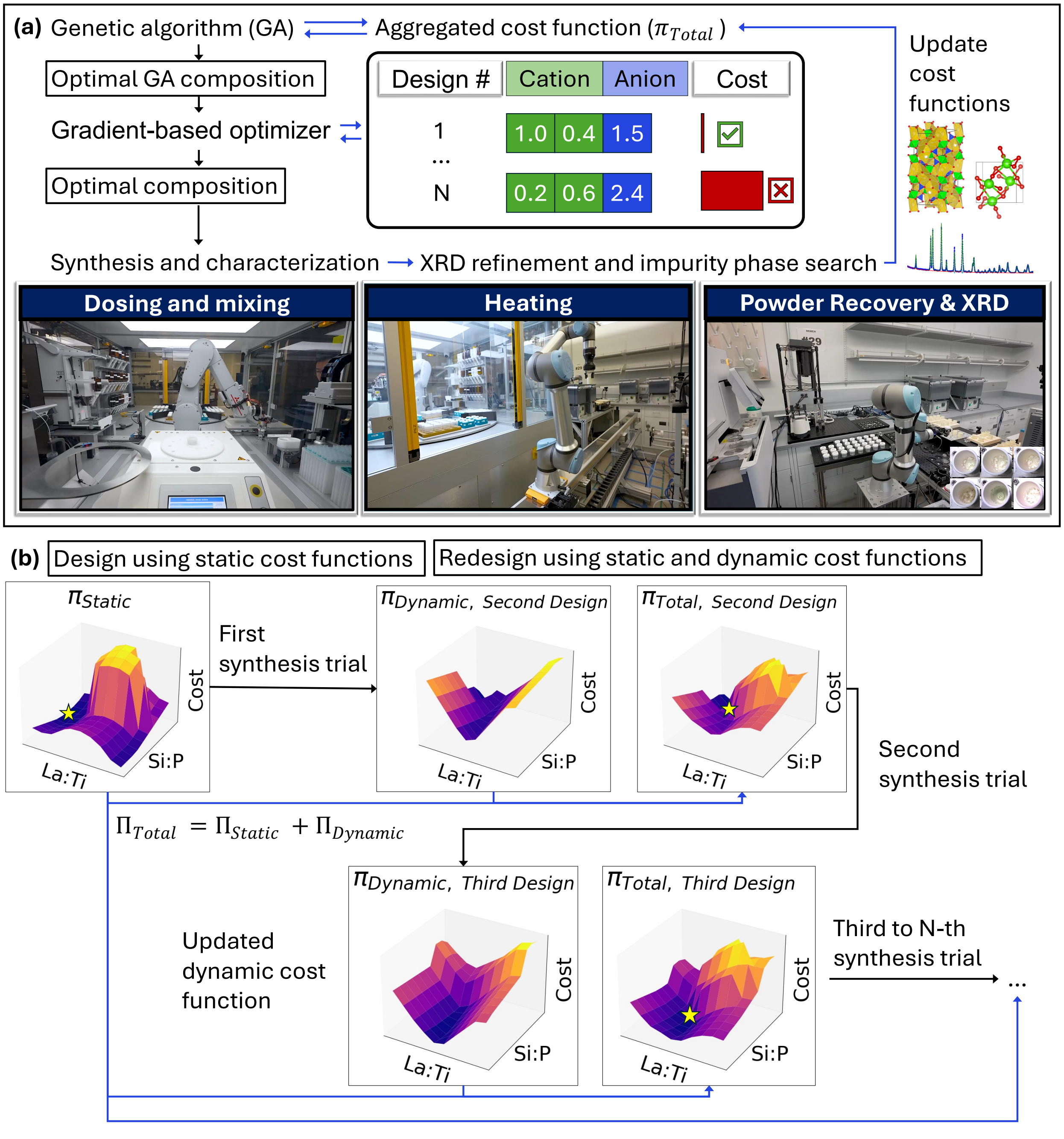

**Figure 1. CASS framework overview. (a)** Closed-loop workflow integrating design, synthesis, and XRD characterization. Feedback from phase analysis updates the dynamic components of the aggregated cost function for subsequent iterations. **(b)** Evolution of the cost function. Initial designs rely on static priors ($\Pi_{Static}$), while subsequent iterations incorporate characterization-derived dynamic costs ($\Pi_{Dynamic}$) into a total cost function ($\Pi_{Total}$). Color gradients indicate relative cost magnitude.

## 2.2. Static and dynamic cost functions for CASS

The aggregated cost function comprises three static terms and two dynamic terms. The static cost functions encode prior physical knowledge available before any experiment is performed and initialize the search toward compositions that are both synthetically accessible and likely to exhibit high ionic conductivity. Specifically, we defined static terms, explained in detail below, to represent thermodynamic stability, structural stability, and ionic conductivity, respectively.

Thermodynamic stability ($\pi_{Thermo}$) is configured to rule out compositions that do not form single-phase materials, classifying them as synthetically inaccessible. As shown in **Figure 2a**, $\pi_{Thermo}$ employs a linear support vector classifier (SVC) to distinguish synthetically accessible (A) from inaccessible (I) compositions. We utilized previously published stability descriptors capable of separating stable and unstable NASICONs[48] to build the feature space. To enable rapid design, we excluded the Ewald energy ($E_{Ewald}$) from this feature space, as its calculation can require the combinatorially expensive generation and energy evaluation of large, ordered supercells for certain compositions[48–50]. Because the exact thermodynamic limit for synthetic accessibility is unknown, and the prior knowledge is dynamically modified with observations in the synthesis campaign, we modified the synthesizability model from reference [48] to make it more inclusive, ensuring that all experimentally observed compositions are classified as synthetically accessible. This was done by retraining the linear SVC with the original methodology and the 3,881 computed NASICON compositions of reference [48]. The energy above hull ($E_{hull}$) limit was increased to a value, $E_{threshold}$, at which all 524 experimentally verified, literature-mined NASICON compositions were correctly classified as synthetically accessible (**Supplementary Figure 1b**). This value of $E_{threshold}$ was 62.22 meV/atom, which we adopted as the synthetic accessibility threshold for this study. We ensured that this reduced descriptor set and final updated classifier retained comparable accuracy, recall, and F1-scores **(see Methods)**. Consequently, the final decision boundary (**Figure 2a**) relies on four remaining descriptors from the prior study[48]: the electronegativity difference between Na and the cations ($X_M^{Na}$), cation radius deviation ($R_M^{STD}$), sodium content ($N_{Na}$), and anion charge deviation ($Q_A^{STD}$). Based on this boundary, the $\pi_{Thermo}$ cost is set to zero for synthetically accessible compositions and scales proportionally with the Platt-scaled instability probability[48,51] for synthetically inaccessible regions.

As low energy above hull is only a required condition but not a sufficient condition for synthesizability[52], we add more information through a structural stability ($\pi_{Distortion}$) cost function (**Figure 2b**) which enforces a ceiling on the local lattice distortion[53–55] quantified by the root-mean-square (RMS) variation of the cation sizes, $R_M^{STD}$, that tightens with increasing compositional complexity. This limit is crucial, as excessive distortion of the cation polyhedra imposes a severe enthalpic penalty that overcomes the system's entropic stabilization, destabilizing the NASICON framework and driving the formation of competing phases. Derived from the 524 experimentally verified compositions mined from the literature, this empirical boundary assigns zero cost to synthetically accessible (A) compositions and a linear penalty to those exceeding the threshold (synthetically inaccessible (I)).

The model for ionic conductivity ($\pi_{IC}$) (**Figure 2c**) defines a likely high ionic conductivity domain[25] (LHIC) surrogate model based on average metal cation radius ($\bar{R}_M$), sodium content ($N_{Na}$), and average polyanion center ion radius ($\bar{R}_A$), adapting data from 484 published NASICON composition-conductivity pairs[25]. Compositions within this domain incur zero cost, while outliers face a linear penalty proportional to their distance from the boundary.

To incorporate the feedback from experiments in this campaign, we additionally define two dynamic cost functions, one related to impurity phases, and the other related to the lattice parameters of the obtained NASICON phase, based on the outcome of prior synthesis trials. Impurity minimization ($\pi_{Impurity}$) constrains the design space by penalizing new candidate compositions if their concentration of ions (except Na) that are found in impurity phases exceeds that of the previous impurity-forming target composition (**Figure 2d**). For example, in the Na-Zr-La-Ti-Si-P-O NASICON system, detection of a $Na_3La(PO_4)_2$ impurity phase penalizes candidates with higher La or P contents than in the prior composition that generated the $Na_3La(PO_4)_2$ impurity. Such feedback allows the system to correct for any deficiencies in its prior model.

When a NASICON phase forms during a synthesis attempt, identifying its composition expands the known library of synthesizable NASICONs. To approximate this NASICON composition in a sample that is not phase-pure, we extract its lattice parameters by automated XRD refinement and map them to candidate solid-solution compositions using Vegard's approximation. In the subsequent iterations, the lattice-parameter cost function ($\pi_{LP}$) steers subsequent candidates toward compositions consistent with the experimentally realized NASICON phase (**Figure 2e**). $\pi_{LP}$ applies a linear penalty to deviations between the experimentally observed lattice parameters and those predicted by a fine-tuned model for the corresponding chemical space (**Methods**).

Consequently, the cost minima explicitly pinpoint candidate compositions whose predicted lattice parameters match the experimental NASICON phase.

These static and dynamic terms are combined into a single aggregated cost function that guides the global search toward synthetically accessible compositions with high ionic conductivity while incorporating feedback from prior synthesis trials. The total cost is constructed as a weighted sum of $w_1\pi_{Thermo} + w_2\pi_{Distortion} + w_3\pi_{IC} + w_4\pi_{Impurity} + w_5\pi_{LP}$, with weights ($w_i$) assigned empirically to ensure comparable magnitudes across the five terms (**Methods**).

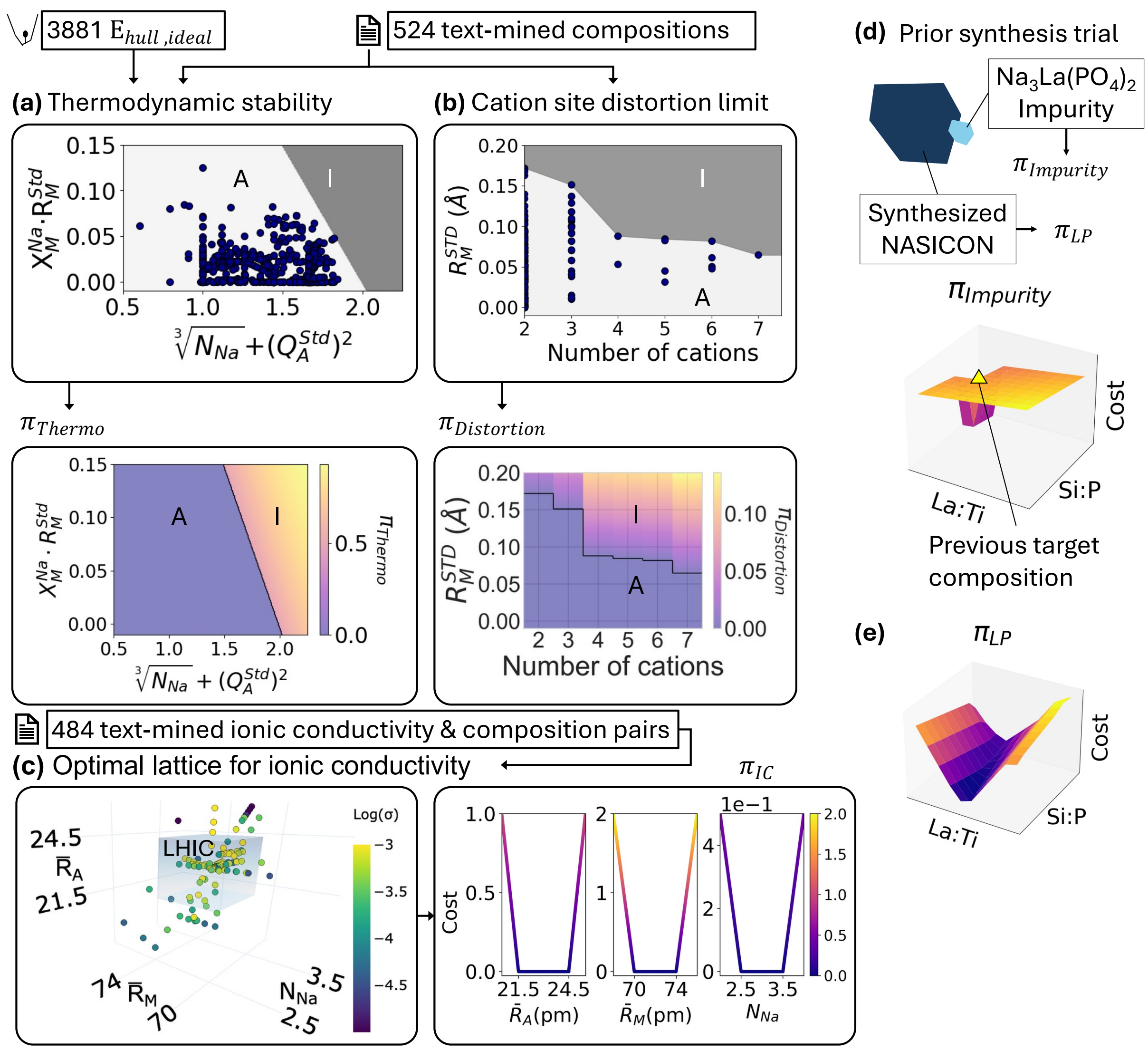

**Figure 2. Static and dynamic cost functions for CASS. (a-b)** Underlying model (top) and cost function (bottom) of the static synthetic accessibility costs. **(a)** Thermodynamic stability ($\pi_{Thermo}$)

utilizes a four-descriptor classifier to penalize compositions in synthetically inaccessible (I) regions based on their probability of instability, while applying no penalty to those within the accessible (A) region based on reference [48]. **(b)** Structural stability ($\pi_{Distortion}$) applies a linear penalty to candidates exceeding an empirical lattice distortion ($R_M^{STD}$) ceiling that tightens with compositional complexity. **(c)** Static ionic conductivity cost function ($\pi_{IC}$) penalizes deviations from a likely high-ionic conductivity (LHIC, inside) domain defined by $\bar{R}_M$, $N_{Na}$, and $\bar{R}_A$. **(d, e)** Dynamic costs derived from synthesis feedback. **(d)** Detection of an impurity phase (e.g., $Na_3La(PO_4)_2$) and NASICON primary phase in a prior trial parameterizes $\pi_{Impurity}$ and $\pi_{LP}$ dynamic cost functions. Impurity minimization ($\pi_{Impurity}$) penalizes candidates with impurity-forming ion concentrations exceeding the prior target (yellow triangle). **(e)** Lattice parameter feedback ($\pi_{LP}$) minimizes the deviation between theoretical (Vegard's approximation) and observed lattice parameters to match the next design with the synthesized stoichiometry. Color gradients indicate cost magnitude.

## 2.3. CASS synthetic accessibility optimization

We first evaluated the capability of the CASS framework in optimizing synthetic accessibility using synthesis-related cost functions (static cost functions $\pi_{Thermo}$, $\pi_{Distortion}$, and dynamic cost functions $\pi_{Impurity}$, $\pi_{LP}$). We initialized the synthesis with compositions containing equimolar cations in the Hf-Zr-Ge pseudoternary (**Figure 3a-b**) and Hf-Zr-Ti-Ge pseudoquaternary (**Figure 3c-d**) phosphate ($PO_4$) systems, intentionally selected for their plausibility to exceed solubility limits considering the large size difference between cations such as $Zr^{4+}$ and $Ge^{4+}$. In this accessibility demonstration the initial equimolar compositions were chosen deliberately rather than by minimizing the static cost functions. In each column of **Figure 3**, the top subfigure (a, c) shows the XRD pattern of the first synthesis attempt while the bottom subfigure (b, d) shows the result after a single iteration in which the composition was adjusted following the guidance of the cost functions. For these two systems, the CASS algorithm was able to reduce the impurity peaks substantially in a single iteration.

For example, in the Hf-Zr-Ge system, the composition $NaHf_{0.67}Zr_{0.67}Ge_{0.67}(PO_4)_3$ was initially attempted. The XRD pattern of the obtained material (**Figure 3a**) indicates a NASICON phase purity of approximately 53 wt%, accompanied by impurity phases containing $NaGe_2(PO_4)_3$ and $ZrP_2O_7$. In the next iteration, the dynamic cost function $\pi_{Impurity}$ guided the design of new compositions to contain lower concentrations of the identified impurity-forming cations, in this case, Ge and Zr. Simultaneously, the $\pi_{LP}$ cost function required the new composition to possess lattice parameters similar to the NASICON phase observed in the first trial, a constraint that prefers higher Zr and Hf contents. Balancing these competing objectives, CASS proposed $NaHf_{1.0}Zr_{0.89}Ge_{0.11}(PO_4)_3$ for the second trial. The XRD pattern of this updated composition (**Figure 3b**) demonstrates a successfully synthesized, single-phase NASICON product. Similarly, in a higher-dimensional Hf-Zr-Ti-Ge system, starting with equimolar cations yielded about 90 wt% NASICON, accompanied by $TiO_2$ and $GeO_2$ impurities (**Figure 3c**). After just one iteration of CASS guidance, a single-phase NASICON with the composition $NaHf_{0.76}Zr_{0.89}Ti_{0.23}Ge_{0.12}(PO_4)_3$ (**Figure 3d**) was successfully synthesized. Here, the Ti and Ge contents were specifically reduced due to the $\pi_{Impurity}$ constraints.

Additional examples of phase purity enhancement across one to three iterations are provided in the Supplementary Information for various NASICON systems. These include $PO_4$-based NASICONs (Zr-Sn-Ti-Ge, Zr-Ti-Ge, Zr-Sn-Ge, and Zr-Nb-Ge) and an $SiO_4$-based NASICON (Zr-Sn-Ti-Ge) (**Supplementary Figures 2-6**). Notably, these improvements occurred even when the automated refinement occasionally failed to identify specific minor impurity phases. These results demonstrate that redesigning compositions based on both synthetic accessibility models and characterized synthesis outcomes effectively optimizes target phase purity. Note that even when the impurity detection cannot provide quantitative information, the effort to find compositions that reproduce the lattice parameter of the observed NASICON may drive the synthesis towards single phases, though with a more ambiguous signal.

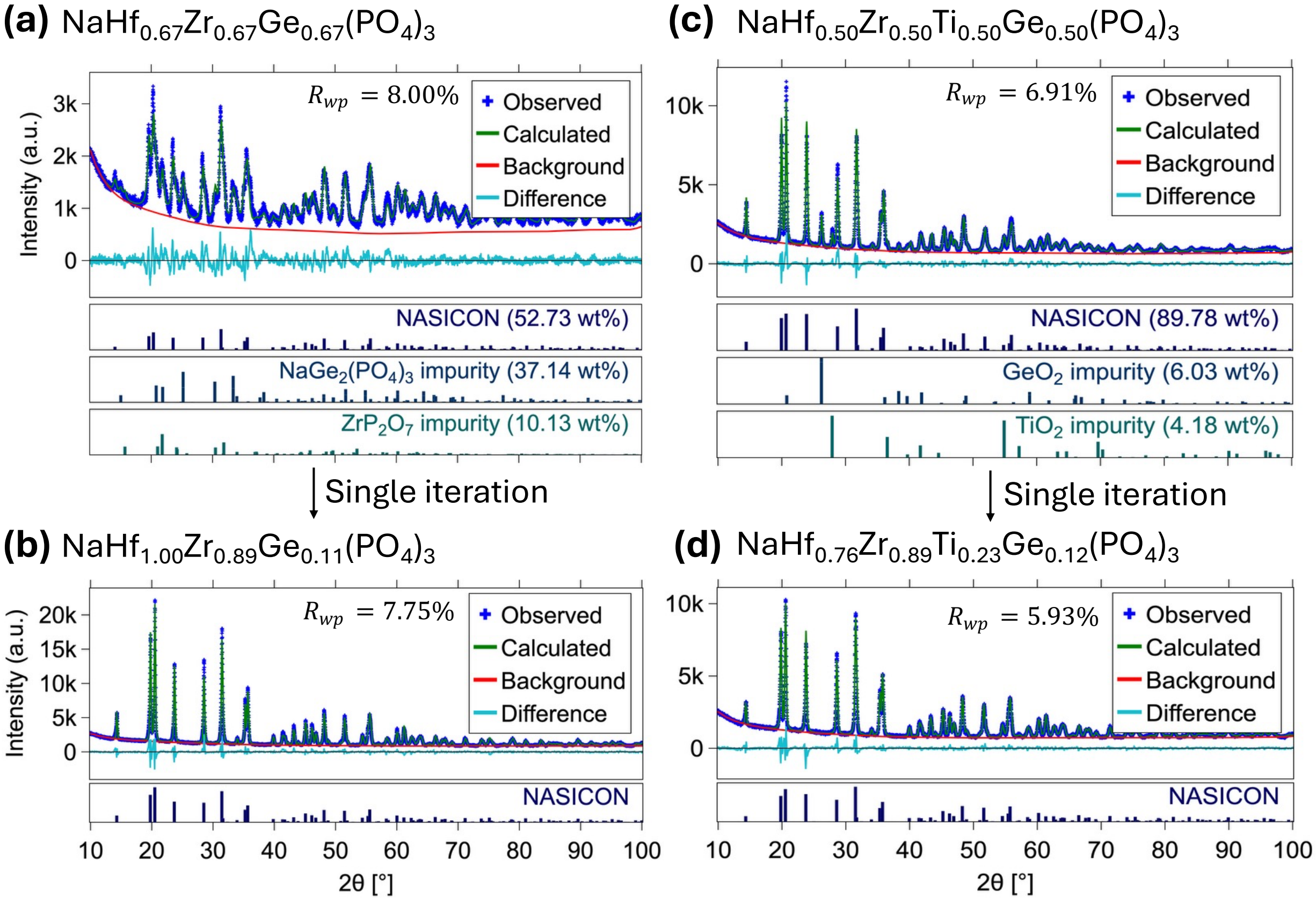

**Figure 3. Validation of synthetic accessibility cost functions.** Comparative Rietveld refinements of initial (top) versus CASS-optimized (bottom) synthesis trials for the **(a, b)** Hf-Zr-Ge and **(c, d)** Hf-Zr-Ti-Ge systems of $PO_4$ polyanion-NASICON after a single iteration. The CASS-optimized trials exhibit substantially decreased impurity peak intensities relative to the primary NASICON phase.

## 2.4. CASS iterative design for fast NASICON ionic conductors

After confirming that CASS is capable of performing synthetic accessibility optimization, we further demonstrate joint optimization of single-phase synthesis with ionic conductivity as an additional target. To demonstrate optimization against these multiple (sometimes conflicting) requirements, we deployed the CASS framework using all cost functions (static cost functions $\pi_{Thermo}$, $\pi_{Distortion}$, $\pi_{IC}$ and dynamic cost functions $\pi_{Impurity}$, $\pi_{LP}$) to explore the high-dimensional NASICON space $Na_wM_2(PO_4)_{3-z}(SiO_4)_z$ for single phases with high ionic conductivity. M contains a mix of at least three types of cations chosen from the following: $Mg^{2+}$, $Zn^{2+}$, $Al^{3+}$, $Sc^{3+}$, $Y^{3+}$, $In^{3+}$, $Ga^{3+}$, $La^{3+}$, $Ti^{4+}$, $Zr^{4+}$, $Hf^{4+}$, $Ge^{4+}$, $Sn^{4+}$, $Nb^{5+}$, and $Ta^{5+}$. The Na content is determined by charge neutrality. Success of a trial was assessed using a Purity Score, a quantifiable metric derived from XRD peak intensities that assesses the relative content of a NASICON phase against impurity phases. Benchmarking of the purity score is discussed in **Methods**. On this calibrated scale, a score > 0.7 indicates NASICON as the major phase with minor impurities, while 1.0 refers to a single-phase NASICON with negligible impurities. To ensure that the NASICON phase likely has a composition close to the targeted composition, a check on the lattice parameters of the phase was done. We classified target compositions as failures (Purity Score = 0.0) if the deviation between the experimentally measured lattice parameters and those predicted by Vegard's approximation exceeded the model's inherent uncertainty, plus an additional tolerance of 15% of the chemical space range (defined by the predicted lattice parameters of the NASICON end members). This extra tolerance of deviation accounts for potential slight off-stoichiometry in the synthesized products and was calibrated against pseudobinary phosphate NASICONs with independently synthesized end members (**Methods, Supplementary Information 1**). For a subset of samples that contain NASICON as the major phase (Purity Score > 0.7), Electrochemical impedance spectroscopy (EIS) was employed to measure their ionic conductivities.

To test whether the CASS framework could leverage immediate experimental feedback to design higher-purity and high-ionic-conductivity compositions, up to two synthesis trials (an initial design and a single optimization) were executed across 54 NASICON compositional spaces. To start, 62 compositions were generated, constrained to having the $(SiO_4)_2(PO_4)$ anion composition, using the static cost functions and subject to automated synthesis in the A-Lab. The results are summarized in **Figure 4**. In the first trial, 24 of 62 compositions (39%) produced an observable NASICON phase, with 12 forming NASICON as the major phase (Purity Score > 0.7). In the second trial, which incorporated updated dynamic cost functions, 16 compositions from those containing an observable NASICON phase were subject to further optimization. These compositions were chosen for practical reasons (excluding Sc to minimize experimental cost) and in multiples of 8 to maximize experimental efficiency (as each box furnace in the A-Lab has a capacity of heating 8 samples per batch). Of these, 11 compounds maintained or enhanced their phase purity, 3 showed poorer purity, and 2 failed to form a NASICON phase close to the target composition. Cumulatively, after two iterations, 23 of the 78 total attempted compositions yielded NASICON as the major phase. These 23 compositions were evaluated against the targeted domain for ionic conductivity (**Figure 2c**, quantified by $\pi_{IC} \leq 0.025$, where $\pi_{IC} > 0$ indicates a composition falls outside the strict target boundary, with the penalty value scaling by distance from that boundary). Eighteen (78%) fell within or near this domain. Ultimately, these 78 trials produced a robust set of major-phase NASICON candidates that can serve as excellent starting points for subsequent Bayesian optimization[41,42].

The formation of the NASICON phase in the first trial and the subsequent phase-purity improvement in the second trial highlight CASS's dual capability: generating initial designs from static priors and refining them by dynamically integrating experimental feedback. For example, the initial static design for the Sn-Zr-In-Ti system, $Na_{3.5}Zr_{0.6}Ti_{0.2}In_{0.5}Sn_{0.7}(SiO_4)_2(PO_4)$, yielded a modest Purity Score of ~0.53. By dynamically adjusting the target to $Na_{3.6}Zr_{0.9}Ti_{0.1}In_{0.8}Sn_{0.2}(SiO_4)_{1.8}(PO_4)_{1.2}$ based on the identified phases and NASICON lattice parameters feedback, the second iteration successfully suppressed competing impurities, forming NASICON as the major phase with a Purity Score of ~0.81. However, occasional regressions in phase-purity, such as the three compositions that decreased in purity and the two that failed to form the target phase, highlight the limitations of the current method. These optimization failures likely originate from the inability of $\pi_{LP}$ to uniquely determine the composition of the NASICON when it is first found in a multi-phase sample given the wide range of solutions allowed by Vegard's approximation in these high-dimensional composition spaces. Furthermore, the drift of composition designs from within to outside the target domain for ionic conductivity ($\pi_{IC}$) during optimization, even as purity increased, reveals a key tradeoff. Returning to the Sn-Zr-In-Ti system, although iterative optimization successfully elevated the Purity Score to ~0.81, the algorithm achieved this increased purity by pushing the target composition severely outside the optimal ionic conductivity domain (drifting from $\pi_{IC} = 0$ to $\pi_{IC} \approx 0.14$). This demonstrates that while new synthesizable compositions exist within the chemical spaces, their resulting structures and Na content may not yield high ionic conductivity, highlighting the competition between property optimization and synthesizability.

Overall, CASS successfully designed and synthesized 18 major-phase NASICON compositions within or near the $\pi_{IC}$ target domain ($\pi_{IC} \leq 0.025$), confirming its ability to balance synthetic accessibility with ionic conductivity screening. A complete record of all 78 synthesis trials is provided in the **Supplementary Data**. To illustrate the broad chemical space and complex stoichiometries accessible by the framework, **Table 1** highlights a subset of successful multi-cation designs that were not selected for the full evaluation detailed below.

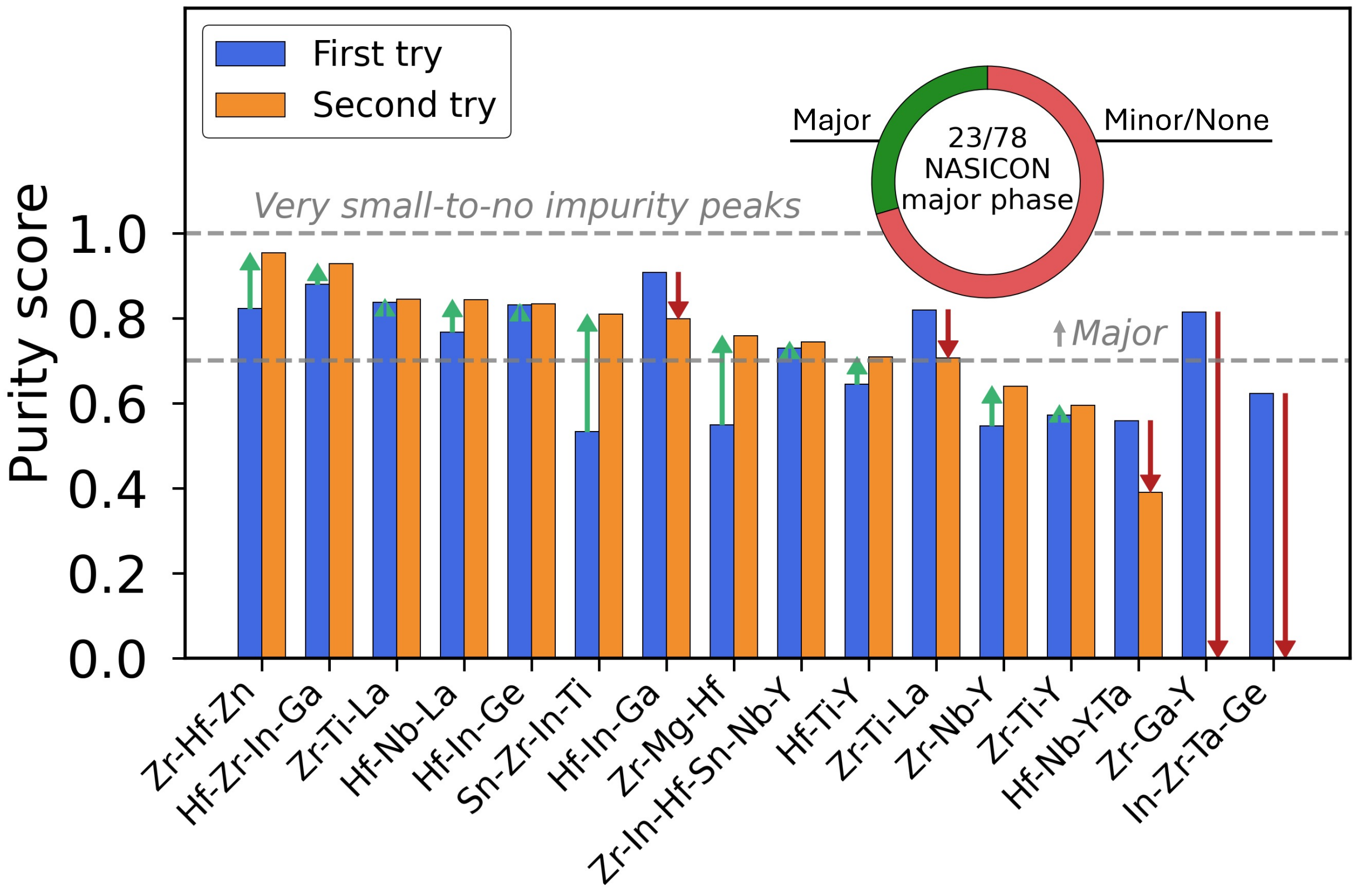

**Figure 4. Assessment of iterative design optimization in multicomponent phosphosilicate NASICONs.** Comparison of Purity Scores between initial (blue) and CASS-redesigned (orange) synthesis trials for 16 selected samples. Green and red arrows denote improved/maintained or declined purity, respectively. Dashed lines define thresholds for negligible impurities (1.0) and major NASICON product phase formation (> 0.7). **Donut chart inset:** Cumulative campaign success, showing 23 of 78 trials yielded a major NASICON phase.

**Table 1.** Nominal compositions of samples containing NASICON as the major phase, with nominal compositions near or within the target ionic-conductivity domain, as quantified by $\boldsymbol{\pi_{IC}} \leq \mathbf{0.025}$

| Nominal global composition | Purity score | $\pi_{IC}$ |
|---|---|---|
| $Na_{3.5}Hf_{0.58}Zr_{1.28}In_{0.08}Ga_{0.06}(SiO_4)_{2.36}(PO_4)_{0.64}$ | 0.93 | 0 |
| $Na_{3.5}Hf_{1.3}In_{0.5}Ge_{0.2}(SiO_4)_2(PO_4)$ | 0.83 | 0 |

| Nominal global composition | Purity score | $\pi_{IC}$ |
|---|---|---|
| $Na_{3.4}Hf_{0.4}Zr_{1.4}Zn_{0.2}(SiO_4)_2(PO_4)$ | 0.82 | 0 |
| $Na_{3.5}Y_{0.2}Zr_{1.5}Ga_{0.3}(SiO_4)_2(PO_4)$ | 0.81 | 0 |
| $Na_{3.4}Hf_{0.4}Zr_{0.3}Sc_{0.3}Ta_{0.2}Ti_{0.1}Nb_{0.2}In_{0.5}(SiO_4)_2(PO_4)$ | 0.77 | 0 |
| $Na_{2.5}La_{0.2}Hf_{1.1}Nb_{0.7}(SiO_4)_2(PO_4)$ | 0.77 | 0 |
| $Na_{3.49}Hf_{0.15}Mg_{0.45}Zr_{1.4}(SiO_4)_{1.59}(PO_4)_{1.41}$ | 0.76 | 0 |
| $Na_{2.9}Y_{0.2}Hf_{0.4}Zr_{0.3}Nb_{0.4}In_{0.6}Sn_{0.1}(SiO_4)_{1.5}(PO_4)_{1.5}$ | 0.74 | 0.0014 |
| $Na_{3.11}Y_{0.06}Hf_{0.4}Zr_{0.42}Nb_{0.35}In_{0.4}Sn_{0.37}(SiO_4)_2(PO_4)$ | 0.73 | 0 |
| $Na_{3.5}Hf_{0.41}Mg_{0.09}Zr_{0.26}Sc_{0.25}Ta_{0.26}Ti_{0.06}Nb_{0.17}In_{0.5}(SiO_4)_2(PO_4)$ | 0.72 | 0 |
| $Na_{3.5}La_{0.2}Zr_{1.3}Ti_{0.5}(SiO_4)_{2.3}(PO_4)_{0.7}$ | 0.71 | 0.00053 |

To evaluate the efficacy of the target ionic conductivity domain, we measured the ionic conductivity of six NASICON samples newly synthesized in this study (**Figure 5a**). These samples contain NASICON as the primary phase, and they featured nominal target compositions representing various distances from the target domain (i.e., different $\pi_{IC}$ values). The stoichiometry of the primary NASICON phase of each sample was assessed by combining Rietveld refinement of XRD (**Supplementary Figures 7-9**), $\pi_{Thermo}$ analysis, lattice parameters verification, and compositions measured via scanning electron microscopy and energy-dispersive X-ray spectroscopy (SEM-EDS) (**Methods**). Deviations between the assessed composition of the primary NASICON phase and the nominal composition were observed, which we attributed to the presence of small amounts of secondary phases. Elemental homogeneity of the NASICON phase was verified, though it was slightly less homogeneous than common mineral standards (**Supplementary Information 2**). Before ionic-conductivity measurements, the as-synthesized powders were cold-pressed and sintered near their synthesis temperatures to over 85% of theoretical density (**Methods**, **Supplementary Information 3, Supplementary Data**). Note that this post synthetic process may decrease the presence of minor impurities (**Supplementary Figure 10**).

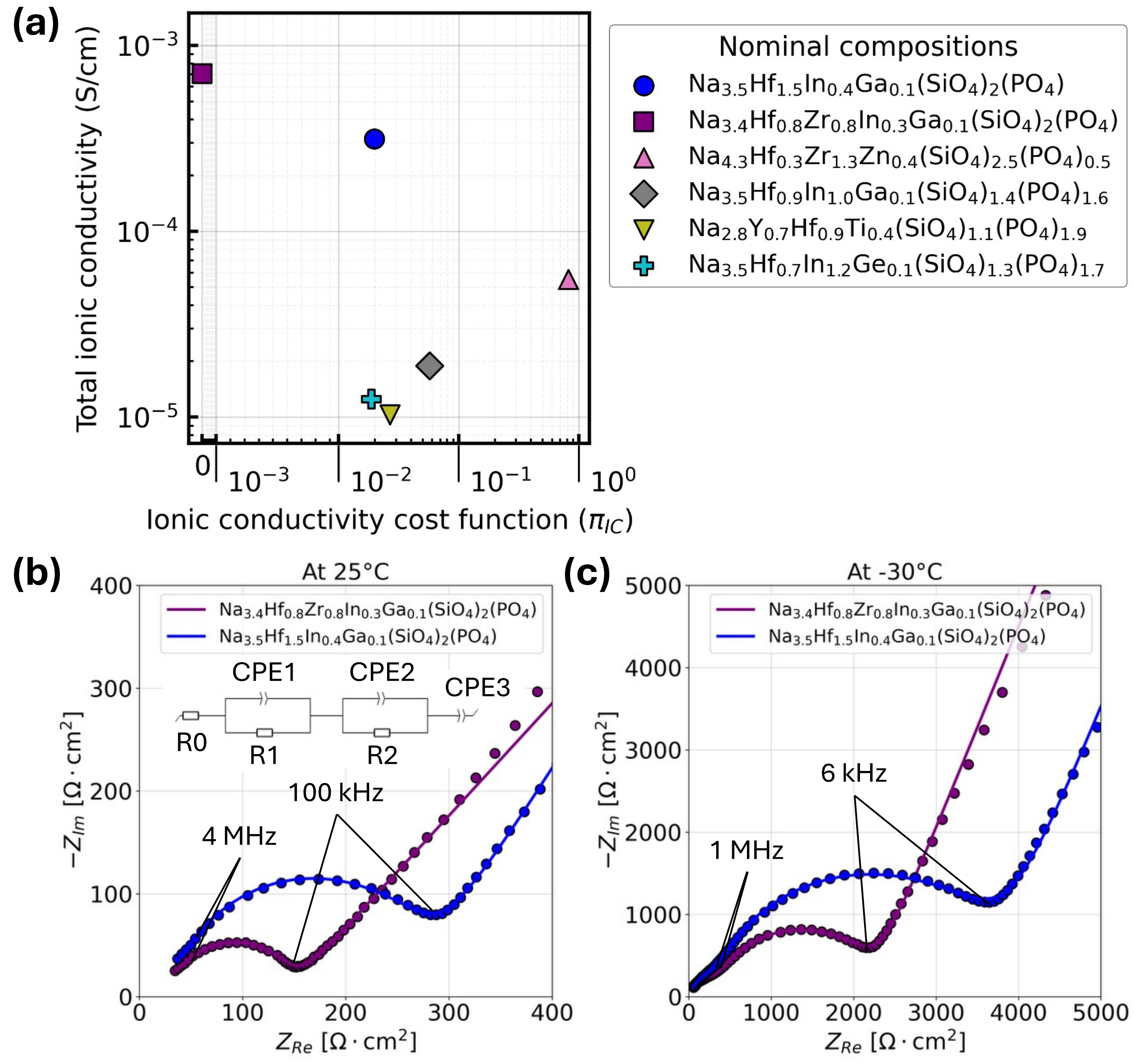

**Figure 5. Total ionic conductivity as a function of $\pi_{IC}$ and electrochemical characterization of the promising NASICON ionic conductors. (a)** Various nominal target compositions containing NASICON as the major phase are plotted according to their calculated $\pi_{IC}$ and measured total ionic conductivities. A lower cost function value indicates a closer proximity to the target domain for ionic conductivity, while a value of zero denotes that the target composition falls inside the target domain. **(b-c)** Nyquist plots for two of the promising NASICON ionic conductors (purple and blue data points) measured at **(b)** 25 °C and **(c)** -30 °C. Characteristic frequencies and the equivalent circuit model (inset, **b**) are shown; solid lines represent fits to the experimental data (circles).

Of the six samples, three compositions were positioned near or within the target domain for ionic conductivity ($\pi_{IC} \leq 0.025$), while the other three were located farther outside the domain ($\pi_{IC} > 0.025$). Among the compositions near or within the target domain, $Na_{3.4}Hf_{0.8}Zr_{0.8}In_{0.3}Ga_{0.1}(SiO_4)_2(PO_4)$ and $Na_{3.5}Hf_{1.5}In_{0.4}Ga_{0.1}(SiO_4)_2(PO_4)$ exhibited room-temperature total ionic conductivities of 0.7 and 0.3 mS/cm, with activation energies of 0.34 and 0.33 eV, respectively (**Supplementary Figure 11a-b**). Their corresponding bulk ionic conductivities reached 3.96 and 3.17 mS/cm consistent with their low activation energies. Low-temperature EIS resolved the bulk and grain boundary contributions by shifting the relaxation frequency into a more measurable range, from ~4 MHz (at room temperature, **Figure 5b**) down to ~1 MHz (at -30 °C, **Figure 5c**), visibly revealing the bulk semicircle. The bulk ionic conductivities of these compositions are comparable to state-of-the-art Na NASICON[56] (~5 mS/cm for optimized $Na_{3.4}Zr_2(SiO_4)_{2.4}(PO_4)_{0.6}$) and other polycrystalline oxide Na solid-state conductors (0.23 to 2.9 mS/cm[57–60]). This is particularly promising given that post-synthetic processing (e.g., densification, sintering conditions, and binder selection) has not been optimized in the current study. However, the third sample, $Na_{3.5}Hf_{0.7}In_{1.2}Ge_{0.1}(SiO_4)_{1.3}(PO_4)_{1.7}$, is also very near the high conductivity domain and yielded only 0.01 mS/cm total ionic conductivity. Conversely, samples located farther outside the domain, $Na_{4.3}Hf_{0.3}Zr_{1.3}Zn_{0.4}(SiO_4)_{2.5}(PO_4)_{0.5}$, $Na_{3.5}Hf_{0.9}In_{1.0}Ga_{0.1}(SiO_4)_{1.4}(PO_4)_{1.6}$, and $Na_{2.8}Y_{0.7}Hf_{0.9}Ti_{0.4}(SiO_4)_{1.1}(PO_4)_{1.9}$, exhibited low ionic conductivities of 0.01 to 0.06 mS/cm, and the latter two showed correspondingly higher activation energies of 0.40 and 0.42 eV, respectively (**Supplementary Figure 11c-d**). These results support the conclusion that minimizing the ionic conductivity cost function ($\pi_{IC}$), that is, designing the nominal global composition to fall close to or within the target domain, generally leads to higher total ionic conductivity, despite the occasional false positives.

## 2.5. Assessment of synthetic accessibility estimators

To assess whether the synthetic accessibility estimators deployed in CASS reflect experimental synthesizability, we evaluated them against a dataset comprising results from the synthesis of 281 unique nominal compositions of mixed-anion phosphosilicate NASICONs ($Na_wM_2(PO_4)_{3-z}(SiO_4)_z$, $0 < z < 3$) (**Supplementary Data**). This dataset includes both the synthesis trials mentioned in previous sections and additional data generated during preliminary testing of each CASS cost function in its design phase. **Figure 6a-b** show how these compositions are distributed across the parameter spaces of our cost functions $\pi_{Thermo}$ and $\pi_{Distortion}$. These compositions populate both the positive and the negative data domain, representing compositions that are penalized or prohibited by the cost functions (e.g., $\pi_{Thermo} > 0$, $\pi_{Distortion} > 0$). The domains shaded in lighter colors in **Figure 6a-b** indicate higher cost and hence less likely synthetic accessibility. Because the $\pi_{Thermo}$ and $\pi_{Distortion}$ feature spaces were chosen to capture ranking of energetic and structural distortion as evaluation metrics for synthesizability, the efficacy of both can be evaluated against the success rate of forming a primary NASICON phase across different regions of the feature space. This region-wise evaluation follows the established practice of binning a continuous predictor and comparing empirical success rates across bins to gauge its predictive power[61]. We bin each feature space by the stability factor it is meant to rank. For $\pi_{Thermo}$, regions are defined by the lowest value of $E_{hull}$ threshold in the SVC (See **5.12**) at which the phase is predicted as stable ($E_{first\ stable\ threshold}$, details in **Methods**). Hence, compositions with a high value of this $E_{first\ stable\ threshold}$ are less likely to be stable. The fact that this is a good predictor is confirmed by the data in **Figure 6c,** which shows the percentage of compositions in each bin that led to samples with a major phase NASICON. The data further reveals that thermodynamically stable compositions ($E_{first\ stable\ threshold}$ = 0 meV/atom above the hull, or "on the hull") achieved a 90% success rate of forming a major NASICON phase indicating the success of our synthesis campaign in actually synthesizing these compounds. Compositions with intermediate stability ($E_{first\ stable\ threshold}$ > 0 to 41 meV/atom) were successful in 37-40% of trials. The success rate dropped to 21% beyond 41 meV/atom, interpreted here somewhat arbitrarily as a practical synthesis limit, and to 0% when exceeding the $\pi_{Thermo}$ synthetic accessibility threshold (62.22 meV/atom) that we inclusively approximated by encompassing 524 experimentally verified compositions (as described in **section 2.2**). Changing the evaluation temperature T to 1373 K does not alter this trend or the practical metastability limit (**Supplementary Figure 12a**).

For $\pi_{Distortion}$, we binned the cation size mismatch relative to the maximum mismatch ($R_M^{STD} - R_{M,max}^{STD}$). These binned regions form the horizontal axis of **Figure 6d**. The data in Figure 6d shows that the distortion cost function $\pi_{Distortion}$ is also a reasonable measure of synthesizability, though its outcome is more binary than the $E_{hull}$ value. For $\pi_{Distortion}$, success rates declined slightly as values approached the maximum mismatch limit, before dropping to just 6% for samples exceeding this ceiling (**Figure 6d**). Ultimately, these results confirm that establishing an inclusive synthetic accessibility threshold based on experimentally verified compositions within an optimized stability feature space creates a successful estimator. Further sampling within this feature space can narrow down the practical limits, specifically adjusting for the exact synthesis methods employed in a given study.

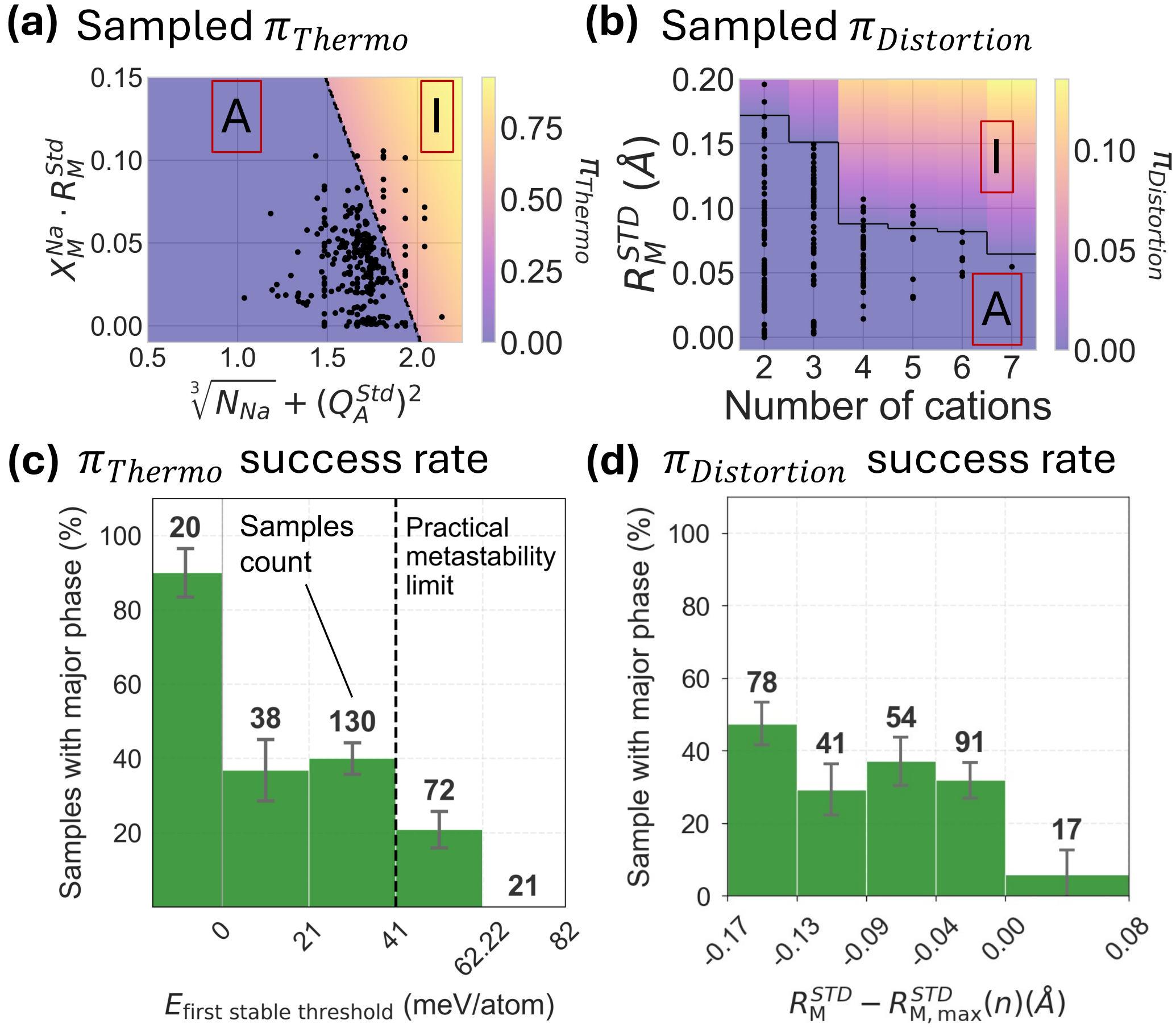

**Figure 6. Validation of static synthetic accessibility priors via large-scale sampling. (a-b)** Distribution of 281 synthesis trials (black dots) across **(a)** thermodynamic ($\pi_{Thermo}$) and **(b)** structural stability ($\pi_{Distortion}$) cost landscapes, where lighter colors indicate higher cost. Synthetically accessible (A) and inaccessible (I) regions are sampled with various compositions. **(c)** Success rate of forming major-phase NASICON versus the minimum threshold required for a composition to be considered stable ($E_{first\ stable\ threshold}$). The dashed vertical line marks the practical metastability limit, and numbers above the bars indicate the sample count within each bin. **(d)** Success rate versus the deviation from the maximum allowable cation size mismatch ($R_M^{STD} - R_{M,max}^{STD}(n)$). Error bars denote the standard error calculated via bootstrap/Monte Carlo methods.

To more accurately understand the thermodynamic synthetic accessibility factors behind NASICON synthesis, we investigated the single-cation and equimolar two-cation NASICON subspaces in phosphates, as well as single-cation silicate NASICONs. The relatively simple formulas of these subspaces allow for the straightforward identification of the lowest-energy ordering in the NASICON phase, enabling a more accurate approximation of $E_{hull,ideal}$ (T = 1000 K was chosen to maintain magnitude comparability with the prior study[48]; calculation details are provided in **Supplementary Information 4**). We attempted to synthesize various compositions by varying the cation pairs (**Supplementary Data**). **Figure 7** correlates these synthesis outcomes with $E_{hull,ideal}$, demonstrating that variations in the metal cations strongly correlate with synthesis success. Successful phosphate syntheses were confined to compositions at or below approximately 15 meV/atom above the hull. However, some compositions below this value, including $NaSnHf(PO_4)_3$ and $Na_3Al_2(PO_4)_3$, did not form the targeted NASICON phase. $Na_3Al_2(PO_4)_3$ formed an amorphous phase, which is consistent with prior studies conducted at ambient pressure[62]. For single-cation silicate NASICONs, successful syntheses were similarly restricted to $E_{hull,ideal}$ < 15 meV/atom. If the ideal configurational entropy is evaluated at the synthesis temperature of T = 1373 K rather than at 1000K, this 15 meV/atom empirical metastability limit further decreases to 9.5 meV/atom (**Supplementary Figure 12b**). Note that, particularly for silicates, our current sparse sampling prohibits establishing a definitive ceiling without more extensive testing. The corresponding comparison of these phosphate and silicate trials with the phosphosilicates in the machine-learned stability feature space is provided in **Supplementary Information 5**.

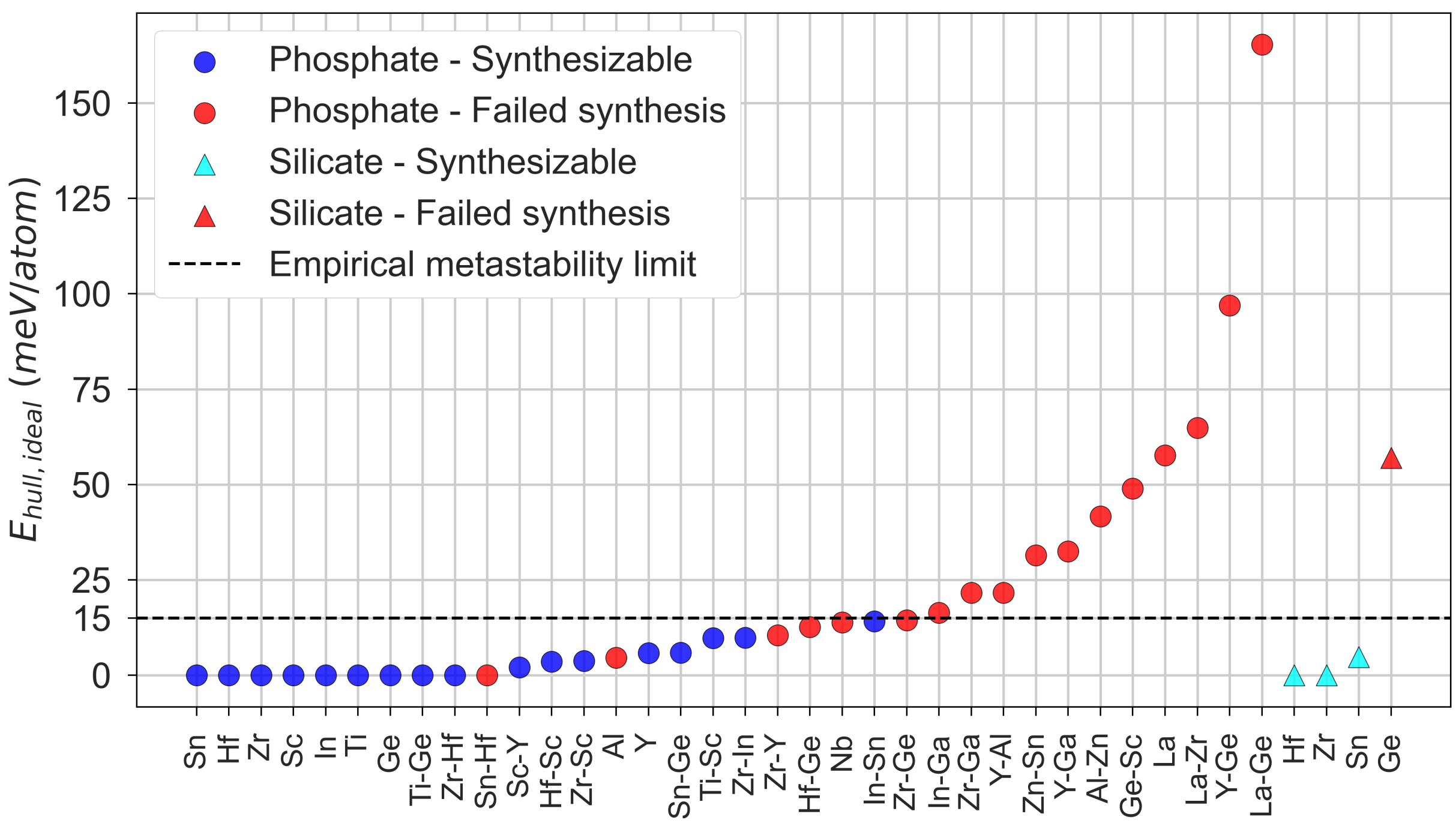

**Figure 7. Empirical metastability limit for phosphate and silicate NASICONs.** Density functional theory (DFT)-calculated energy above the hull augmented with ideal configurational entropy evaluated at 1000 K ($E_{hull,ideal}$) for various phosphate (circles) and silicate (triangles) NASICON compositions. Blue and cyan markers denote successful synthesis, whereas red markers denote failure. The horizontal dashed line at approximately 15 meV/atom marks an empirical upper bound for successful syntheses under the conditions tested; a few unsuccessful trials also occur below this value.

# 3. Discussion

## 3.1. Autonomous synthetic accessibility-guided navigation in multicomponent design space

The CASS framework navigates complex multicomponent NASICON spaces by integrating static ($\pi_{Thermo}$, $\pi_{Distortion}$) models that combine prior physics-based information and dynamic ($\pi_{LP}$, $\pi_{Impurity}$) models that incorporate the data obtained in the experimental campaign. We demonstrated that with this approach CASS successfully steers designs toward high phase-purity compositions. While Gaussian process (GP)-based Bayesian optimization could also offer rigorous global search capabilities, deploying it effectively often requires substantial initial sampling to establish useful priors[41,42], and, in this setting, complex data pipelines to probabilistically model noisy phase-purity estimates from automated XRD refinements. In contrast, by directly embedding physical domain knowledge into its cost functions, CASS sidesteps the abrupt phase-boundary modeling challenges that can complicate standard stationary GP kernels. Consequently, CASS typically requires only one to three experimental iterations to achieve satisfactory phase purity. This efficiency relies on the synergy between anchoring the search in theoretical priors and iteratively refining local solubility maps via immediate experimental feedback. Broadly, this positions CASS as a pragmatic active learning framework capable of identifying single-phase regions within high-dimensional chemical spaces, enabling the exploration of complex multicomponent materials[53,63]. Furthermore, the system's adaptive learning provides resilience against systematic uncertainties, such as variations in precursor properties that shift solubility limits[64,65], by dynamically recalibrating the search to implicitly account for these unmodeled factors. This combination of prior knowledge and in-context experimental data has recently been employed in large language model (LLM)-driven campaigns[66,67]. However, as evidenced by the few iterations needed to achieve improved phase purity in this study, we expect explicit models that map prior knowledge and new experimental data onto the experimental input space, such as the CASS approach, to be more efficient, albeit biased toward regions where the mapping holds. Configuring the models to be more inclusive systematically decreases this bias, a strategic dial for balancing exploration and exploitation within a fixed experimental budget[68], potentially useful as a tool for agentic LLM-driven campaigns[67,69].

### 3.2. Multi-objective design via modular property surrogates

By integrating synthetic accessibility estimators with an ionic conductivity surrogate ($\pi_{IC}$), CASS designed and synthesized 23 major-phase NASICONs across 78 total trials in the A-Lab. Notably, 18 of these compositions fell within or near the high-conductivity target domain. This success is driven by a unified, computationally efficient architecture that couples a gradient-free evolutionary solver with a gradient-based optimizer to navigate complex cost functions. Because the framework is highly modular, it allows for rapid adaptation as physical understanding improves, permitting the testing of new hypotheses or the updating of existing models. More broadly, this capability unlocks multi-objective optimization within complex solid-solution frameworks. Equipped with proven synthetic accessibility optimizers, CASS can readily integrate additional design objectives, such as precursor cost, elemental scarcity, and interfacial stability[21,22], transforming it into a customizable platform for multi-objective solid electrolyte design.

### 3.3. Improvement via interpretable models

Conductivity measurements validated our design strategy, confirming that compositions falling within the $\pi_{IC}$ target domain can yield high ionic conductivity. However, exceptions such as $Na_{3.5}Hf_{0.7}In_{1.2}Ge_{0.1}(SiO_4)_{1.3}(PO_4)_{1.7}$, which sits close to the target domain but fails to achieve high conductivity, highlight the limitations of the current predictive model. Despite its favorable placement, it is possible that excessive cation size mismatch broadens the site energy distribution beyond the percolation optimum[63], a factor not yet explicitly modeled. Future framework improvements can mitigate this by explicitly selecting compositional descriptors to explain site energy distributions or by directly learning their mapping. Another approach would be to bring molecular dynamics modeling of ionic conductivity directly in-line with the optimization framework, though this would require very significant computing resources to deliver the conductivity results in a reasonable time over large parameter spaces. When coupled with an autonomous synthesis laboratory[32], such modified approaches would transform the framework into a powerful tool for real-time hypothesis testing and materials discovery.

Beyond the capability to easily integrate new physical models, CASS also enables the refinement of existing priors by recalibrating them against experimental outcomes. For instance, our synthesis data revealed that the practical thermodynamic stability thresholds for NASICONs are anion-dependent: the threshold for NASICONs containing only phosphate ($PO_4$) or silicate ($SiO_4$) anion groups (~15 meV/atom) is significantly lower than that for mixed-anion phosphosilicates ($(PO_4)_{3-z}(SiO_4)_z$) (~41 meV/atom). This agrees well with a previous study on the thermodynamic limits for metastable inorganic materials, which showed that stoichiometry and constituent elements affect these boundaries[70]. Notably, while these empirical limits significantly exceed a conservative heuristic of 0 meV/atom[48], the observed success rates align with the prediction that synthetic inaccessibility scales with energy above the hull. Fundamentally, shifting from pure silicate ($SiO_4$) or phosphate ($PO_4$) anion groups into mixed-anion phosphosilicates ($(PO_4)_{3-z}(SiO_4)_z$) could increase the diversity of synthesis pathways due to the incorporation of new anion groups that would enable the elements within the system to form various intermediates. This supports theoretical models linking metastability and selectivity to specific synthesis pathways[32,53,70–73]. Meanwhile, the formation of amorphous $Na_3Al_2(PO_4)_3$ prohibits the crystalline version from forming, supporting the understanding that the metastability limit is also governed by the relative stability of the amorphous phase[70]. These findings expose a critical design trade-off: leveraging lattice distortion via elemental substitution (e.g., in high entropy materials) improves ionic conductivity by controlled flattening of site energy landscapes[63], but doing so risks violating thermodynamic and lattice stability limits, especially when changing anion groups. Consequently, rather than rigid cutoffs, the probability distributions of success rate to form a major NASICON phase established here could serve as more flexible proxies for synthetic accessibility to expand the accessible design space.

### 3.4. Bridging discovery and scale-up through automated bulk synthesis

This work employs bulk powder solid-state synthesis[32,35], closely emulating current industrial conditions[74] while enabling a more representative evaluation of ionic conductivity with the differentiation of intragrain versus intergrain transport. A current limitation is the production of loose powder, which requires manual densification (~700 MPa pressing and sintering) for ionic conductivity testing within most current laboratories. However, the system's modular architecture (spanning CASS, AlabOS[75], and hardware[32]) supports the future integration of automated grinding and pressing units, paving the way for autonomous scalable solid electrolyte engineering beyond compositional discovery.

Advancements in automated characterization are also critical to identify the composition of a NASICON when it is first detected in a multi-phase sample. Some estimates of the observed

NASICON composition satisfied lattice parameters constraints ($\pi_{LP}$) but missed the actual compositions due to the mathematical non-uniqueness of Vegard's law in high-dimensional spaces, where lattice parameters map to solution surfaces rather than unique points. Addressing this requires integrating automated phase-specific elemental analysis (e.g., SEM-EDS[76]). Coupling chemical data with structural refinement enables multimodal cost functions that precisely pinpoint phase composition and capture key microstructural features, such as elemental distribution and particle size, providing deeper insights into processing-performance relationships.

## 4. Conclusion

This work establishes a robust system for the accelerated discovery of multicomponent solid-state electrolytes by coupling an automated robotic solid-state synthesis laboratory with the Cost-guided Autonomous Solid-state Synthesis (CASS) framework. By anchoring the autonomous search in interpretable physical descriptors, we enabled model refinement where failed syntheses proved as valuable as successes. This closed-loop approach actively recalibrated synthesizability boundaries, navigating the high-dimensional phosphosilicate NASICON space to overcome traditional bottlenecks of inaccessibility and competing phases, achieving high purity designs in fewer trials than those necessary for rigorous global search algorithms to warm up useful initial priors. Consequently, the framework discovered 23 unique compositions yielding NASICON as the major phase, 18 of which are located within or near the target domain for potentially high ionic conductivity. These discoveries include promising fast ionic conductors, such as $Na_{3.4}Hf_{0.8}Zr_{0.8}In_{0.3}Ga_{0.1}(SiO_4)_2(PO_4)$, which exhibited measured total and bulk ionic conductivities of 0.7 mS/cm and 3.96 mS/cm, respectively.

We demonstrated that optimizing for synthetic accessibility via interpretable cost functions yields improved phase purity, effectively guiding the system toward thermodynamically stable regions even though purity was optimized only indirectly, through the synthetic-accessibility and impurity-feedback terms rather than an explicit phase purity (wt%) quantified from XRD. Integrating experimental feedback allowed us to empirically quantify distinct metastability limits for phosphates (~15 meV/atom) and mixed anion phosphosilicates (~41 meV/atom). These refined priors provide data-driven design rules, underscoring the necessity of balancing the pursuit of high ionic conductivity, often driven by lattice distortion, with thermodynamic stability.

The modular architecture comprising CASS, AlabOS orchestration, and flexible hardware offers a scalable blueprint for an autonomous materials acceleration platform. This framework is readily expandable to address higher-order multi-objective challenges, such as particle morphology, grain boundary optimization, and electrode compatibility. Ultimately, the successful deployment of CASS reinforces the transformative potential of synergistic collaboration between autonomous laboratories and domain experts, moving beyond static high-throughput screening toward intelligent laboratories capable of navigating the complex landscape of energy materials discovery.

## 5. Methods

### 5.1. Automated robotic solid-state synthesis

Synthesis trials were performed using the automated robotic solid-state synthesis and characterization platform within the A-Lab[32], orchestrated by AlabOS[75], which integrates modular stations for precursor dosing and mixing, heating, and solid recovery and XRD sample preparation for characterization. The following summarizes the synthesis orchestration steps performed by the A-Lab.

- **Precursor dosing and mixing:** The powder feeding and slurry handling system was custom-built by Labman Automation. A Mitsubishi robotic arm handled sample transfers. The following steps were performed sequentially for each individual sample, with multiple samples processed in parallel depending on instrument availability. Stoichiometric masses of precursors (Sigma-Aldrich, **Supplementary Information 6**) were dispensed using a Mettler Toledo Quantos balance. A 10 wt% excess of $Na_2CO_3$ was added to all samples to compensate for high-temperature sodium loss[77]. The powders were wet-mixed in ethanol using ten 5 mm zirconia milling balls within a plastic vial in a Hauschild Smart DAC250 centrifugal mixer. The resulting slurry was then transferred to alumina crucibles via a robotic pipettor (Sartorius) and dried at 80°C.
- **Heating:** A Mitsubishi robotic arm transferred the dried samples to a transfer station. From there, a UR5e robotic arm (Universal Robots) mounted on a linearly actuated rail loaded them into box furnaces, subsequently unloading them upon completion of the heating profile. The thermal protocol comprised an initial ramp to 300°C (2°C/min), a second ramp to 1100°C (5°C/min), a 12-hour dwell at this peak temperature, and natural cooling within the closed furnace down to 300°C. At this threshold, the furnace door automatically opened to facilitate faster cooling.
- **Characterization:** A third robotic arm performed the following steps sequentially for each individual sample, with multiple samples processed in parallel depending on instrument availability. Sintered samples were recovered and ground using a modified Fritsch Mini-mill Pulverisette 23. They were then prepared on a clean (verified by scanning prior to preparation) XRD sample holder (> 100 mg of powder) and loaded into the diffractometer, followed by automated unloading upon scan completion. Diffraction scans were performed using an Aeris Minerals diffractometer (Malvern Panalytical).

## 5.2. Composition design algorithm

The total cost function ($\Pi_{Total}$) evaluated by CASS presents a potentially rugged, non-convex landscape, defined as:

$$\Pi_{Total} = w_1\pi_{Thermo} + w_2\pi_{Distortion} + w_3\pi_{IC} + w_4\pi_{Impurity} + w_5\pi_{LP}$$

$$w_1 = w_2 = w_3 = w_5 = 50; w_4 = 1$$

Each component ($w_i\pi_i$) of the total cost function is constructed to be as simple as possible. Mimicking a rectified linear unit (ReLU), each term incorporates a linear penalty that evaluates to 0 when a design falls within the preferred domain and increases linearly as it deviates. We deliberately chose this formulation over rigid step functions so that individual components still provide non-zero directional gradients to help guide the solver toward preferred regions.

To navigate this complex terrain, CASS utilizes a hybrid optimization architecture that couples a Real-Coded Genetic Algorithm (RCGA) with a Broyden-Fletcher-Goldfarb-Shanno (BFGS) gradient-based solver. Purely gradient-based methods are prone to "greedy" local convergence in such landscapes, where the solver becomes trapped in the nearest local minimum rather than exploring the broader search space. To overcome this, the RCGA provides essential global exploration. Unlike purely random sampling, which lacks memory and search direction, the RCGA leverages information persistence across G=4 generations. By employing deterministic truncation to preserve the top P=30 candidates per generation and randomized blended crossover for the reproduction of K=30 offspring, the algorithm effectively focuses computational effort on promising domains rather than independent, blind guesses. To prevent premature convergence in local minima, stochastic injection (N=40 random candidates per generation) maintains the global exploration benefits of random sampling. This evolutionary warm start identifies the most viable region for the global minimum, which is then refined via BFGS to ensure rapid local convergence and high numerical precision.

## 5.3. $\pi_{Thermo}$ formulation

For each candidate composition (ϒ), where “I” and “A” represent synthetically inaccessible and accessible respectively, the $\pi_{Thermo}$ cost function is defined as:

$$\pi_{Thermo}(\Upsilon) = \begin{cases} P(y = \text{"I"}|\Upsilon), & \text{if Class } (\Upsilon) = \text{"I"} \\ 0, & \text{if Class } (\Upsilon) = \text{"A"} \end{cases}$$

where Class $(\Upsilon)$ is determined by a support vector classifier (SVC), and $P(y = "\mathrm{I}" | \Upsilon)$ is the Platt-scaled probability[48,51] from the SVC output.

The $\pi_{Thermo}$ SVC was defined by adjusting NASICON thermodynamic stability descriptors and energy threshold ($E_{threshold}$) from a previously published study[48]. This adjustment ensured that 524 experimental NASICON compositions, sourced from text-mined literature and the Inorganic Crystal Structure Database (ICSD)[78], were correctly classified as synthetically accessible, while also enabling rapid computational evaluation. Specifically, the Ewald energy ($E_{Ewald}$) was excluded from the descriptor set and the $E_{threshold}$ was increased from 0 to 62.22 meV/atom. To validate the classifier's ability to accurately separate synthetically accessible from inaccessible NASICON compositions, we evaluated its performance scores before and after the descriptor modifications, as well as after the $E_{threshold}$ increased. Using five-fold cross-validation, the final classifier achieved an average test accuracy of 87.25%, a recall of 86.91%, and an F1-score of 84.78%, demonstrating that robust performance was maintained following these changes (details in **Supplementary Information 7**).

## 5.4. $\pi_{Distortion}$ formulation

Recognizing that local lattice distortion governs the stability of multicomponent systems like high-entropy alloys[55] and solid solutions[53,54], we bound this factor by using $R_M^{STD}$ (standard deviation of the metal cation radii) as a quantitative descriptor. Because the reference dataset of 524 crystalline compositions extends only to 7-cation systems, we conservatively apply the stability limit observed for 7 cations to all compositions with higher complexity. Formally, $\pi_{Distortion}$ is defined as:

$$\pi_{Distortion}(\Upsilon) = \begin{cases} 0, & \text{if } R_M^{STD} \leq R_{M,max}^{STD}(n(\Upsilon)) \\ R_M^{STD} - R_{M,max}^{STD}\big(n(\Upsilon)\big), & \text{otherwise} \end{cases}$$

where $n(\Upsilon)$ is the number of cations of the candidate composition $(\Upsilon)$ and $R_{M,max}^{STD}(n(\Upsilon))$ is the maximum limit of $R_M^{STD}$.

### 5.5. $\pi_{IC}$ formulation

The ionic conductivity cost function ($\pi_{IC}$) sums linear penalties for mean cation radius, sodium content, and polyanion center ion radius (detailed in **Supplementary Information 8**) as adapted from a boundary proposed in a previous study[25].

$$\pi_{IC} = \pi_{\bar{R}_M} + \pi_{N_{Na}} + \pi_{\bar{R}_A}$$

Where $\pi_{\bar{R}_M}$, $\pi_{N_{Na}}$, and $\pi_{\bar{R}_A}$ represent the cost components bounding the average cationic radius, sodium content, and average polyanion center ion radius, respectively.

### 5.6. $\pi_{Impurity}$ formulation

Following each synthesis trial, the XRD characterization identifies the set of cations present in detected impurities ($V_{imp}$), including P and Si but excluding Na since it is defined by the cation and anion ratios alongside the charge neutrality constraint, based on XRD phase analysis. We define a Boolean indicator function, $m(\Upsilon, V_{imp}, \Upsilon_{prior})$, which evaluates to True if the normalized concentration of any element $i \in V_{imp}$ in the candidate composition $\Upsilon$ exceeds its concentration in the prior synthesis attempt $\Upsilon_{prior}$. The cost function is formally defined as:

$$\pi_{Impurity}(\Upsilon) = \begin{cases} 3 + \Sigma_{\mathrm{i}} c_i, & \text{If } m(\Upsilon, V_{imp}, \Upsilon_{\mathrm{prior}}) \\ \Sigma_{\mathrm{i}} c_i, & \text{otherwise} \end{cases}$$

where $c_i$ represents the concentration of element $i$ normalized by its crystallographic site multiplicity. The static penalty of 3 serves as a barrier against increasing the concentration of impurity-prone elements, while the gradient term $\sum_i c_i$ drives the optimization toward their minimization or complete removal. If no impurities are detected in the prior trial, $\pi_{impurity}$ is set to 0. We understand that this is a conservative approximation as it is possible that the presence of an element $i$ in the impurities is strongly influenced by another element $j$.

### 5.7. Fine-tuned Vegard's approximation and $\pi_{LP}$ formulation

Two multiple linear regression models predict lattice parameter $a$ and cell volume $V$, respectively, based on $\bar{R}_M$ (average cationic radius), $\bar{R}_A$ (average polyanion center ion anionic radius), and $N_{Na}$ (Na content) as the input features (predictors). The decision to employ two separate models is based on empirical observations of the correlations between these predictors and the respective outputs (responses). To improve accuracy within a specific chemical space, we applied adaptive fine-tuning by assigning a weight of $w = 10{,}000$ to local compositions during fitting. This large weighting essentially forces the model to fit the compositions inside the target chemical space, while utilizing out-of-space compositions for general guidance when local data is sparse. The cost function is defined as the mean percentage difference between predicted and experimentally refined parameters ($a$, $c$, $V$).

$$\pi_{LP} = \frac{1}{3}\left(\frac{|a_{XRD} - a_{\Upsilon}|}{a_{XRD}} + \frac{|c_{XRD} - c_{\Upsilon}|}{c_{XRD}} + \frac{|V_{XRD} - V_{\Upsilon}|}{V_{XRD}}\right) \times 100\%$$

where $a_{XRD}$, $c_{XRD}$, and $V_{XRD}$ are the unit cell $a$ and $c$ lattice parameters and volume, respectively, obtained from the experimentally refined values. The variables with the $\Upsilon$ subscript denote the corresponding structural values predicted by the regressors. Training and testing details are provided in **Supplementary Information 9**.

### 5.8. XRD measurement, refinement, and phase analysis

Powder XRD measurements were conducted using Cu-Kα radiation over a 2θ range of 10°-100° (8 minutes duration, ~0.01° step size, and a detector active length of 5.542° in 2θ), with diffractograms automatically uploaded to a central server for analysis. For structural refinement, a set of candidate rhombohedral NASICON structures with varying Na-site occupancies was generated, utilizing initial lattice parameters predicted by the fine-tuned Vegard's approximation. Rietveld refinement to extract the lattice parameters of the NASICON phase was performed for each candidate using BGMN software[79], and the solution yielding the lowest weighted residual error ($R_{wp}$) was selected. To ensure the refined phase corresponded to the target NASICON structure, its validity was confirmed by comparing the extracted $a$ lattice parameter and unit cell volume against model predictions. A phase was flagged as unlikely to form if deviations exceeded a composite threshold of the model's intrinsic uncertainty (95% confidence interval) plus 15% of the model-predicted end-member range within the chemical space (**Supplementary Information 1**).

Subsequently, phases were identified using the DARA[80] algorithm. Non-NASICON phases were classified as impurities; if multiple NASICON phases were detected, those with lower weight percentages were also deemed impurities. The compositions of these impurity phases are the input for the $\pi_{Impurity}$ cost function. The phase search space comprised generated NASICON structures alongside known phases from the ICSD[78] and Crystallography Open Database (COD)[81]. However, to mitigate combinatorial explosion in complex compositional spaces (≥ 7 elements), the search pool was restricted to common phases identified during attempts in lower-dimensional subspaces. The samples selected for ionic conductivity measurements underwent comprehensive manual phase identification and refinement using GSAS-II[82] software. The following parameters were refined: background, sample displacement and transparency, phase fractions, isotropic crystallite size and microstrain, lattice parameters, and surface roughness.

### 5.9. Purity Score and calibration

Because automated phase search and refinement often fail to capture all impurity phases, we assessed phase purity using the Estimated Purity (EP). This metric is defined as the ratio of the integrated intensity of the refined NASICON phase to the total background-subtracted pattern intensity[83] (which captures residuals due to both unidentified impurities and suboptimal refinement parameters):

$$\mathrm{EP}(\%) = \frac{|\mathrm{I}_{NASICON}|}{|I_{NASICON}| + |I_{Residual}|} \times 100\%$$

To normalize this metric against experimental baselines and decouple physical purity from automated refinement artifacts, we convert EP into a Purity Score. Using a calibration set of high-purity samples (**Supplementary Information 10**), we established a reference baseline ($\mathrm{EP_{upper}} = 78.33\%$) representing negligible impurity content. To ensure the synthesized phase matches the design, we enforce a strict stoichiometric validation check. The Purity Score is calculated as:

$$\text{Purity Score} = \begin{cases} 0.0, & \text{If } |\Delta \mathrm{L}| > (\delta_{\mathrm{model}} + 0.15 \cdot \mathrm{R_{space}}) \\ \min\left(1.0, \dfrac{\mathrm{EP}}{\mathrm{EP_{upper}}}\right), & \text{otherwise} \end{cases}$$

A score of 1.0 indicates a high-purity sample, while a score > 0.7 identifies a major NASICON phase. However, a sample is reclassified as a failure (Purity Score = 0.0) if the absolute deviation ($|\Delta \mathrm{L}|$) of the $a$ lattice parameter or the unit cell volume from its predicted value exceeds the composite threshold defined by the model's intrinsic uncertainty ($\delta_{\mathrm{model}}$) plus 15% of the parameter's maximum range ($R_{space}$) within the chemical space, as defined by the predicted lattice parameters of the NASICON end members. The value of 15% follows from the calibration against the pseudobinary phosphate NASICONs whose single-cation end members were themselves synthesized in this work (**Supplementary Information 1**).

### 5.10. Multimodal NASICON composition analysis

NASICON phase composition for the selected samples was resolved by coupling lattice-parameter and thermodynamic stability screening with automated SEM-EDS (Thermo Fisher Phenom XL G2) using the AutoEMX framework[76]. The analysis followed a three-step process: (1) automated SEM-EDS data acquisition, filtering, and $k$-means clustering; (2) identification of the most likely experimental NASICON composition via hybrid optimization minimizing $\pi_{Thermo}$ and Euclidean distance to centroids, constrained by lattice parameter validity; and (3) final phase identification, including impurities, conditional on the XRD refinement. Details in **Supplementary Information 11**.

### 5.11. Ionic conductivity measurement

Ionic conductivity was measured by electrochemical impedance spectroscopy (EIS) on sintered pellets of the as-synthesized powders. Powders were ball-milled, cold-pressed into green bodies, and sintered near their synthesis temperature to over 85% of theoretical density (**Supplementary Data**), then rapidly cooled to preserve the target phase. Indium electrodes were applied and impedance recorded from 7 MHz to 100 mHz. Spectra were fit to an equivalent circuit from which bulk, grain-boundary, and total ionic conductivities were extracted. Activation energies were obtained from Arrhenius fits of subsets of temperature-dependent measurements from -30 to 85 °C. Full instrument settings, pellet-processing parameters, the equivalent-circuit definition, the fitting procedure, and the activation-energy analysis are provided in **Supplementary Information 3.**

### 5.12. Determining the stabilizing energy above hull threshold ($E_{first\ stable\ threshold}$)

The $\pi_{Thermo}$ linear SVC uses machine-learned descriptors[48] to rule out synthetically inaccessible compositions. It uses a large training data set of NASICONs for which $E_{hull}$ is computed and augmented with the ideal configurational entropy ($E_{hull,ideal}$)[48]. The SVC used in this work used an $E_{threshold}$ of 62.22 meV/atom for the hull energy to divide synthesizable NASICONs from non-synthesizable ones. To evaluate if this value is optimal, we can retrain the SVC with different values for this threshold on the hull energy. We trained 83 individual SVCs using the 3,881 unique NASICON composition and $E_{hull,ideal}$ pairs from the previous study[48] using $E_{threshold}$ values ranging from 0 to 82 meV/atom in increments of 1 meV/atom (T = 1000 K was chosen to maintain magnitude comparability with the prior study[48]). We trained the classifiers using five-fold cross-validation, selecting the models with an accuracy closest to the fold average as the final classifiers. These 83 SVCs act as variants of $\pi_{Thermo}$ that disqualify compositions exceeding their specific $E_{threshold}$. For each target composition in our experimental campaign, $E_{first\ stable\ threshold}$ was obtained by identifying the SVC with the lowest $E_{hull}$ threshold used in its training that identified the phase as stable.

## Author contributions

B.R., Y.Z., and G.C. developed the theoretical formalism. B.R. and Y.Z. developed the experimental formalism. B.R. developed the CASS design system (optimization scheme, design rules adjustment, and cost function definition); conceived and planned synthesis experiments; developed NASICON structure generation and lattice parameter-composition relation; executed and verified the automated XRD analysis results; analyzed the synthesizability data; and wrote the manuscript. B.R. and Y.F. developed the synthesis and XRD characterization automation and data handling. B.R., Y.F., D.M., and M.J.M. developed the automation for solid handling. T.H. developed the text-mining pipeline for NASICON compositions and ionic conductivity. B.R., X.Y., and H.Q. prepared the as-synthesized samples for EIS characterization. B.R. and A.G. carried out the multimodal XRD, SEM, and EDS measurement and analysis. B.R., Y.F., L.N.W., D.M., and M.J.M. executed synthesis experiments. Y.F. and M.J.M. developed the automated XRD refinement and phase search. B.O. contributed to the theoretical formalism for synthetic accessibility of NASICONs. Y.Z. supervised the development of synthesis and XRD characterization automation and execution; and supervised the sample preparation for EIS measurement. G.C. supervised the project. All authors contributed to the manuscript revision.

## Data availability

The comprehensive dataset supporting this study is publicly available in the figshare repository at https://doi.org/10.6084/m9.figshare.31829305. This deposit includes the refined XRD patterns (incorporating both NASICON and impurity phases), SEM images and EDS clustering results, and EIS fitting data for the six selected samples. Additionally, the repository contains comprehensive synthesis results for 345 samples, comprising 281 phosphosilicates (78 of which were designed with ionic conductivity considerations) and 64 phosphates and silicates, indexed by their global target composition. These high-throughput data include expected versus refined lattice parameters, Purity Scores, lattice parameter validations, and refined X-ray diffractograms (NASICON phase only), alongside the refined XRD patterns of the high-purity samples used for Purity Score calibration. All other raw data are available from the corresponding authors upon reasonable request.

## Code availability

Code for the design algorithm, cost function implementations, Vegard's approximation, and composition verification is available at https://github.com/CederGroupHub/cass. This repository also includes the scripts used for the multimodal (XRD and SEM-EDS) NASICON composition analysis and general plotting. Additional scripts for raw data visualization are available in the figshare repository at https://doi.org/10.6084/m9.figshare.31829305.

## Supporting Information

Supporting Information and Supplementary Data accompany this manuscript.

## Acknowledgement

This research was primarily supported as part of the Energy Storage Research Alliance (ESRA), an Energy Innovation Hub funded by the U.S. Department of Energy (DOE), Office of Science, Basic Energy Sciences (BES), under Contract No. DE-AC02-06CH11357 for the development of the Cost-guided Autonomous Solid-state Synthesis (CASS) framework, synthesis planning, electrochemical characterization, and overall project leadership. Additional support was provided by the D2S2 program within the U.S. DOE, Office of Science, BES, Materials Sciences and Engineering Division under Contract No. DE-AC02-05-CH11231 for the development of automated solid-state synthesis and XRD analysis workflows, and multimodal phase analysis using SEM-EDS and XRD. Implementation of the ideas in this work as an AI-campaign was supported by the U.S. Department of Energy, Office of Science, Basic Energy Sciences, Materials Sciences and Engineering Division, and the Office of Advanced Scientific Computing Research (ASCR) under Contract No. DE-AC02-05-CH11231 within the LBNL GENESIS program. B.R. acknowledges support from the Kavli ENSI Graduate Student Fellowship. L.N.W. acknowledges funding support from the BIDMaP postdoctoral fellowship. The authors thank Marilyn Sargent at Berkeley Lab for capturing photos of the A-Lab.

## Conflict of Interest

The authors declare no conflict of interest.

# Combining physical models with dynamically acquired experimental information for the optimization of multicomponent NASICON fast ionic conductors in a self-driving laboratory (Supplementary Information)

Bernardus Rendy[1,2,3], Yuxing Fei[1,2], Tanjin He[4], Xiaochen Yang[1,2], Andrea Giunto[1,2], Lauren N. Walters[5,1], David Milsted[2], Hao Qiu[1], Matthew J. McDermott[2], Bin Ouyang[2,6], Yan Zeng[2,7,*], Gerbrand Ceder[1,2,3,*]

## Table of Contents

[1] Department of Materials Science & Engineering, University of California Berkeley, Berkeley, CA 94720, USA
[2] Materials Sciences Division, Lawrence Berkeley National Laboratory, Berkeley, CA 94720, USA
[3] Energy Storage Research Alliance, Argonne National Laboratory, 9700 South Cass Avenue, Lemont, IL 60439, USA
[4] Argonne National Laboratory, 9700 South Cass Avenue, Lemont, IL 60439, USA
[5] Bakar Institute of Digital Materials for the Planet, UC Berkeley, Berkeley, CA 94720, USA
[6] Department of Chemical and Biomolecular Engineering, Vanderbilt University, Nashville, TN 37235, USA
[7] Department of Mechanical Engineering, Vanderbilt University, Nashville, TN 37235, USA
[*] Corresponding authors: gceder@berkeley.edu (G.C.), yan.zeng@vanderbilt.edu (Y.Z.)

## 1. Provenance of the 15% off-stoichiometry tolerance

The validity check compares the refined $a$ lattice parameter and unit-cell volume V with the fine-tuned model (**Supplementary Information 9**) predictions, and flags the target phase as unlikely to have formed when either deviates by more than the model uncertainty plus 15% of the range between the model-predicted end-member values (**Methods**). The tolerance accommodates slight off-stoichiometry, potentially arising from the loss of volatile components at synthesis temperatures[1]. This section reports the measurements from which the 15% was set.

The position of the refined phase relative to the two single-cation end members A and B is given by the fractional coordinate:

$$x(q) = \frac{q_{refined} - q_{\mathrm{A}}}{|q_{\mathrm{B}} - q_{\mathrm{A}}|} = \frac{q_{refined} - q_{\mathrm{A}}}{\mathrm{R}_{\mathrm{space}}}$$

where $q$ is either the $a$ or V, and $q_A$ and $q_B$ (with $q_B > q_A$) are its measured values at the end members A and B. $\mathrm{R}_{\mathrm{space}} = |q_{\mathrm{B}} - q_{\mathrm{A}}|$ is the range of $q$ across the pair.

For an equimolar target following Vegard's law[2], $x(q) = 0.5$, and the distance from the target along the pair is:

$$\Delta x_{equimolar\ target} = |x\,(q) - 0.5|$$

The Vegard's-approximation models for $a$ and V carry an intrinsic uncertainty $\delta_{\mathrm{model}} = 2\sigma_q$, taken as the 95% confidence interval. Here $\sigma_q$ is the root-mean-square residual of the fine-tuned regression for $q$ over the NASICON structures within the chemical system of that pair, so $\delta_{\mathrm{model}}$ differs from pair to pair. Accounting for this uncertainty gives:

$$\Delta x_{equi,UA} = \max\left(0, \Delta x_{equimolar\ target} - \frac{2\sigma_q}{R_{space}}\right)$$

$\Delta x_{equi,UA}$ is zero when the refined value lies within $\delta_{\mathrm{model}}$ of the midpoint between the end members, and otherwise is the part of that distance lying beyond $\delta_{\mathrm{model}}$, expressed as a fraction of $R_{space}$. Values are reported as percentages (×100).

The calibration set is the thirteen equimolar two-cation phosphate pairs for which both single-cation end members were also synthesized and refined in this work, so that $q_{\mathrm{A}}$ and $q_{\mathrm{B}}$ are measured rather than extrapolated. Four of these trials did not form their target phase, either because the target reflection is split or because the refined cell is incompatible with the equimolar phase. In $NaSnHf(PO_4)_3$, the (110) reflection splits into two components and the single-phase fit sits between them. The refined $a$ values of $NaHfGe(PO_4)_3$ and $NaZrGe(PO_4)_3$ fall within 0.016 and 0.021 Å of measured $NaHf_2(PO_4)_3$ and $NaZr_2(PO_4)_3$, respectively, on pairs spanning 0.664 and 0.708 Å, placing both at $x(q) \geq 0.967$ on $a$ and V; this is consistent with little or no germanium entering the structure, and $GeO_2$ was identified in $NaZrGe(PO_4)_3$. In $Na_2ZrY(PO_4)_3$, the distance from the equimolar target, $\Delta x_{equimolar\ target}$, is 0.08 for $a$ and 0.39 for V, compared with normalized model uncertainties of 0.01 and 0.05, respectively. In the remaining nine trials the target phase formed without showing any split reflections.

**Supplementary Figure 13** shows $\Delta x_{equi,UA}$ for all thirteen pairs, on $a$ (circles) and V (diamonds), against the adopted 15% tolerance. Taking the larger of the two values for each pair, the largest among trials that formed their target phase is 11.5%, and the smallest among those that did not is 18.2%. Any tolerance between these values separates the two groups, and the adopted 15% lies within this window. **Supplementary Figures 14 and 15** show each pair for $a$ and V, respectively. Each panel plots the fine-tuned model across the pair, with $\delta_{\mathrm{model}}$ as the dark band and $\delta_{\mathrm{model}}$ plus 15% of the model-predicted end-member range as the light band. The measured end members are squares, the model value at the equimolar target is an open circle, and the refined value is a filled circle. The percentage on each panel is $x(q) \times 100$ measured from A, so an ideal equimolar product reads 50%.

## 2. Multimodal NASICON composition analysis results

All possible phases and confidences from AutoEMX are accessible in **Supplementary Data**. The refined XRD patterns are shown in **Supplementary Figure 7-9**. Cluster homogeneity in these synthetic samples, quantified by the root mean square of point-to-centroid distances ($d_{RMS}$)[3] of the EDS compositional spectra ($3.89<d_{RMS}<6.64$ at%), is slightly worse than that of mineral standards ($d_{RMS}<3$ at%)[3]. Below are the phase compositions per target global composition:

- $Na_{3.4}Hf_{0.8}Zr_{0.8}In_{0.3}Ga_{0.1}(SiO_4)_2(PO_4)$

- NASICON: $Na_{3.30}Hf_{0.71}Zr_{0.97}In_{0.22}Ga_{0.10}(SiO_4)_{1.98}(PO_4)_{1.02}$
  - Impurities:
    - $HfZrO_4$
    - $Na_2HfSi_2O_7$
    - $Na_{2.22}Hf_{0.71}Zr_{0.94}In_{0.34}Ga_{0.01}(SiO_4)_{0.87}(PO_4)_{2.13}$
- $Na_{3.5}Hf_{1.5}In_{0.4}Ga_{0.1}(SiO_4)_2(PO_4)$
  - NASICON: $Na_{3.46}Hf_{1.74}In_{0.19}Ga_{0.07}(SiO_4)_{2.20}(PO_4)_{0.8}$
  - Impurities:
    - $Na_2HfSi_2O_7$
    - $HfO_2$
    - $Na_{3.23}Hf_{0.59}In_{1.33}Ga_{0.08}(SiO_4)_{0.82}(PO_4)_{2.18}$
- $Na_{3.5}Hf_{0.7}In_{1.2}Ge_{0.1}(SiO_4)_{1.3}(PO_4)_{1.7}$
  - NASICON: $Na_{3.16}Hf_{0.76}In_{0.88}Ge_{0.36}(SiO_4)_{1.29}(PO_4)_{1.71}$
  - Impurities:
    - $In_2O_3$
    - $HfO_2$
    - $Na_{3.21}In_2(SiO_4)_{0.21}(PO_4)_{2.79}$
- $Na_{4.3}Hf_{0.3}Zr_{1.3}Zn_{0.4}(SiO_4)_{2.5}(PO_4)_{0.5}$
  - NASICON: $Na_{3.57}Hf_{0.33}Zr_{1.67}(SiO_4)_{2.57}(PO_4)_{0.43}$
  - Impurities:
    - $HfZrO_4$
    - $Na_2Zn(Si_2O_6)$
- $Na_{3.5}Hf_{0.9}In_{1.0}Ga_{0.1}(SiO_4)_{1.4}(PO_4)_{1.6}$
  - NASICON: $Na_{3.27}Hf_{0.95}In_{0.86}Ga_{0.19}(SiO_4)_{1.22}(PO_4)_{1.78}$
  - Impurities:
    - $In_2O_3$
    - $HfO_2$
    - $Na_{3.33}In_2(SiO_4)_{0.33}(PO_4)_{2.67}$
- $Na_{2.8}Y_{0.7}Hf_{0.9}Ti_{0.4}(SiO_4)_{1.1}(PO_4)_{1.9}$
  - NASICON: $Na_{2.87}Y_{0.19}Hf_{1.14}Ti_{0.67}(SiO_4)_{1.68}(PO_4)_{1.32}$
  - Impurities

- $Na_4Hf_2(SiO_4)_3$
- $Na_{3.05}Y_{1.60}Hf_{0.28}Ti_{0.12}(SiO_4)_{0.45}(PO_4)_{2.55}$

## 3. Measurement of ionic conductivity and activation energies

Ionic conductivities were determined via electrochemical impedance spectroscopy (EIS). The as-synthesized powders were ball-milled (SPEX 8000M, 30 min), cold-pressed into 6 mm pellets under ~700 MPa, sintered at 1050-1100 °C for 1-2 hours (**Supplementary Data**), and fast-cooled, all under air atmosphere. Indium electrodes were applied via cold isostatic pressing at 30 MPa. EIS spectra were collected using a BioLogic VMP-300 potentiostat (7 MHz - 100 mHz, 10 mV amplitude). The EIS spectra were fitted using the *impedance.py* library[4] with an equivalent circuit of series setup resistance ($R_0$), bulk ($R_1||CPE_1$), grain boundary ($R_2||CPE_2$), and diffusion element ($CPE_3$). Parameter initialization was automated via a valley search algorithm applied to Savitzky-Golay smoothed data, followed by differential evolution optimization[5].

To determine activation energies ($E_a$) for four of the six samples, temperature-dependent AC impedance spectroscopy was conducted at a subset of discrete intervals: -30, -10, 25, 35, 45, 55, 65, 75, and 85 °C (**Supplementary Figure 11**). To preserve high-frequency signal integrity (~ 7 MHz) within the thermal chamber during low-temperature measurements, the cell was mechanically isolated using a thick foam base to dampen vibrations, and physical contact between cabling and chamber walls was also minimized. The activation energy ($E_a$) was derived by performing a linear regression of the data according to the linearized Arrhenius equation:

$$\ln(\sigma_{total}T) = m\left(\frac{1000}{T}\right) + c$$

where $m$ is the slope, $c$ is the intercept, $\sigma_{total}$ is the total ionic conductivity, and $T$ is temperature in Kelvin. The activation energy is then calculated as:

$$E_a = -1000 \cdot m \cdot k_B$$

where $k_B$ is the Boltzmann constant.

## 4. Ideal configurational entropy formulation and $E_{hull,ideal}$

The rhombohedral NASICON unit cell contains five distinct crystallographic sites: Na (6b, 18e, 36f), M (12c), and P/Si (18e).

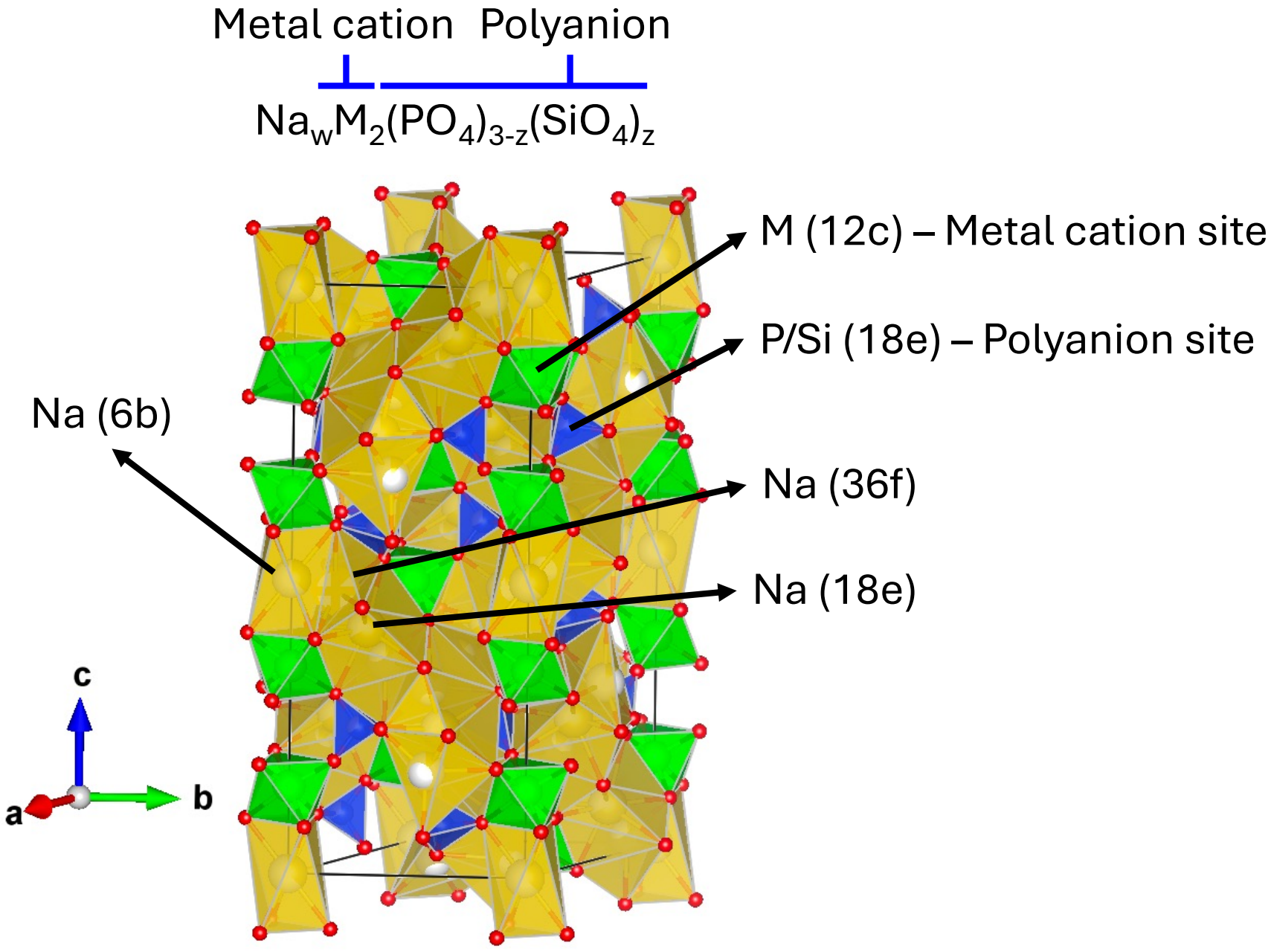

Assuming a fully random configuration for each site, NASICONs with a sodium content of $w \leq 4$ only occupy the 6b and 18e sites, whereas those with $w > 4$ access all three Na sites. This assumption provides an upper bound, as partial or full cation ordering typically decreases the true configurational entropy. Based on this framework, the ideal configurational entropy per atom ($s_{ideal,config}$) is defined as:

$$s_{ideal,config} = -\frac{k_B}{w+17}\left(w\tilde{s}_{Na} + 2\tilde{s}_M + 3\tilde{s}_A\right)$$

$$\tilde{s}_{Na} = \begin{cases} \frac{w}{4}\ln\left(\frac{w}{4}\right) + \left(1-\frac{w}{4}\right)\ln\left(1-\frac{w}{4}\right), & 0 \leq w \leq 4 \\ \frac{w}{10}\ln\left(\frac{w}{10}\right) + \left(1-\frac{w}{10}\right)\ln\left(1-\frac{w}{10}\right), & w > 4 \end{cases}$$

$$\tilde{s}_M = \sum_i \frac{y_i}{2}\ln\left(\frac{y_i}{2}\right)$$

$$\tilde{s}_A = \frac{z}{3}\ln\left(\frac{z}{3}\right) + \left(1-\frac{z}{3}\right)\ln\left(1-\frac{z}{3}\right)$$

where $k_B$ is the Boltzmann constant, $w$ is the sodium content, $y_i$ is the stoichiometric content of metal cation $i$, and $z$ is the silicon content per normalized formula unit of $Na_wM_2(PO_4)_{3-z}(SiO_4)_z$.

Note that $w$ is used as the prefactor for $\tilde{s}_{Na}$ (rather than the total site multiplicity of 4 or 10) to act as an empirical dampening factor. A fully random distribution across all Na sites is physically unlikely, as the sites fill sequentially (6b, then 18e, then 36f); substituting $w$ prevents artificial inflation of the sodium entropy contribution. To calculate the contribution of this entropy to the energy above the hull at temperature $T$, the augmented energy above the hull ($E_{hull,ideal}$) is evaluated as:

$$E_{hull,ideal} = E_{hull} - Ts_{ideal,config}$$

where the $E_{hull}$ of the NASICON phase is obtained from the Materials Project[6,7].

## 5. Empirical metastability limits of phosphate, silicate, and phosphosilicate NASICONs

To compare the synthesis outcomes of the phosphates, silicates, and phosphosilicates with their thermodynamic stability, we mapped the phosphate, silicate, and phosphosilicate trials into the $\pi_{Thermo}$ feature space (**Supplementary Figure 16**). We overlaid the synthetic accessibility boundaries derived from the SVCs across four distinct $E_{threshold}$ limits: the thermodynamically stable limit (0 meV/atom), the empirical metastability limit for phosphates and silicates (15 meV/atom), the practical metastability limit for phosphosilicates (41 meV/atom), and the absolute $\pi_{Thermo}$ accessibility limit for phosphosilicates (62.22 meV/atom). Synthesizable phosphates and silicates are completely encompassed within the 15 meV/atom boundary, whereas several successful phosphosilicates are not. Distinctively, most (and eventually all) phosphosilicates are captured within the synthetically accessible domains defined by the 41 meV/atom and 62.22 meV/atom thresholds, respectively. While we did not establish a direct $E_{hull,ideal}$ limit for phosphosilicates, these findings indicate that NASICONs with pure phosphate or pure silicate anion groups possess a distinct empirical metastability limit that is significantly lower than that of mixed phosphosilicate systems.

## 6. Precursor details

Prior to use in the A-Lab, the $Na_2CO_3$ and $NH_4H_2PO_4$ powders were each pre-processed separately in 10 g batches to achieve a finer particle size. This milling was performed in a polypropylene vial containing ten 5 mm $ZrO_2$ balls using a Hauschild Smart DAC250 centrifugal mixer.

| Formula | Purity | Notes and product number |
|---|---|---|
| $Na_2CO_3$ | ≥ 99.5% | Anhydrous, 222321 |
| $HfO_2$ | 98% | 202118 |
| $ZrO_2$ | 99% trace metals basis | 5 µm, 230693 |
| $Ta_2O_5$ | 99% trace metals basis | 303518 |
| $Sc_2O_3$ | 99.9% trace rare earth metals basis | 307874 |
| $Nb_2O_5$ | 99.9% trace metals basis | 325 mesh, 208515 |
| MgO | ≥ 99% trace metals basis | 325 mesh, 342793 |
| $TiO_2$ | 99.8% trace metals basis | Anatase, 232033 |
| $In_2O_3$ | 99.99% trace metals basis | 289418 |
| $SnO_2$ | 99.9% trace metals basis | 325 mesh, 244651 |
| $Y_2O_3$ | 99.99% trace metals basis | 205168 |
| $Ga_2O_3$ | ≥ 99.99% trace metals basis | 215066 |
| $Al_2O_3$ | N/A | Nanopowder <50 nm particle size (TEM), 544833 |
| ZnO | N/A | Nanopowder <50 nm particle size, 544906 |
| $GeO_2$ | 99.998% trace metals basis | 199478 |
| $La_2O_3$ | ≥ 99.9% | L4000 |
| $NH_4H_2PO_4$ | ≥ 98% | 216003 |
| $SiO_2$ | 99.5% trace metals basis | 325 mesh, 342890 |

## 7. Support vector classifier (SVC) test set performance comparison (5-fold cross-validation)

| Classifier | Average accuracy (%) | Average recall (%) | Average F1-score (%) | Average training and test data amount (A = Synthetically Accessible, I = Synthetically Inaccessible) |
|---|---|---|---|---|
| Before descriptor change | 82.08 | 83.26 | 74.70 | Training: 512 (A) and 2592 (I)<br>Test: 129 (A) and 648 (I) |

| Classifier | Average accuracy (%) | Average recall (%) | Average F1-score (%) | Average training and test data amount (A = Synthetically Accessible, I = Synthetically Inaccessible) |
|---|---|---|---|---|
| After descriptor change | 81.60 | 82.22 | 74.00 | Training: 512 (A) and 2592 (I)<br>Test: 129 (A) and 648 (I) |
| After descriptor and threshold changes | 87.25 | 86.91 | 84.78 | Training: 2259 (A) and 845 (I)<br>Test: 565 (A) and 212 (I) |

## 8. Ionic conductivity cost-function components

$$\pi_{\bar{R}_M} = \begin{cases} 0.70 - \bar{R}_M, & \text{If } \bar{R}_M < 0.70 \text{ Å} \\ \bar{R}_M - 0.74, & \text{Else If } \bar{R}_M > 0.74 \text{ Å} \\ 0, & \text{otherwise} \end{cases}$$

$$\pi_{N_{Na}} = \begin{cases} 2.5 - N_{Na}, & \text{If } N_{Na} < 2.5 \\ N_{Na} - 3.5, & \text{Else If } N_{Na} > 3.5 \\ 0, & \text{otherwise} \end{cases}$$

$$\pi_{\bar{R}_A} = \begin{cases} 0.215 - \bar{R}_A, & \text{If } \bar{R}_A < 0.215 \text{ Å} \\ \bar{R}_A - 0.245, & \text{Else If } \bar{R}_A > 0.245 \text{ Å} \\ 0, & \text{otherwise} \end{cases}$$

## 9. Vegard's approximation training and $\pi_{LP}$ details

To link composition to structural metrics, we constructed two multiple linear regression models[8] rooted in Vegard's approximation[2,9–11]. The training dataset comprised 102 literature-reported structures from ICSD and 11 in-house synthesized single-cation NASICONs; to ensure consistency, we exclusively selected data collected at ambient pressure and near room temperature (250 K < T < 305 K). Using the average polyanion center ion radius ($\bar{R}_A$), cation radius ($\bar{R}_M$), and sodium content ($N_{Na}$) as predictors, these models independently estimate the $a$-lattice parameter and unit cell volume ($V$), from which the $c$-lattice parameter is geometrically derived. While the global models yield high predictive accuracy ($R^2$ values of 0.93 and 0.89 for $a$ and $V$, respectively; **Supplementary Figure 17a-b**), precise decision-making within specific sub-systems requires minimized local error. To achieve this, we implemented an adaptive fine-tuning strategy for each chemical space: during the fitting process, compositions within the target chemical space are assigned a weight of 10,000. This forces the regressor to prioritize local data fidelity while still leveraging global trends to constrain predictions in sparse regions. An example for NASICONs in the Na-Mn-V-Ti-P-O chemical space is shown in **Supplementary Figure 17c-d**, where the general Vegard's approximation model yields a significantly higher mean absolute error (MAE) and a lower $R^2$ value compared to the fine-tuned model. Finally, the lattice parameters feedback cost function ($\pi_{LP}$) is defined as the mean percentage difference between the model-predicted values for a candidate composition (ϒ) and the experimentally refined values obtained from a prior synthesis trial:

$$\pi_{LP} = \frac{1}{3}\left(\frac{|a_{XRD} - a_{\Upsilon}|}{a_{XRD}} + \frac{|c_{XRD} - c_{\Upsilon}|}{c_{XRD}} + \frac{|V_{XRD} - V_{\Upsilon}|}{V_{XRD}}\right) \times 100\%$$

where $a_{XRD}$, $c_{XRD}$, and $V_{XRD}$ are the unit cell $a$ and $c$ lattice parameters and volume, respectively, obtained from the experimentally refined values. The variables with the ϒ subscript denote the corresponding structural values predicted by the regressors. An example using the as-synthesized $Na_2TiSc(PO_4)_3$ NASICON is shown in **Supplementary Figure 17e**, where the minimum of $\pi_{LP}$ accurately corresponds to $Na_2TiSc(PO_4)_3$ composition. This demonstrates that the experimentally refined lattice parameters and volume of the synthesized material closely match those predicted by the linear regressors, confirming the self-consistency of the predictive model against the experimental result.

**10. Calibration for high purity and major NASICON samples**

To mitigate the influence of automated refinement artifacts on the purity assessment, we calibrated the Purity Score such that a value of 1.0 corresponds to samples with negligible impurity content. We assembled a calibration dataset of 20 high-purity samples (**Supplementary Data**), with representative refined diffractograms shown in **Supplementary Figure 18a-c**. Since the raw Estimated Purity (EP) is sensitive to fit quality, suboptimal automated refinement can result in artificially low scores. This is exemplified in **Supplementary Figure 18c**, where automated refinement yielded a low EP due to poor fit ($R_{wp}$=14.52%). In contrast, manual refinement of the same pattern (**Supplementary Figure 18e**) achieved a significantly better fit ($R_{wp}$=6.78%), confirming the sample's actual high purity and necessitating this calibration.

To establish the threshold defining a "major" NASICON phase, we performed a sensitivity analysis by incrementally increasing the Purity Score threshold in steps of 0.1. The criterion for a major phase was defined as the NASICON maximum peak intensity exceeding that of any unidentified secondary peaks. Representative diffractograms for samples meeting this criterion are shown in **Supplementary Figure 19a-c**, while a sample falling below it is shown in **Supplementary Figure 19d**. This comparison determined that a Purity Score threshold of 0.7 conservatively distinguishes samples with a predominant NASICON phase from those dominated by impurities.

**11. Multimodal NASICON composition analysis**

To resolve the composition of the synthesized NASICON phase, we employed a multimodal approach integrating thermodynamic and lattice-parameter-based screening with automated SEM-EDS characterization. The analysis followed a three-step workflow: (1) Automated SEM-EDS data acquisition, filtering, and *k*-means clustering; (2) Hybrid optimization for composition assignment; and (3) Final phase identification. The specific details of each step are outlined below:

1. **Automated SEM-EDS data acquisition, filtering, and *k*-means clustering**: Compositional measurement was performed using the AutoEMX framework[3] on a Thermo Fisher Phenom XL G2 desktop SEM equipped with an Ametek FAST SDD (controlled via the PyPhenom Python API). Powder samples were dispersed onto carbon tape on aluminum stubs, and for each sample, 200 individual 50,000-count EDS spot spectra were autonomously acquired at 15 kV on randomly selected particles. Following the automated acquisition, principled filtering of unreliable spectra and *k*-means clustering were performed to identify and extract the compositions of the individual constituent phases in each sample, as established in prior work[3].
2. **Hybrid optimization for composition assignment:** Each resulting cluster possesses a centroid representing the mean of its compositional distribution. To identify the best-fitting NASICON composition for each cluster, we employed a hybrid optimization scheme. To ensure an exhaustive evaluation, we utilized the composition-to-structure relationships derived from Vegard's approximation (see **Methods** and **Supplementary Information 9**). This, alongside the uncertainty-aware verification described in the XRD phase analysis (see **Methods**), constrained the viable NASICON candidates to those matching the observed lattice parameters. We performed a global search to identify 100 distinct compositions satisfying this constraint. Simultaneously, a fine-grid search was conducted around the nominal target composition, generating an additional 5,000 proximate candidates. These candidates were screened using the thermodynamic cost function ($\pi_{Thermo}$) to eliminate synthetically inaccessible phases, yielding a final pool of 67-92 globally diverse and 3,974-5,000 local candidate compositions per sample. Initialized with these globally diverse candidates and a subset of the local candidates (totaling 250 points), the algorithm executed differential evolution followed by gradient-based BFGS optimization. The objective function minimized a composite cost comprising $\pi_{Thermo}$, the Euclidean compositional distance to the cluster centroid, and a lattice parameter penalty restricting candidates to within the intrinsic uncertainty of Vegard's approximation plus 15% of the chemical system's total parameter or volume range. By incorporating these constraints, this hybrid optimization resolves an optimal solution that accounts for the inherent experimental uncertainties of both XRD and SEM-EDS techniques.

3. **Final phase identification:** To rigorously determine which candidate compositions best match each measured cluster, the best-fitting NASICON compositions for each cluster, together with impurity phases identified from XRD pattern refinement, were fed as candidate phases into the AutoEMX clustering analysis routine. This routine systematically evaluates whether clusters stem from single-phase or two-phase mixtures of candidate phases to accurately resolve the constituent phases for each cluster. The compositions and phase combinations with the highest statistical confidence were ultimately selected and reported.

## 12. Supplementary Figures 1-19

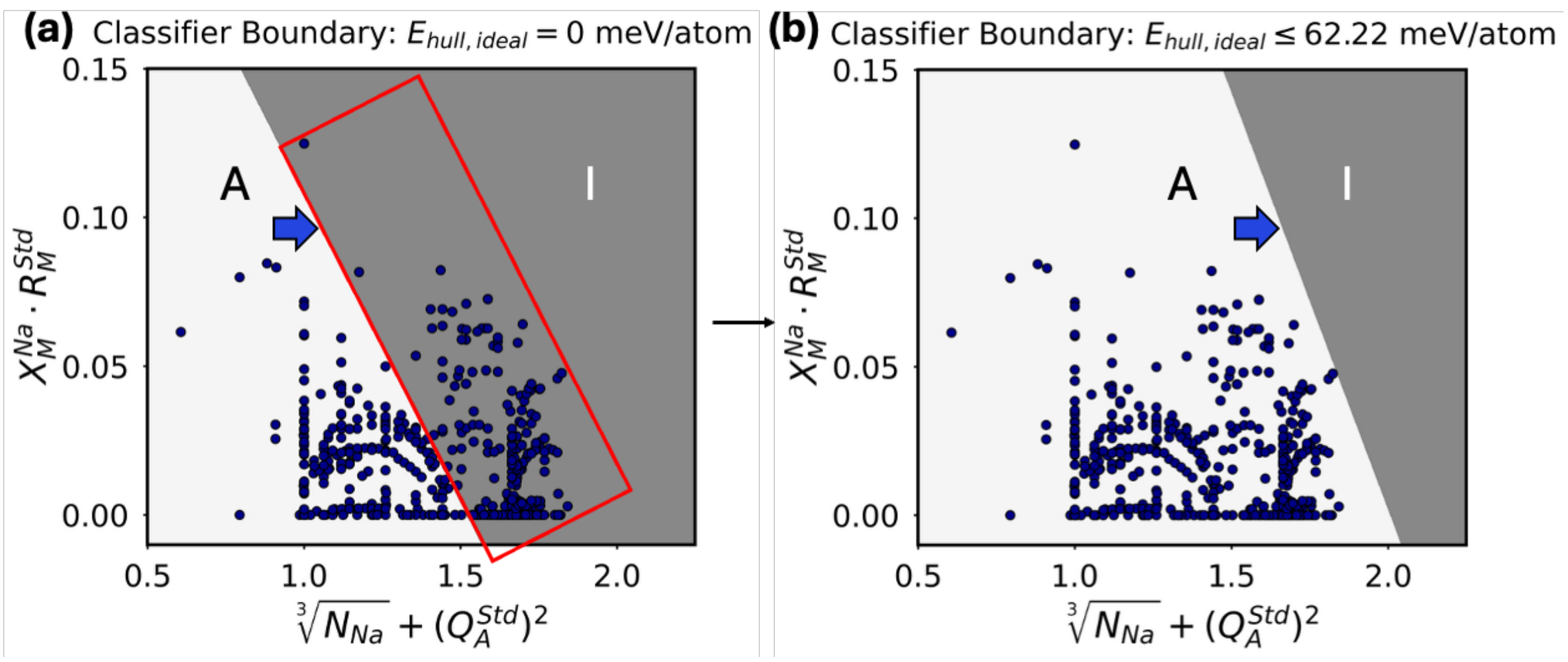

**Supplementary Figure 1. Threshold adjustment to resolve classification discrepancies and define $\pi_{Thermo}$.** Mapping of 524 experimentally verified, literature-mined NASICON compositions (dark blue dots) into the reduced thermodynamic feature space. The background shading delineates the support vector classifier (SVC) prediction regions: synthetically accessible (**A**, light region) and synthetically inaccessible (**I**, dark gray region). **(a)** The SVC decision boundary trained using the original strict stability threshold of $E_{hull,ideal} = 0$ meV/atom. Under this stringent threshold, a significant number of known, experimentally verified compositions are incorrectly classified as inaccessible, as highlighted by the red box. The black arrow between the panels represents the threshold relaxation process. **(b)** The updated SVC decision boundary trained using an adjusted stability threshold of $E_{hull,ideal} \leq 62.22$ meV/atom. The blue arrows in both panels illustrate the resulting rightward shift of the decision boundary. This shifted boundary represents the minimum threshold required to correctly encompass all 524 verified compositions within the accessible (**A**) region, effectively resolving the prior discrepancy. The axes represent combinations of the four retained stability descriptors: the electronegativity difference between Na and the cations ($X_M^{Na}$), cation radius deviation ($R_M^{STD}$), sodium content ($N_{Na}$), and anion charge deviation ($Q_A^{STD}$).

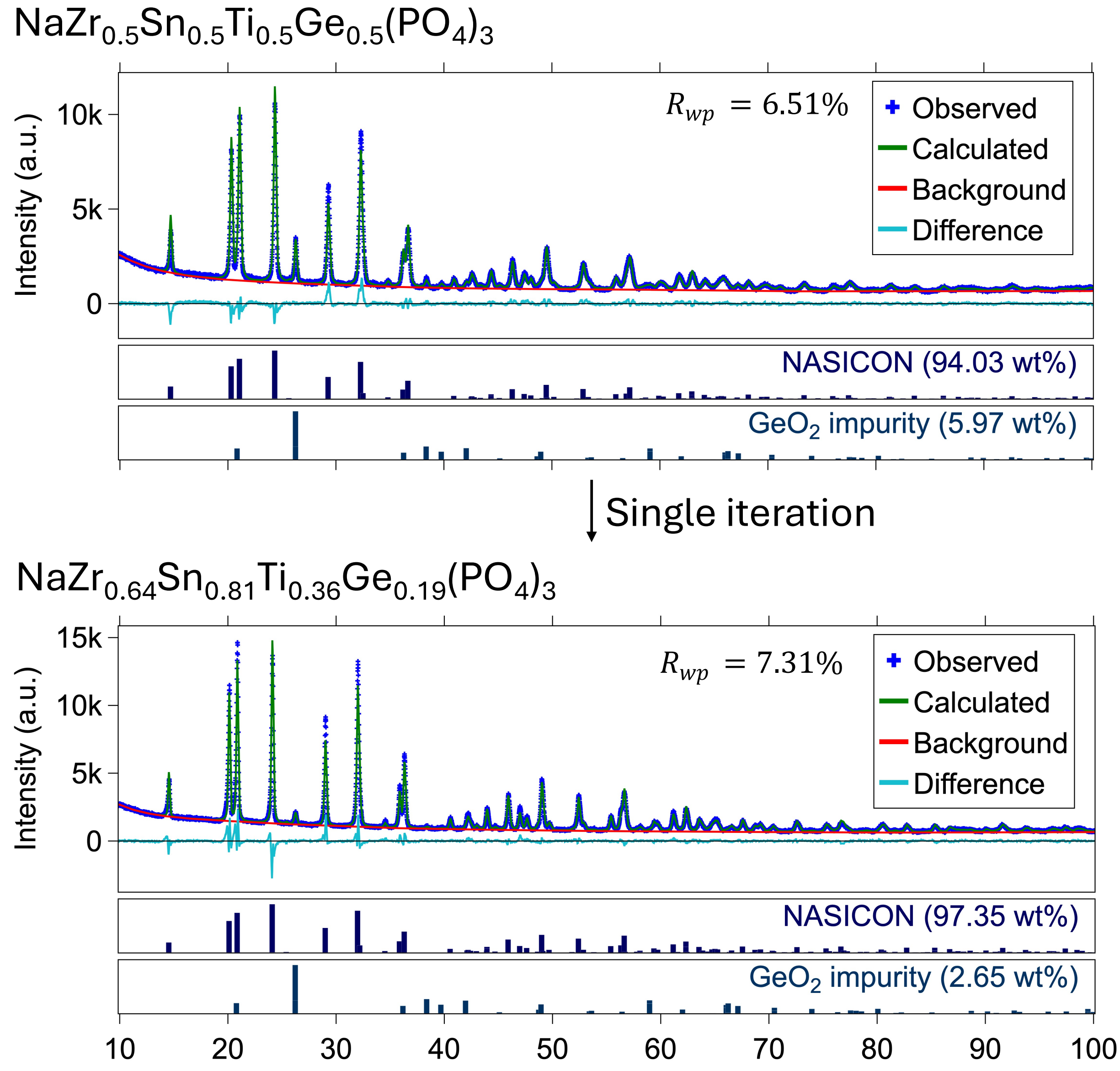

**Supplementary Figure 2. Validation of synthetic accessibility cost functions (Zr-Sn-Ti-Ge phosphate).** Diffractograms from Rietveld refinements of XRD patterns comparing the initial synthesis trial (top) and the subsequent CASS-designed trial (bottom) for the Zr-Sn-Ti-Ge pseudoquaternary phosphate system.

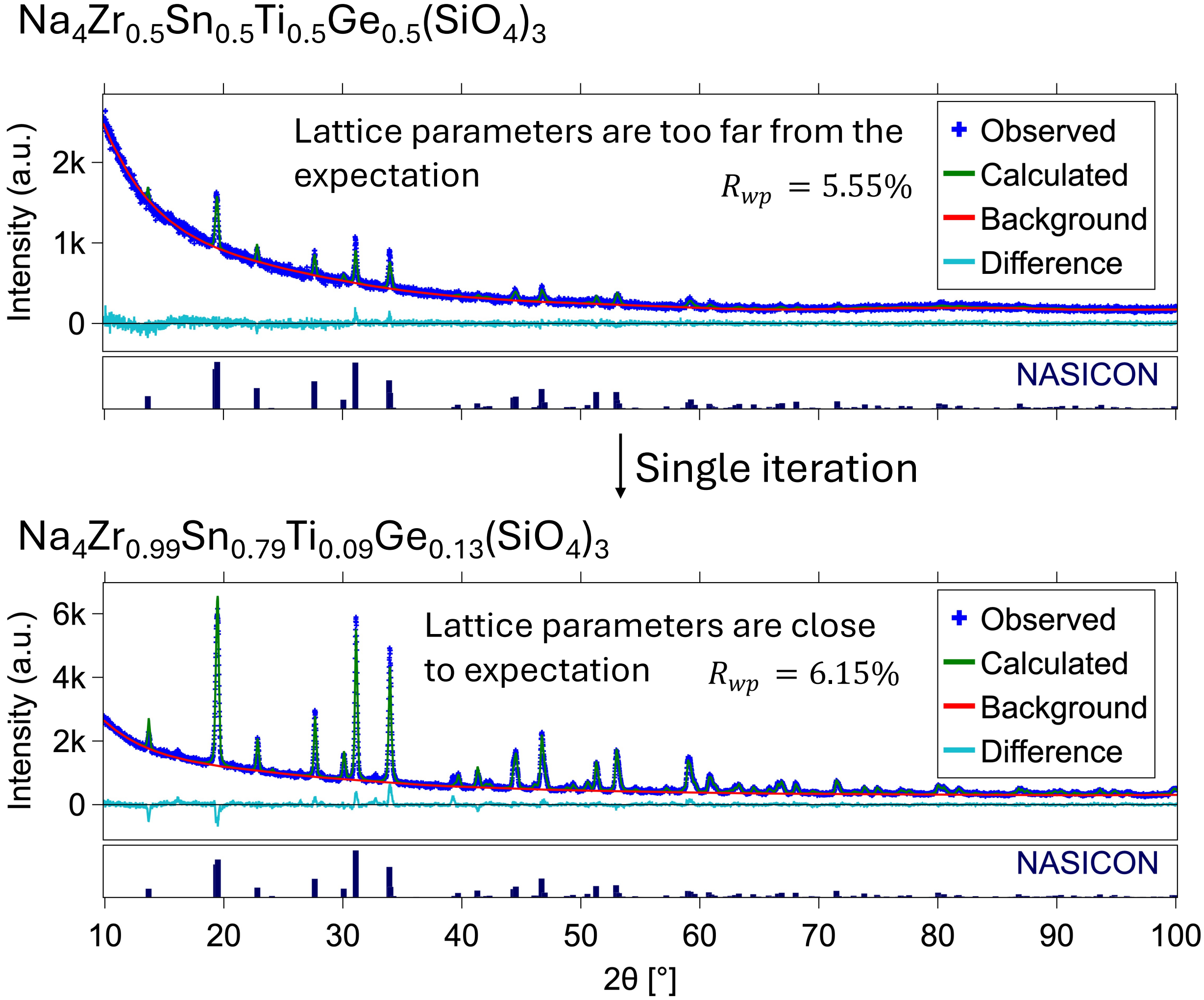

**Supplementary Figure 3. Validation of synthetic accessibility cost functions (Zr-Sn-Ti-Ge silicate).** Comparison of Rietveld-refined XRD patterns for an initial synthesis trial (top) and a subsequent CASS-designed trial (bottom). The initial trial resulted in a phase with lattice parameters deviating significantly from theoretical predictions for the target composition, indicating a failure to form the desired solid solution. In contrast, the CASS-designed trial yielded lattice parameters closely matching expected values of the new design, demonstrating the system's ability to successfully navigate toward synthetically accessible regions of the design space.

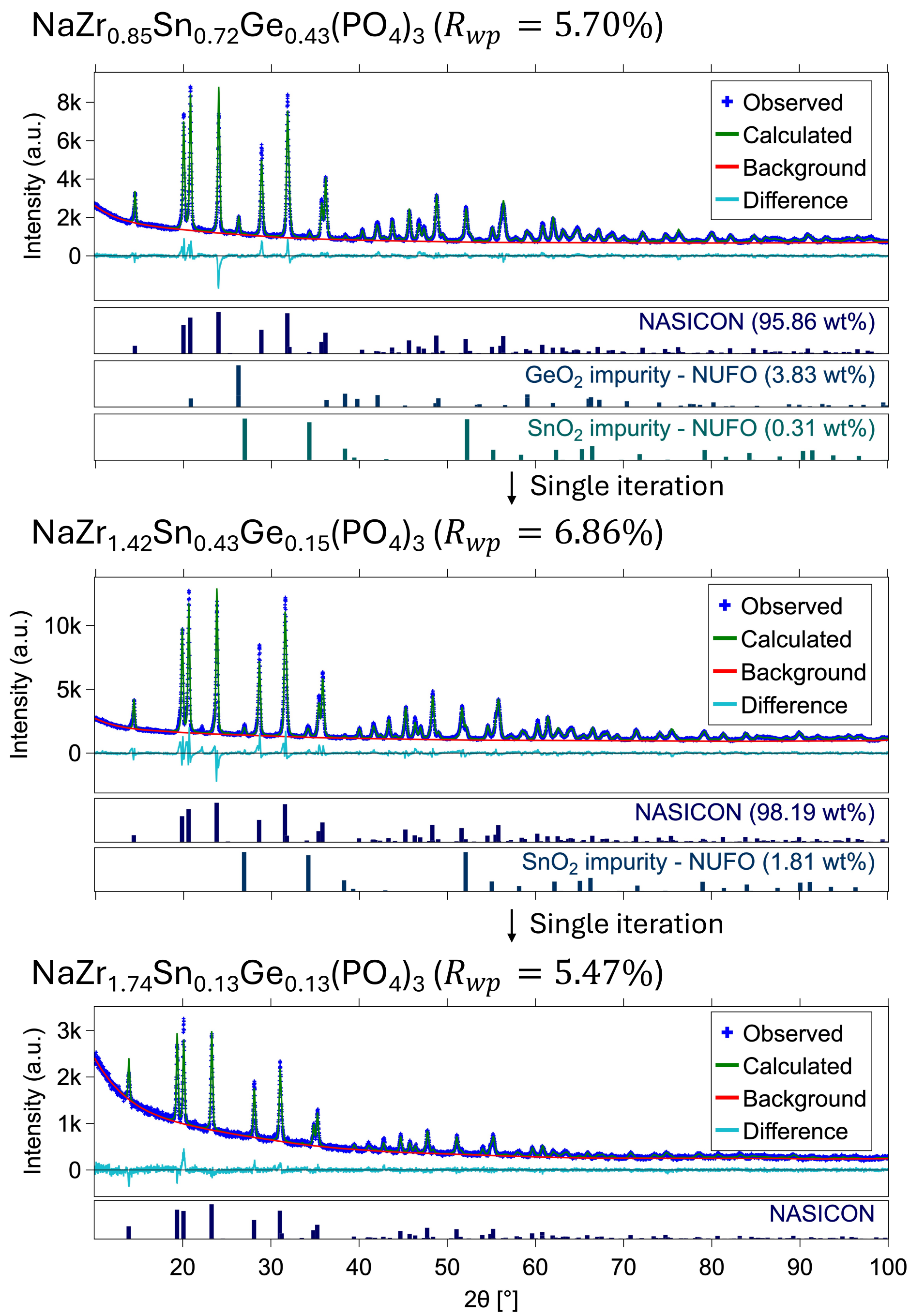

**Supplementary Figure 4. Validation of synthetic accessibility cost functions (Zr-Sn-Ge phosphate).** Diffractograms from Rietveld refinements of XRD patterns comparing the initial synthesis trial (top) and the subsequent CASS-designed trials (middle, bottom) for the Zr-Sn-Ge pseudoternary phosphate system. "NUFO" label indicates that the impurity was not used for optimization since it was later added manually and was not detected within the automated phase search during the CASS loop.

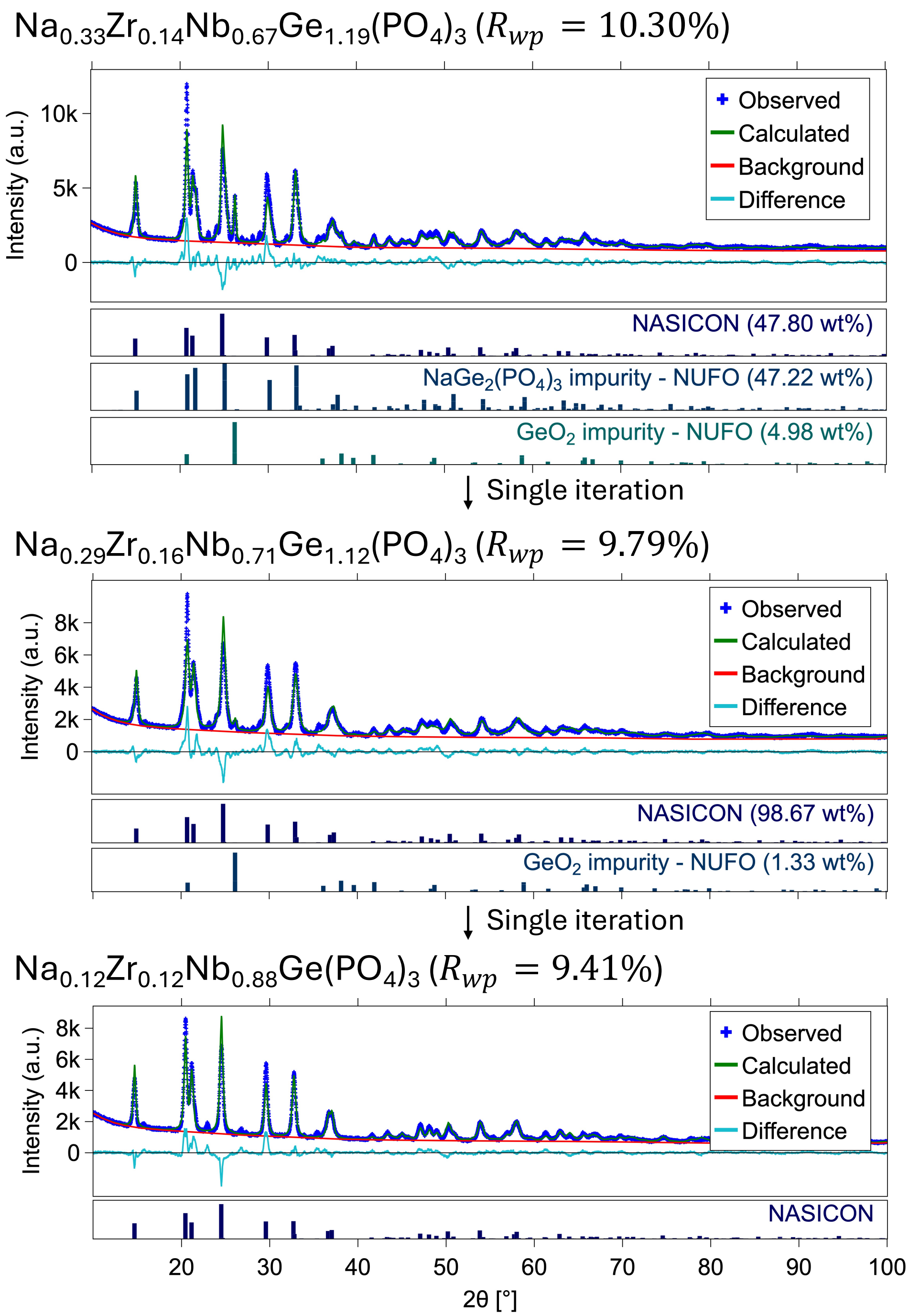

**Supplementary Figure 5. Validation of synthetic accessibility cost functions (Zr-Nb-Ge phosphate).** Diffractograms from Rietveld refinements of XRD patterns comparing the initial synthesis trial (top) and the subsequent CASS-designed trials (middle, bottom) for the Zr-Nb-Ge pseudoternary phosphate system. "NUFO" label indicates that the impurity was not used for optimization since it was later added manually and was not detected within the automated phase search during the CASS loop.

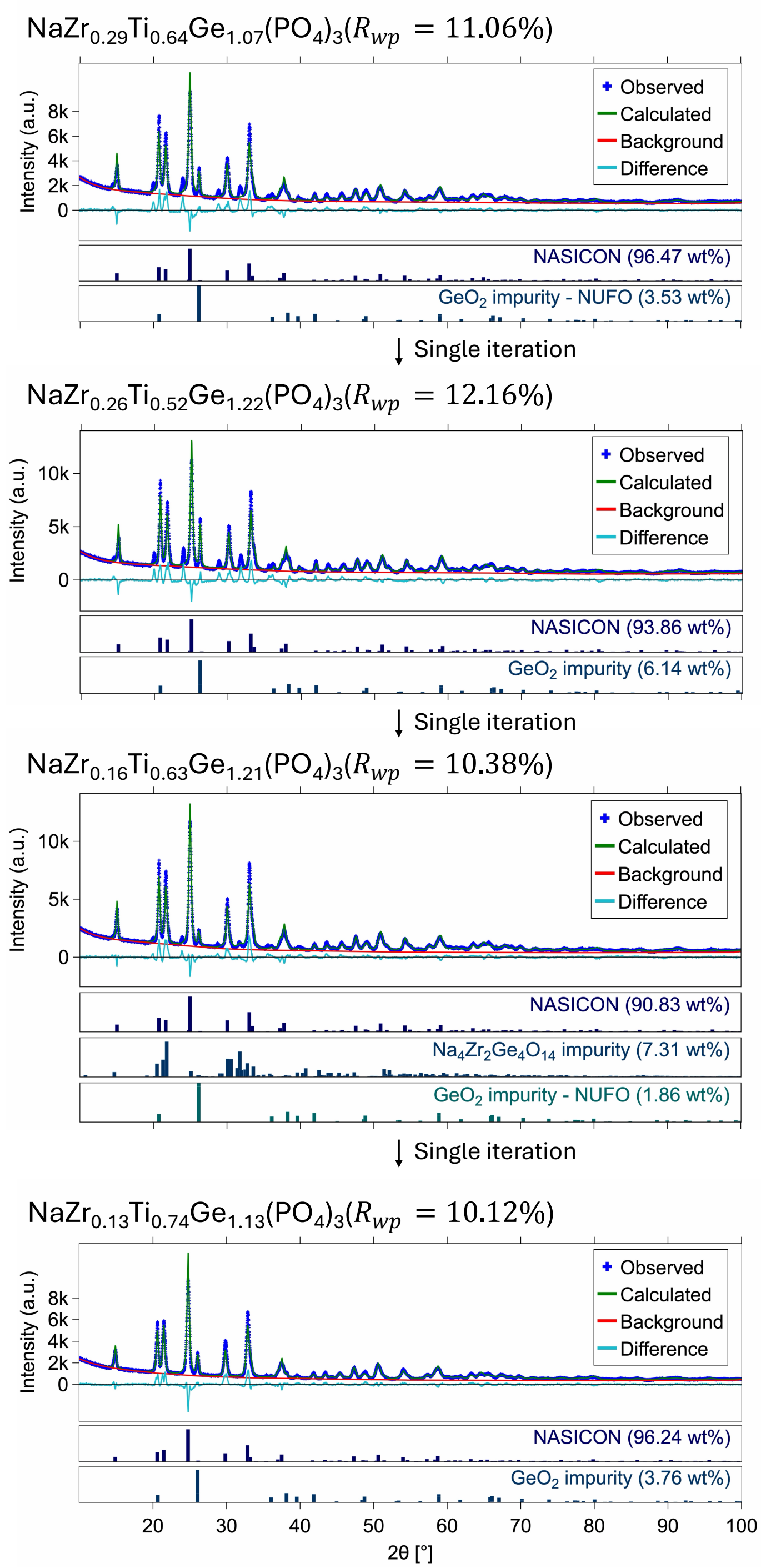

**Supplementary Figure 6. Validation of synthetic accessibility cost functions (Zr-Ti-Ge phosphate).** Diffractograms from Rietveld refinements of XRD patterns comparing the initial synthesis trial (top) and the subsequent CASS-designed trials (second, third, bottom) for the Zr-Ti-Ge pseudoternary phosphate system. "NUFO" label indicates that the impurity was not used for optimization since it was later added manually and was not detected within the automated phase search during the CASS loop.

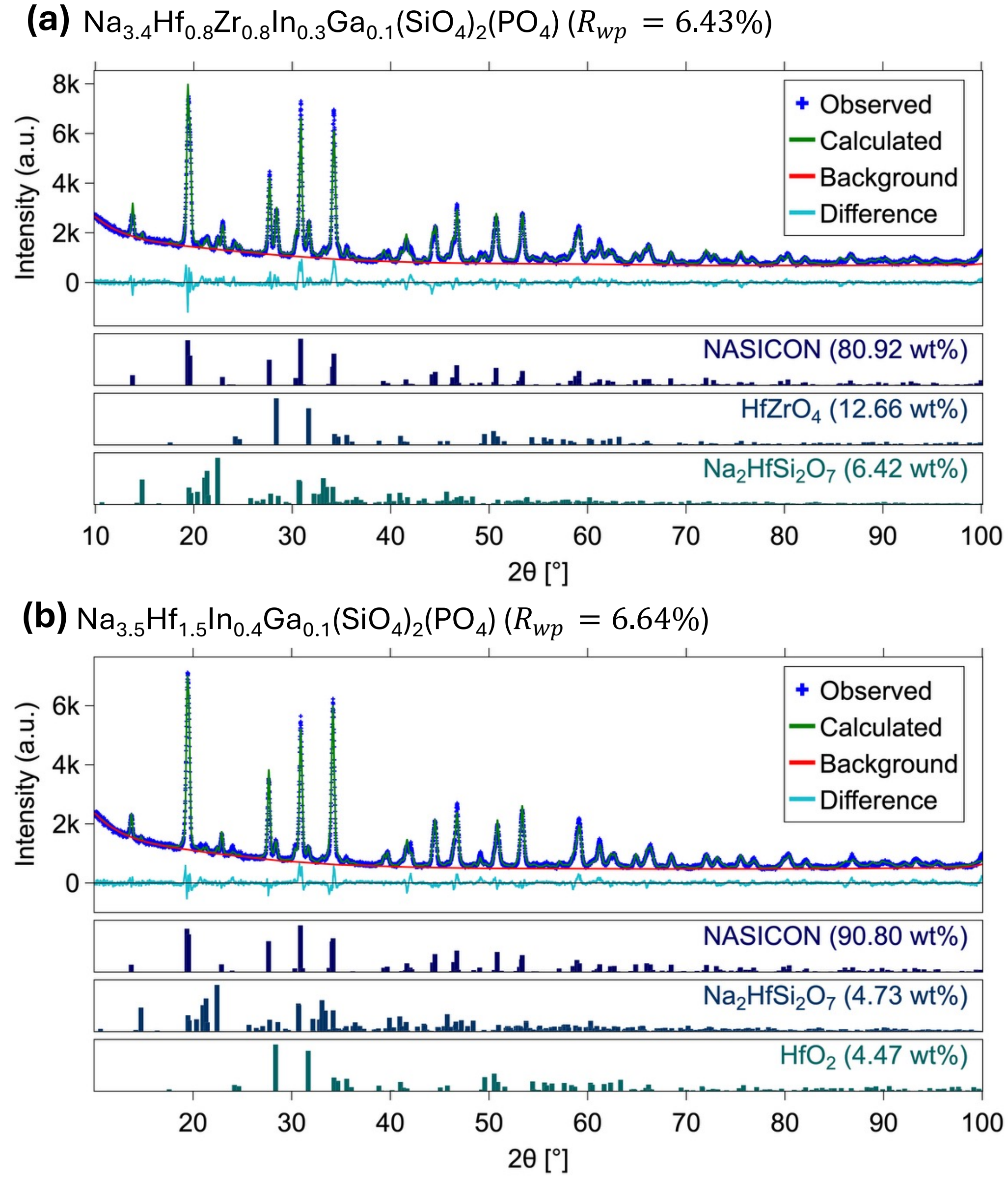

**Supplementary Figure 7. Diffractograms from Rietveld refinements of XRD patterns for the respective as-synthesized promising NASICON ionic conductors.** (a) $Na_{3.4}Hf_{0.8}Zr_{0.8}In_{0.3}Ga_{0.1}(SiO_4)_2(PO_4)$ and (b) $Na_{3.5}Hf_{1.5}In_{0.4}Ga_{0.1}(SiO_4)_2(PO_4)$, displaying observed (blue), calculated (green), background (red), and difference (cyan) profiles alongside weighted profile R-factors ($R_{wp}$).

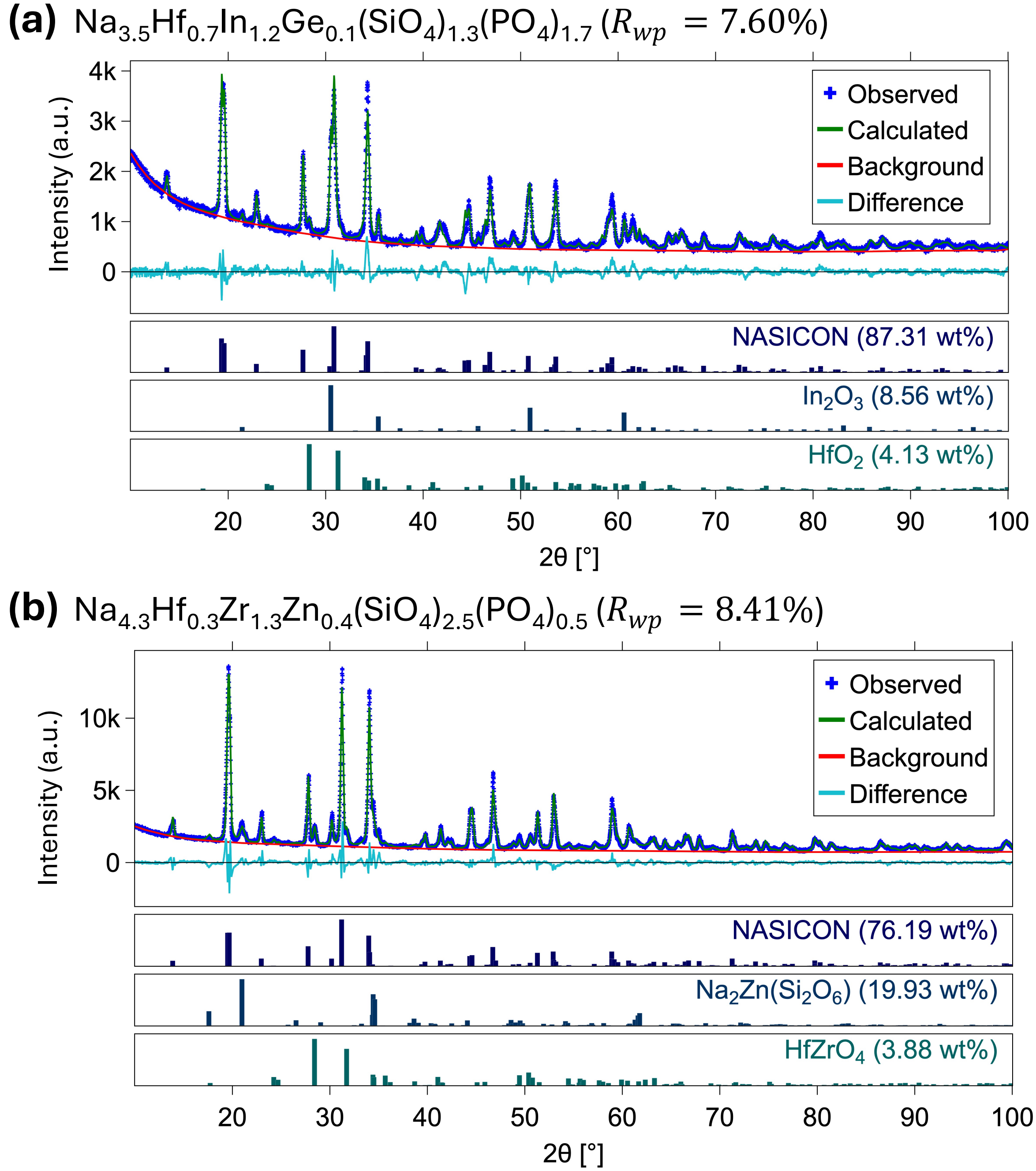

**Supplementary Figure 8. Diffractograms from Rietveld refinements of XRD patterns for the respective as-synthesized less ionically conducting samples**. The observed (blue crosses), calculated (green line), background (red line), and difference (cyan line) profiles are shown. The weighted profile R-factors ($\boldsymbol{R_{wp}}$) are indicated for each refinement. **(a)** $Na_{3.5}Hf_{0.7}In_{1.2}Ge_{0.1}(SiO_4)_{1.3}(PO_4)_{1.7}$ and **(b)** $Na_{4.3}Hf_{0.3}Zr_{1.3}Zn_{0.4}(SiO_4)_{2.5}(PO_4)_{0.5}$.

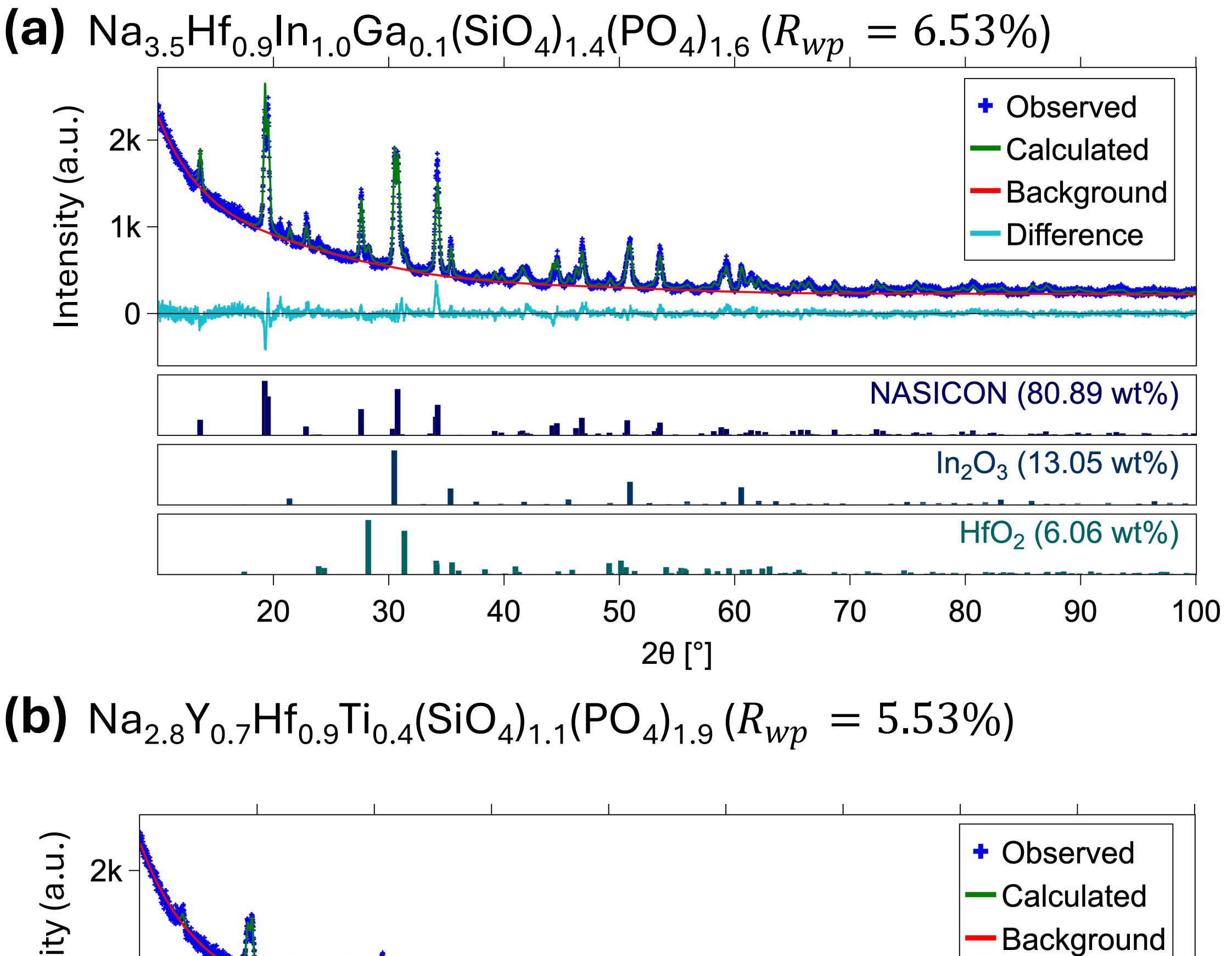

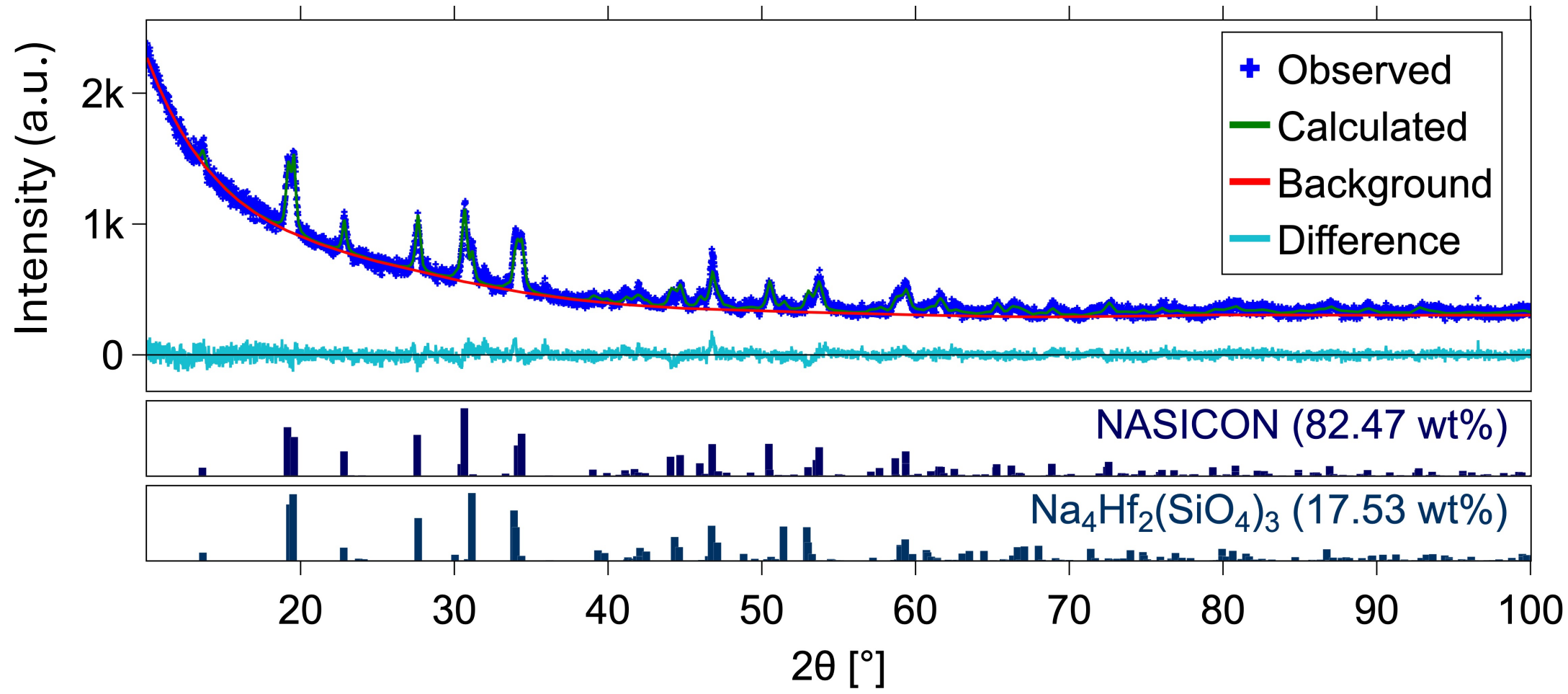

**Supplementary Figure 9. Diffractograms from Rietveld refinements of XRD patterns for the respective as-synthesized less ionically conducting samples (cont.)**. The observed (blue crosses), calculated (green line), background (red line), and difference (cyan line) profiles are shown. The weighted profile R-factors ($\mathbf{R_{wp}}$) are indicated for each refinement. **(a)** $Na_{3.5}Hf_{0.9}In_{1.0}Ga_{0.1}(SiO_4)_{1.4}(PO_4)_{1.6}$ and **(b)** $Na_{2.8}Y_{0.7}Hf_{0.9}Ti_{0.4}(SiO_4)_{1.1}(PO_4)_{1.9}$.

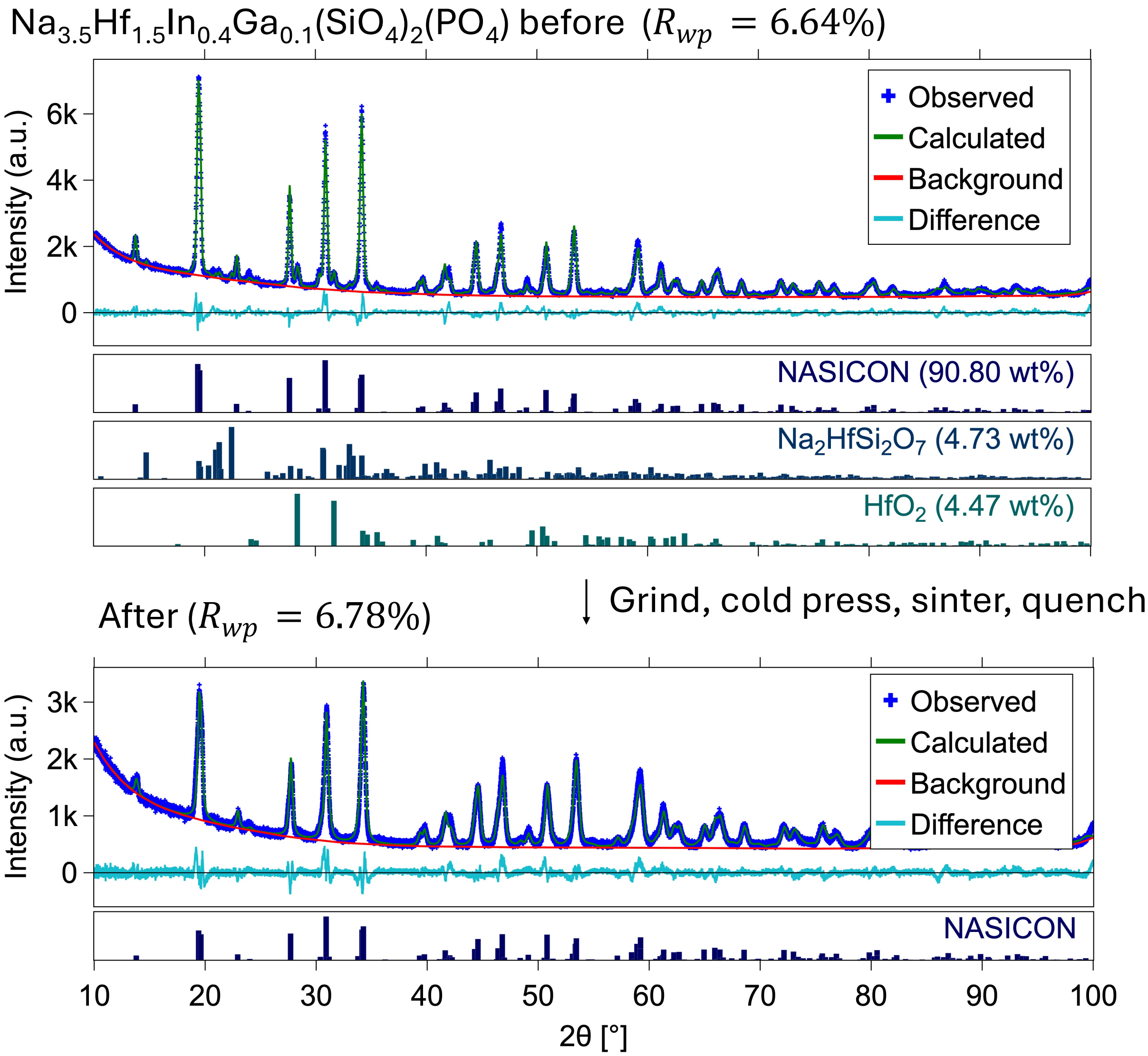

**Supplementary Figure 10. Post-processing result of $Na_{3.5}Hf_{1.5}In_{0.4}Ga_{0.1}(SiO_4)_2(PO_4)$.** Diffractograms from Rietveld refinements of XRD patterns comparing the as-synthesized sample (top) and the subsequent post-processed sample (bottom).

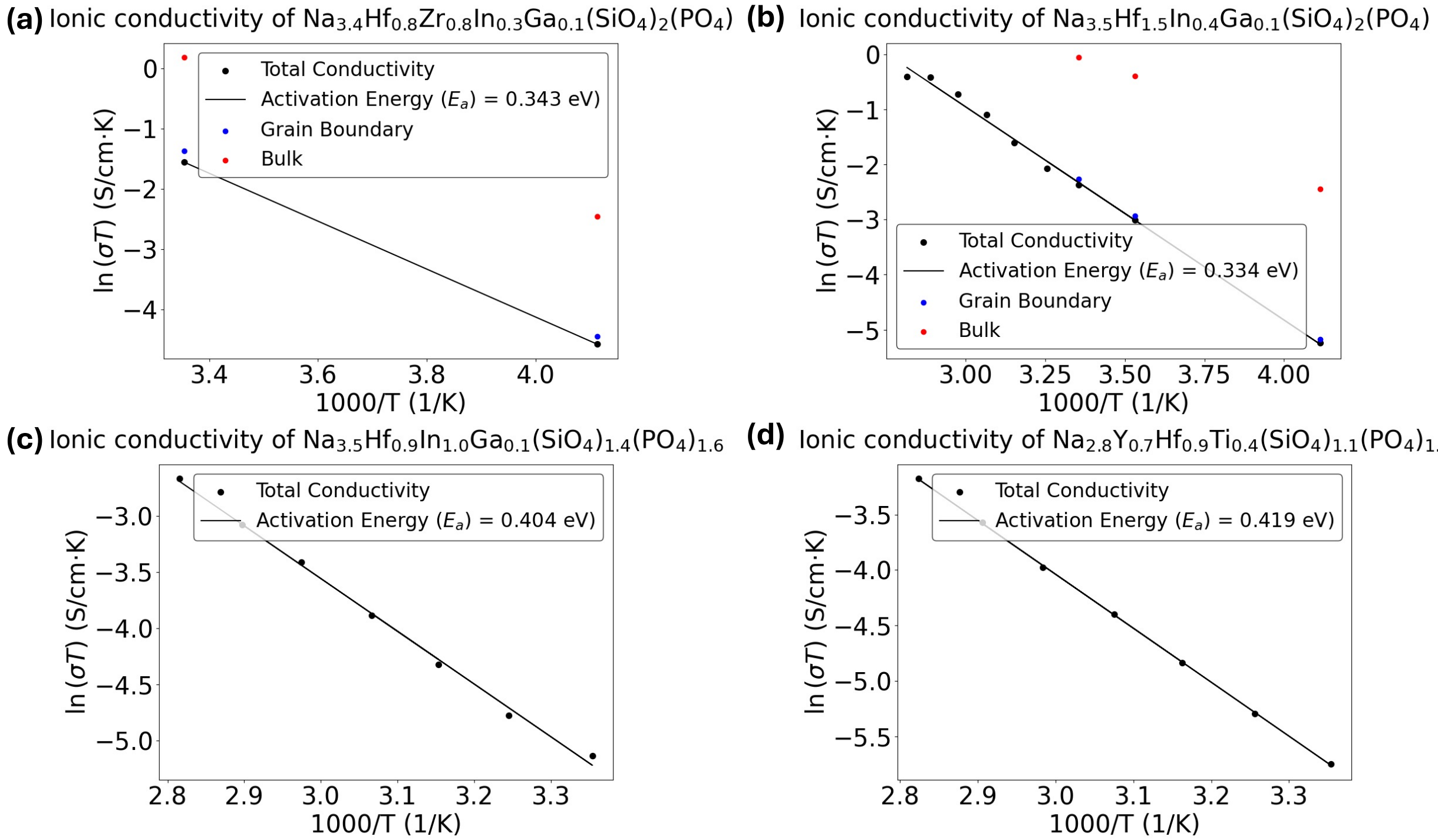

**Supplementary Figure 11. Temperature-dependent ionic conductivity profiles.** Arrhenius plots displaying the total ionic conductivity ($\sigma_{total}$), alongside resolvable bulk ($\sigma_{bulk}$) and grain boundary ($\sigma_{gb}$) components, for: **(a)** $Na_{3.4}Hf_{0.8}Zr_{0.8}In_{0.3}Ga_{0.1}(SiO_4)_2(PO_4)$, **(b)** $Na_{3.5}Hf_{1.5}In_{0.4}Ga_{0.1}(SiO_4)_2(PO_4)$, **(c)** $Na_{3.5}Hf_{0.9}In_{1.0}Ga_{0.1}(SiO_4)_{1.4}(PO_4)_{1.6}$, and **(d)** $Na_{2.8}Y_{0.7}Hf_{0.9}Ti_{0.4}(SiO_4)_{1.1}(PO_4)_{1.9}$.

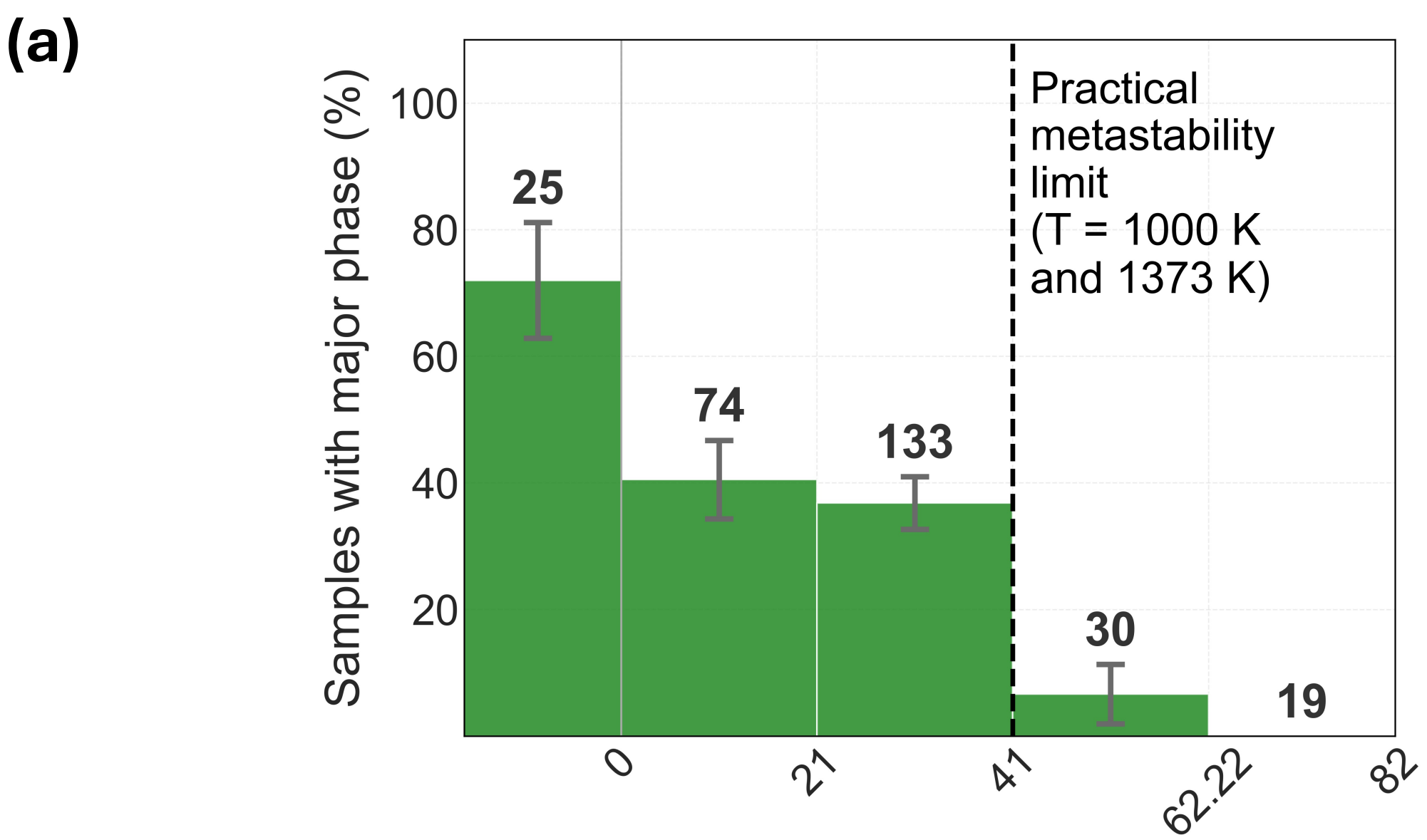
(a)
Samples with major phase (%)
100
80
60
40
20
25
74
133
30
19
Practical metastability limit (T = 1000 K and 1373 K)
0
21
41
62.22
82
$E_{\text{first stable threshold}}$ (meV/atom) at T = 1373 K

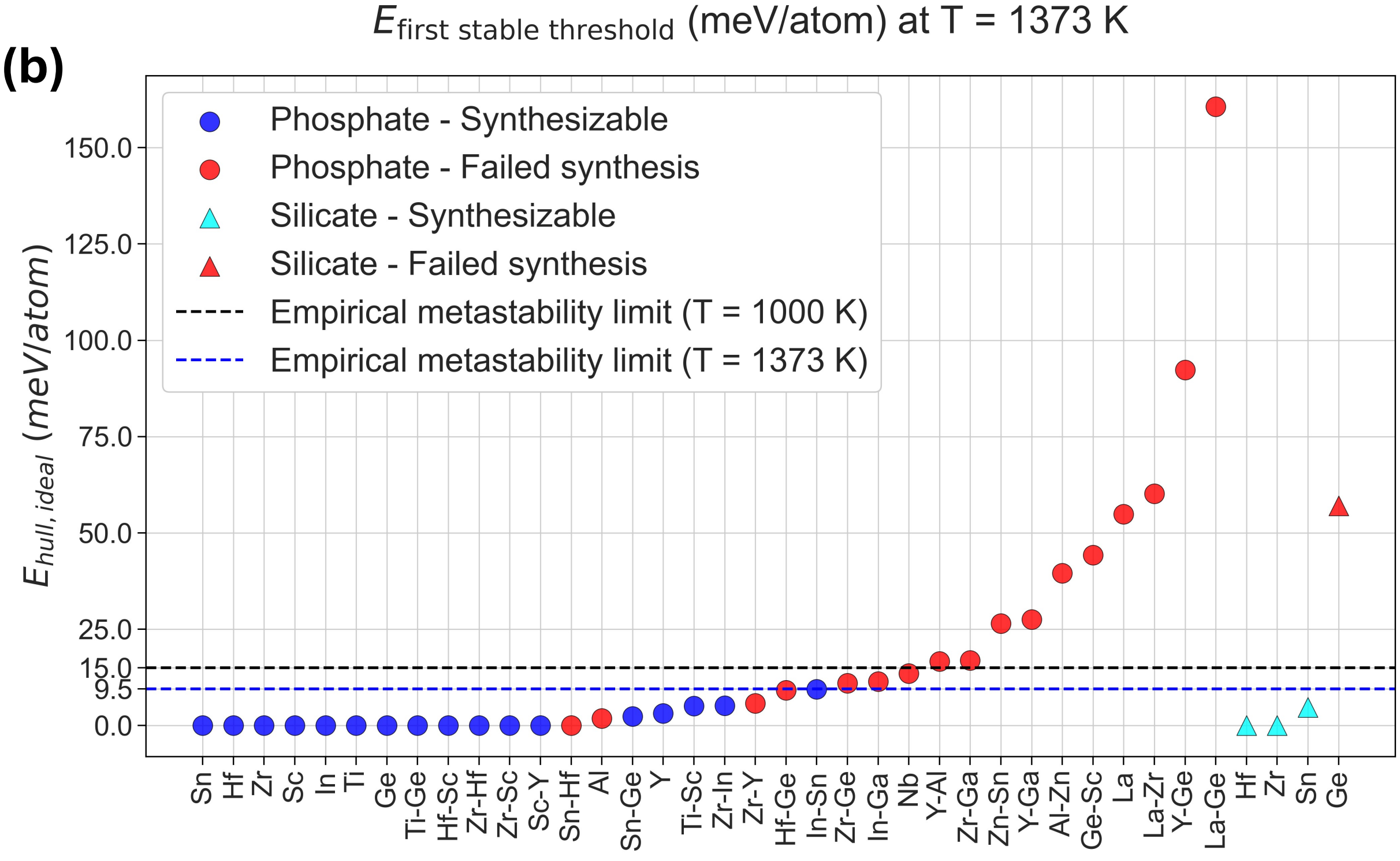
(b)
$E_{hull, ideal}$ (meV/atom)
Phosphate - Synthesizable
Phosphate - Failed synthesis
Silicate - Synthesizable
Silicate - Failed synthesis
Empirical metastability limit (T = 1000 K)
Empirical metastability limit (T = 1373 K)
150.0
125.0
100.0
75.0
50.0
25.0
15.0
9.5
0.0
Sn
Hf
Zr
Sc
In
Ti
Ge
Ti-Ge
Hf-Sc
Zr-Hf
Zr-Sc
Sc-Y
Sn-Hf
Al
Sn-Ge
Y
Ti-Sc
Zr-In
Zr-Y
Hf-Ge
In-Sn
Zr-Ge
In-Ga
Nb
Y-Al
Zr-Ga
Zn-Sn
Y-Ga
Al-Zn
Ge-Sc
La
La-Zr
Y-Ge
La-Ge
Hf
Zr
Sn
Ge

**Supplementary Figure 12. Practical and empirical metastability limits at T = 1000 K and T = 1373 K. (a)** Success rate of forming a major-phase NASICON versus the initial stability threshold ($E_{first\ stable\ threshold}$) calculated at T = 1373 K. The dashed vertical line denotes the practical metastability limit at 41 meV/atom, which remains consistent across both 1000 K and 1373 K evaluations. Numbers above the bars indicate the sample count within each bin, and error bars represent the standard error. **(b)** DFT-calculated energy above the hull augmented with ideal configurational entropy ($E_{hull,ideal}$) evaluated at the synthesis temperature of T = 1373 K for various phosphate (circles) and silicate (triangles) NASICON compositions. Synthesizable compositions (blue and cyan markers) are separated from failed syntheses (red markers) by an empirical metastability limit. The horizontal dashed lines highlight the temperature-dependent shift of this empirical limit: from 15 meV/atom at T = 1000 K (black dashed line) down to 9.5 meV/atom at T = 1373 K (blue dashed line).

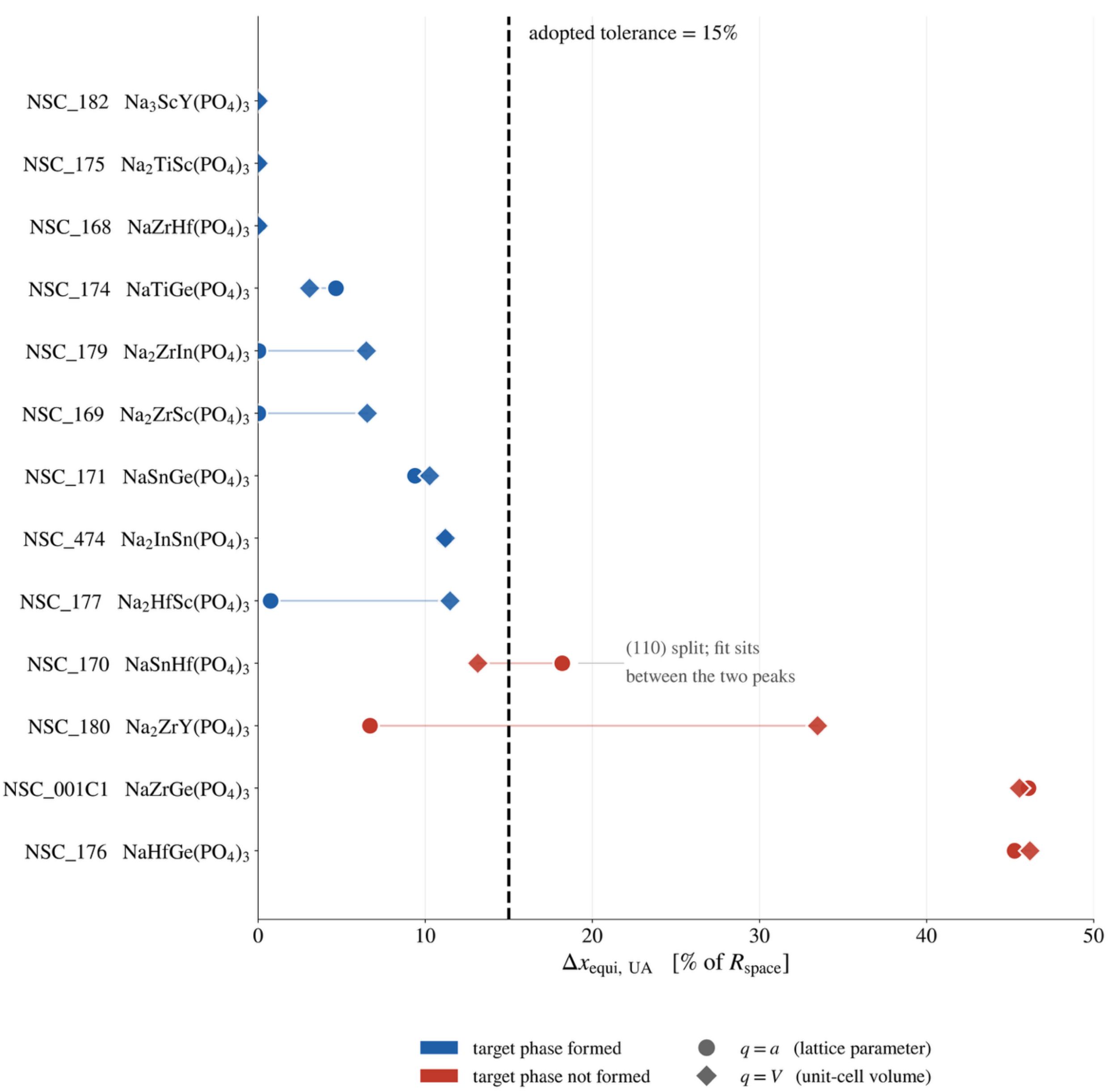

**Supplementary Figure 13.** Distance of each product from its equimolar target, beyond the model uncertainty, $\Delta x_{equi,UA}$, for the thirteen pseudobinary phosphate pairs. Each product is located on its pair by the refined $a$ (circles) and unit-cell volume (V, diamonds), between the measured end members. Blue, target phase formed; red, not formed. Dashed line, the adopted tolerance of 15%. Products that formed reach at most 11.5% on both quantities; those that did not form reach at least 18.2% on at least one. $NaSnHf(PO_4)_3$ exceeds on $a$ only and $Na_2ZrY(PO_4)_3$ on V only.

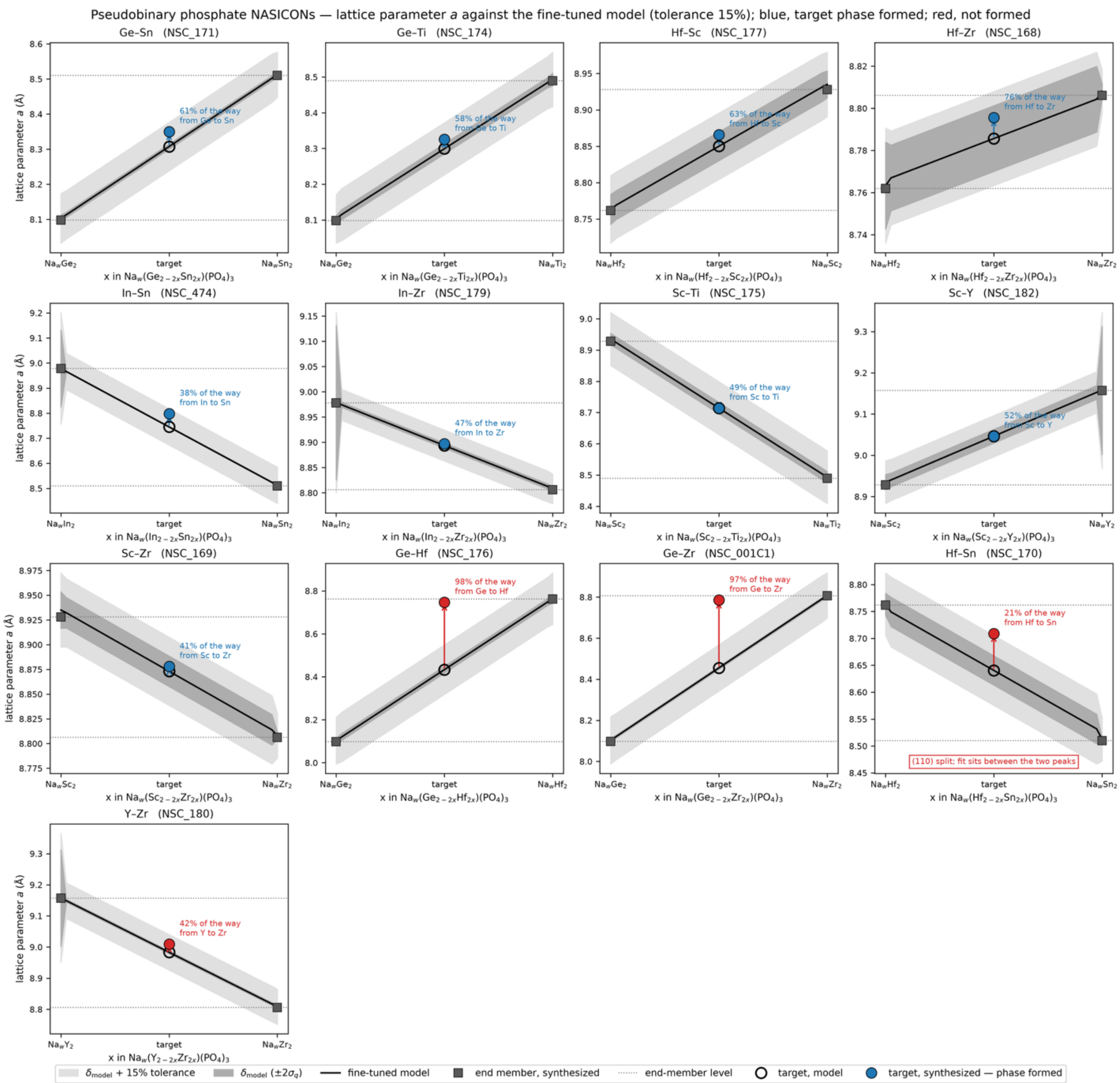

**Supplementary Figure 14.** Pseudobinary sections in the $a$ lattice parameter for the thirteen phosphate pairs. Solid line, fine-tuned model; dark band, $\delta_{\mathrm{model}}$; light band, $\delta_{\mathrm{model}}$ plus 15% of the model-predicted end-member range; squares with dotted levels, measured end members; open circle, model value at the equimolar target; filled circle, refined value, blue where the target phase formed and red where it did not. Percentages are $x(q) \times 100$ measured from the left, so an ideal equimolar product reads 50%. $NaGeHf(PO_4)_3$ and $NaGeZr(PO_4)_3$ read 98% and 97%, placing them at the Hf and Zr end members. $NaSnHf(PO_4)_3$ is marked for its split (110) reflection.

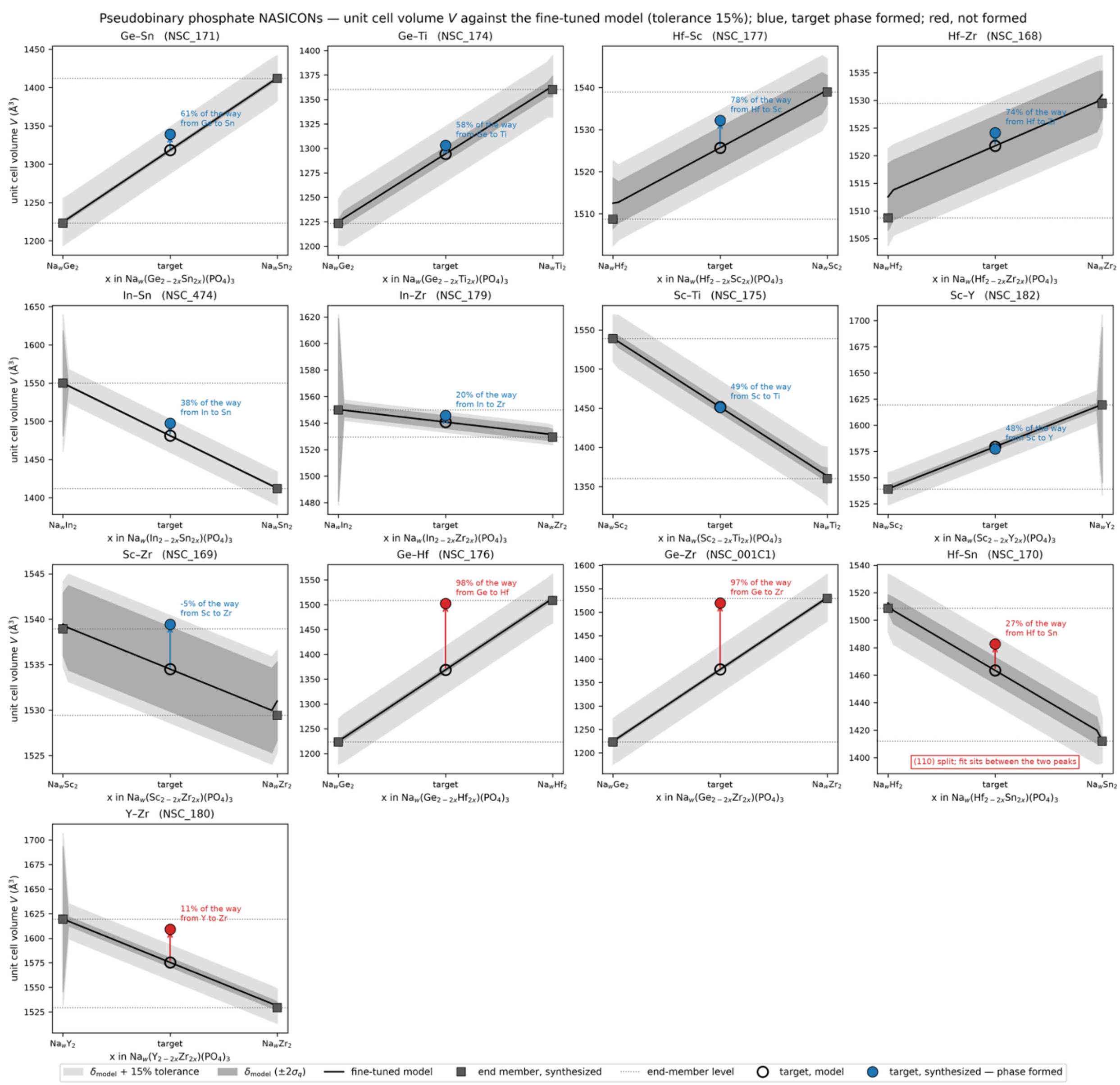

**Supplementary Figure 15.** As Supplementary Figure 14, for the unit-cell volume V. $Na_2ZrSc(PO_4)_3$ reads -5%, its refined volume falling just outside the end-member interval; that pair spans only 9.5 Å³ against $\delta_{model}$= 4.6 Å³, so $x(V)$ is poorly resolved there.

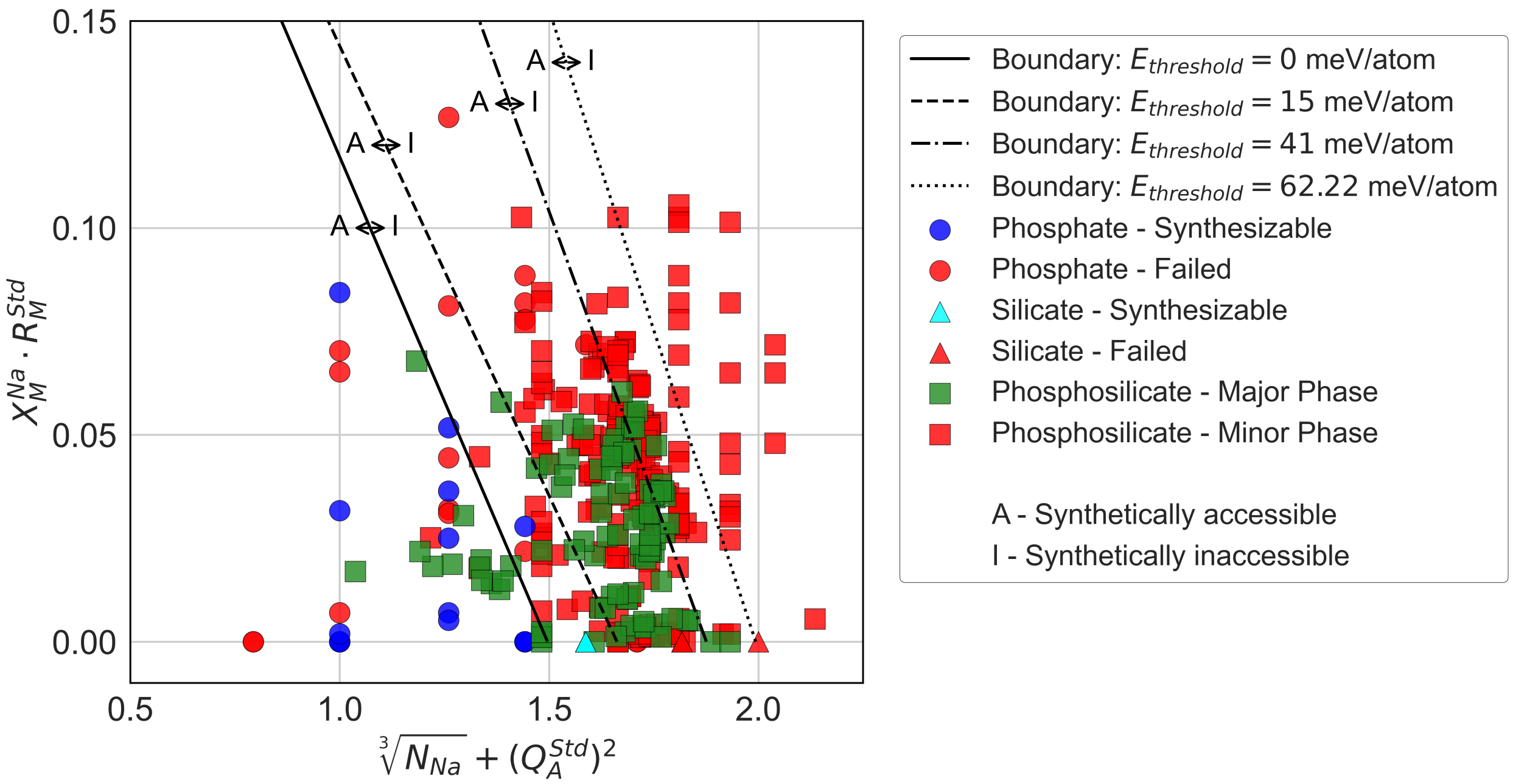

**Supplementary Figure 16.** Stability boundaries in the machine-learned feature space for NASICONs. Distribution of phosphate (circles), silicate (triangles), and phosphosilicate (squares) compositions within the machine-learned stability feature space, defined by descriptors. All compositions are categorized by synthesis success: phosphates and silicates are differentiated by successful (blue/cyan) versus failed (red) syntheses, while phosphosilicates are distinguished by forming a major (green) or minor/no (red) NASICON phase. The plot illustrates four distinct linear decision boundaries corresponding to different energy thresholds ($E_{threshold}$ = 0, 15, 41, and 62.22 meV/atom), which divide the feature space into synthetically accessible (A) and inaccessible (I) domains.

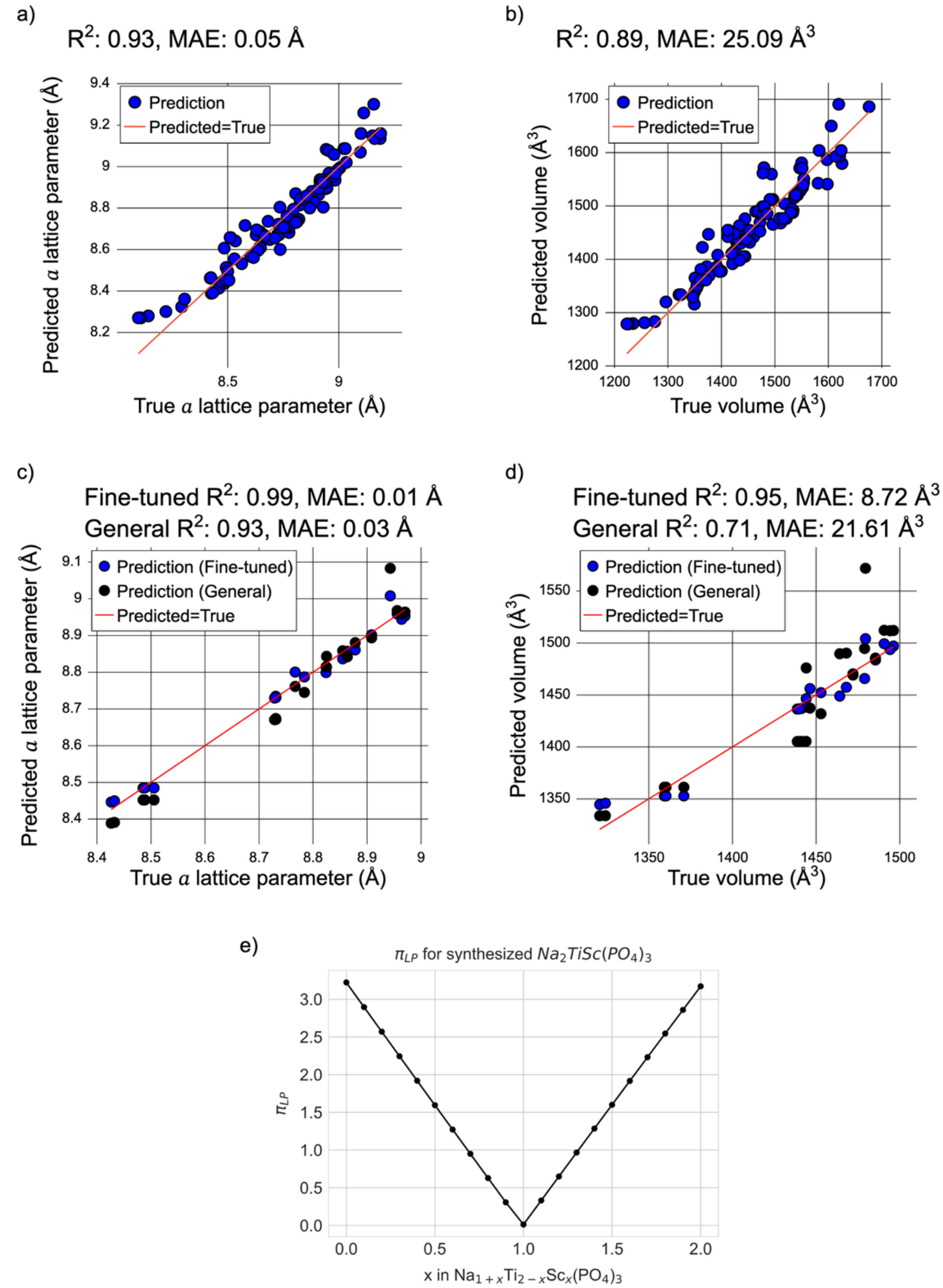

**Supplementary Figure 17. Validation of Vegard's approximation models and $\pi_{LP}$.** Multiple linear regression fits over all structures (ICSD and single-cation phosphate NASICON syntheses) for predicting the unit cell **(a)** $a$ lattice parameter and **(b)** volume. The $R^2$ and mean absolute error (MAE) of each regressor are shown at the top of each plot. Comparison between the fine-tuned and general regressors for NASICON in the Na-Mn-V-Ti-P-O chemical space for (c) $a$ lattice

parameter and (d) volume predictions. The blue (black) dots represent predictions from the fine-tuned (general) regressors, while the red line shows the ideal condition where predicted and true values are equal. **(e)** Visualization of $\pi_{LP}$ for composition variation in $Na_{1+x}Ti_{2-x}Sc_x(PO_4)_3$ evaluated against the experimental volume and lattice parameters of as-synthesized $Na_2TiSc(PO_4)_3$. Each dot represents one generated composition.

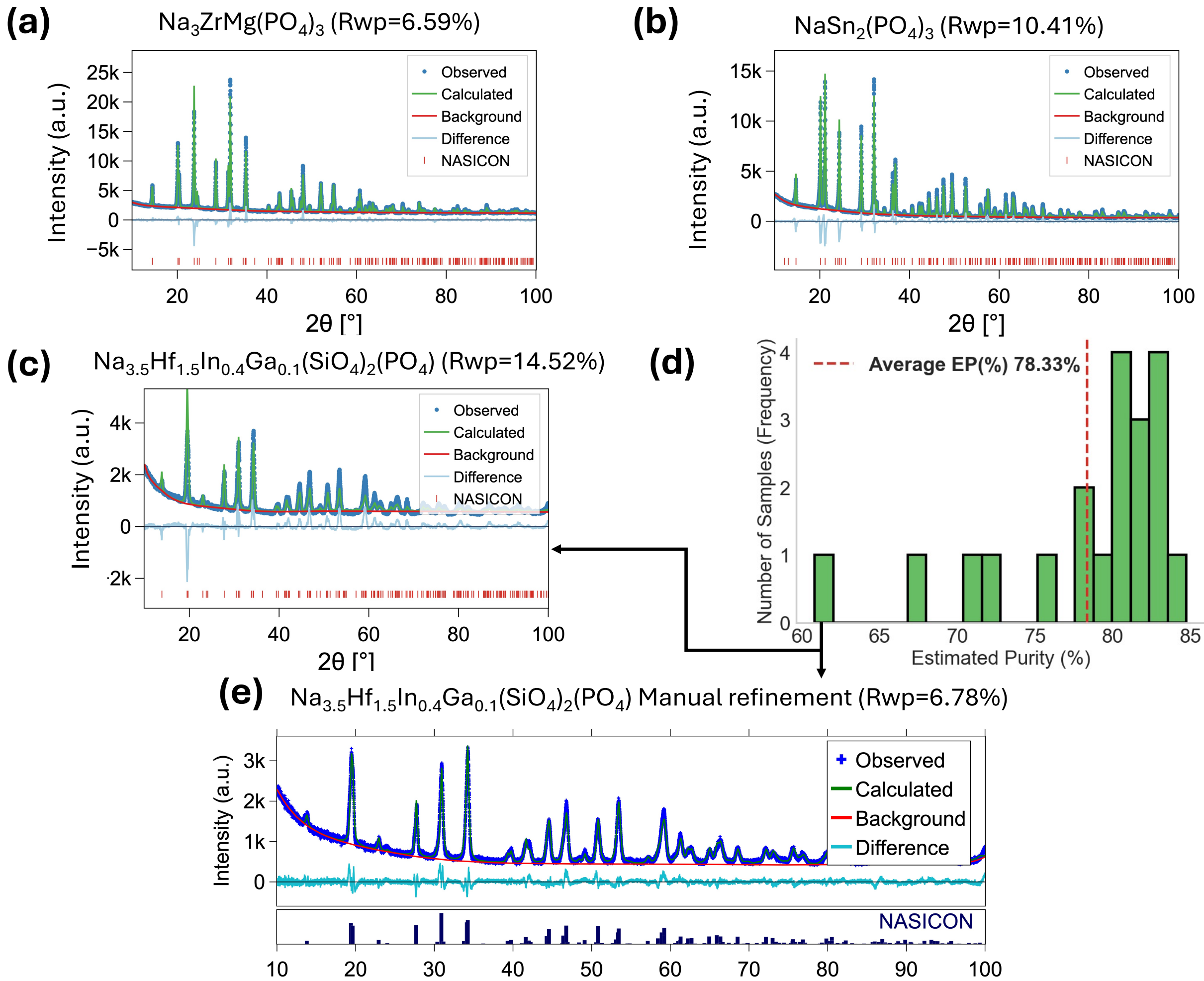

**Supplementary Figure 18. Validation of high-purity samples via X-ray diffraction for Purity Score upper threshold determination. (a-c)** Representative diffractograms and weighted profile R-factors ($R_{wp}$) obtained from the automated Rietveld refinement of samples with no impurity peaks. **(d)** Distribution of Estimated Purity (EP) percentages across the high-purity sample set from the automated refinements; the red dashed line indicates the average EP (78.33%). **(e)** Manual Rietveld refinement of the sample shown in panel **c** ($Na_{3.5}Hf_{1.5}In_{0.4}Ga_{0.1}(SiO_4)_2(PO_4)$). This sample yielded the lowest EP in the automated analysis; the significant improvement in fit quality between the automated (**c**, $R_{wp} = 14.52\%$) and manual (**e**, $R_{wp} = 6.78\%$) refinements highlights the limitations of the automated routine in resolving complex peak profiles. In all plots, experimental data (blue dots in a-c, blue crosses in e), calculated profiles (green lines), background (red lines), and difference (cyan lines) are shown.

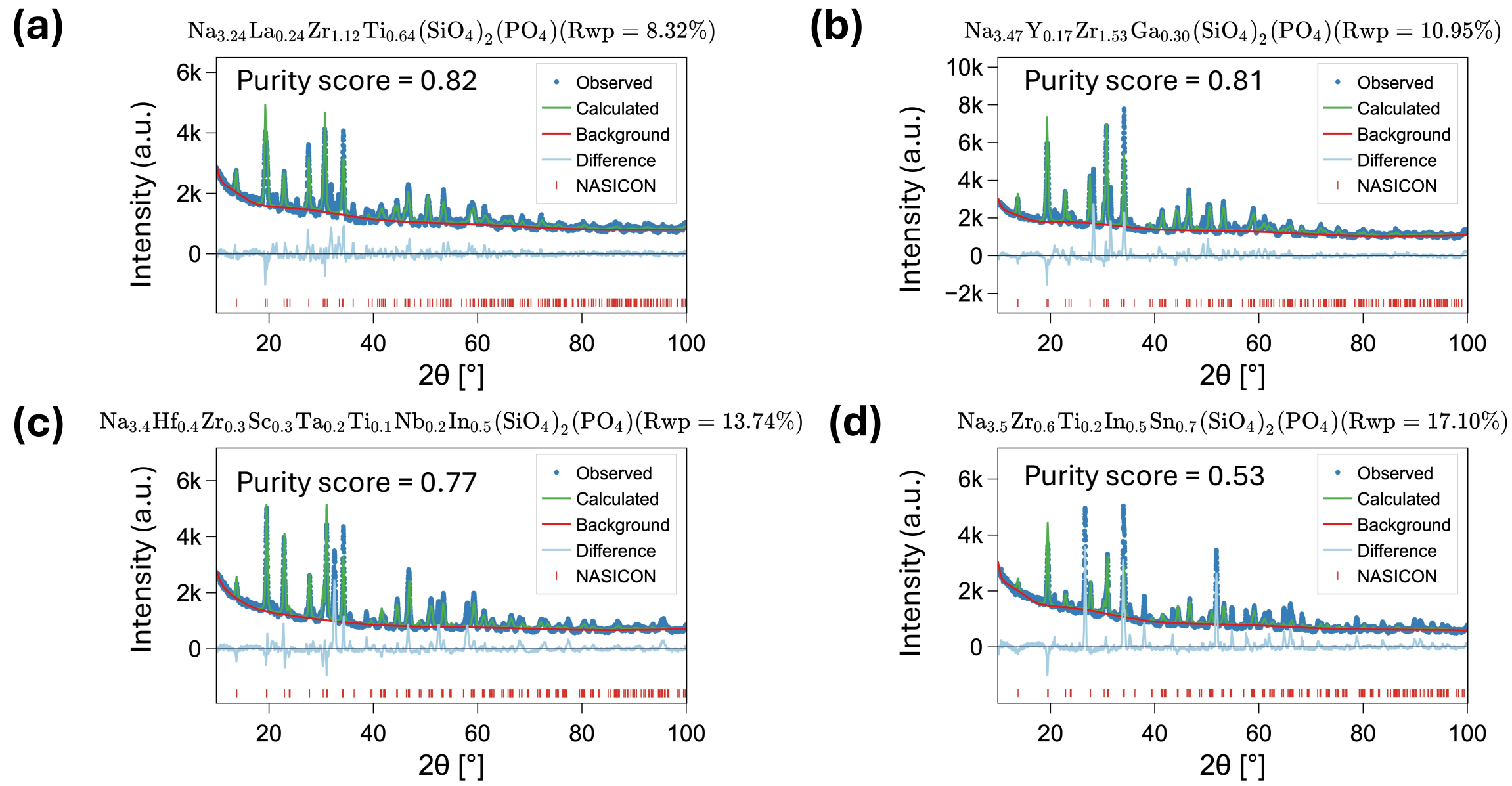

**Supplementary Figure 19. Validation of Purity Score threshold via X-ray diffraction.** Representative diffractograms obtained from automated Rietveld refinement used to calibrate the Purity Score threshold. Panels **(a–c)** display samples identified as having a major NASICON phase, characterized by high Purity Scores (>0.7). Panel **(d)** shows a sample ($Na_{3.5}Zr_{0.6}Ti_{0.2}In_{0.5}Sn_{0.7}(SiO_4)_2(PO_4)$) where the NASICON phase is only a minor phase, resulting in a low Purity Score of 0.53. In all plots, experimental data (blue dots), calculated profiles (green lines), background (red lines), and difference curves (light-blue lines) and NASICON reflection positions (red ticks) are shown.